UNIVERSITY OF BUCHAREST
FACULTY OF MATHEMATICS AND COMPUTER SCIENCE

# Algorithmical Aspects of Some Bio Inspired Operations

**PhD THESIS**

*Supervisor:*
Prof.univ.dr. Victor MITRANA

*PhD Candidate:*
Adrian Marius Dumitran

**Bucharest**
**2015**

## Acknowledgments

First and foremost, I want to thank my advisor Prof. Dr Victor Mitrana, first for accepting me as his PhD student but mostly for his patience and perseverance during the PhD process. Mr. Mitrana spent a lot of time to show me various research directions until we found something that I was really excited about. Later he has given me a lot of precious advice and has worked closely with me to obtain a lot of the results in this thesis. I am really happy I was able to have as a PhD advisor a teacher I have highly admired ever since my first course with him, the second year formal languages course.

My second biggest thank goes to Dr. Florin Manea who has pushed me to be a better researcher. The two months spent in the same office with him at Kiel have hugely helped me in my thesis development. I would like to thank him for the many talks we had, for the advice he has given me but also for the moments where he demanded more of me. I doubt that without Florin's support I would have managed to finish my thesis on time. Keeping to my research period in Kiel I would like to thank Prof. Dr. Dirk Nowotka, the leader of the group, for having me there as a guest in a long research visit. I would also like to acknowledge the entire research group at Kiel for the discussions we had and the ideas we shared. A special thanks in this direction goes to Prof. Dr. James Currie. I would also like to thank Thorsten Ehlers PhD. candidate for his kindness and friendship.

I will forever be thankful to all the members of my Faculty that have given me a place to flourish in the last eleven years. I have always appreciated the spirit of the place as a student and as a member of the teaching stuff. The relaxing and friendly discussions I have had with most of the teachers and students here lured me towards a PhD. I would like to give special thanks to our dean Conf. Dr. Victor Tigoiu for the long discussions we had and the pushes he has given me towards finishing my PhD. In the same direction I would like to thank Prof. Dr. Denis Enachescu, Conf. Dr. Radu Gramatovici, Prof. Dr. Liviu Ornea and Conf. Dr. Rodica Ceterchi. Younger members of the teaching stuff have also given me really good advice and support, and I want to thank Lect. Dr. Denisa Diaconescu, Lect. Dr. Ruxandra Olimid, Lect. Dr. Adela Georgescu and Dr. Robert Mercas for this.

I want to extend my gratitude to the members of my guiding committee: Prof. Dr. Adrian Atanasiu, Conf. Dr. Mugurel Andreica and Conf. Dr. Liviu Dinu for the support they have given me throughout my thesis. The advice they have given me has helped me steer my PhD in the right direction when I was struggling.

I would like to thank my mentor during my internship at Facebook, Dr Kristen Parton, who has taught me a lot and has given me good advice. A big like to my colleagues and friends from Facebook who helped me a lot while I was there and who have given me a lot of motivational talks.

I especially want to thank my family for the unconditional support that they have offered me while writing my thesis but also for the way they have raised me. My brother Daniel has always been a model for me and he has always been there not only with advice but also with technical support.

I would also like to thank my students from who I have learned a lot during the last six years. My students have been my source of inspiration and truly my main motivation. The talks we have had in class and the questions they have asked me made me work harder and learn more. Some of them have even helped me in obtaining some of the results in the thesis, and I would like to stress Adrian Budau's help in this direction. Others like Mihai Calancea, Andrei Grigorean, Vlad and Iulia Duta, Teodor Plop, Dragos Rotaru, Mihai Nitu and so many others have helped improve my algorithmic skills but also took a bit of the burden of teaching from my back when I needed time for my PhD.

I also thank my friends that have supported me during my PhD, some for helping me decide to go towards a PhD, like Radu Dita, some of them for encouraging me to finish my thesis like Toma Ionut but most of them just for simply being there. Special thanks go to the members of my beloved handball team H.C. Victoria, they have been like my second family during all this time.

# Contents

# Chapter 1

# Introduction

## 1.1 Thesis subject in the general context

In most of the thesis we discuss about three operations inspired by biological phenomena, namely the prefix-suffix duplication, the bounded prefix-suffix duplication and the prefix-suffix-square completion operations. Duplication is one of the most frequent and less understood mutations among the genome rearrangements [54]. Roughly speaking, duplication is the process in which a stretch of DNA is duplicated yielding two or more adjacent copies. This process may happen at any position in the chromosome, including its beginning and its end. Actually, the distribution of these tandem repeats varies widely along the chromosomes and some authors consider that approximately 5% of the genome rearrangements are various types of duplications [56].

It is considered that a so-called phylogenetic analysis which might be useful in the investigation of the evolution of species, it would be possible by the study of duplications along the genome, that is by determining the most likely duplication history [59]. The detection of these tandem repeats and algorithms for tandem repeats reconstructing history have received a great deal of attention in bioinformatics [5, 4, 55]. However, a special type of duplications, known as telomeres, appear only at the ends of chromosomes. Generally, telomeres consist of tandem repeats of a small number of nucleotides, specified by the action of telomerase. They are considered to be protective DNA-protein complexes found at the end of eukaryotic chromosomes which stabilize the linear chromosomal DNA molecule [10, 54]. The length of telomeric DNA is important for the chromosome stability: the loss of telomeric repeat sequences may result in chromosome fusion and lead to chromosome instability [52]. In [56] one states that it is a further challenge the sequencing of the 20% of the genome that is formed by repetitive heterochromatin which is implicated in the process of chromosome replication and maintenance.

The interpretation of duplications as a formal operation on words has inspired several works in the areas of Combinatorics on Words and Formal Languages, starting with [7, 28] and continuing in [20, 60, 48] and the references therein. It is worth mentioning that a PhD thesis has been devoted to this topic [47]. These works have been the starting point for this thesis.

In [32] duplications that appear at both ends of the words, namely prefix-suffix duplications, were first considered. The basic motivation of introducing these operations was to mathematically model a special type of duplications within DNA sequences, that appear only at the ends of these sequences, also known as telomeres. Another motivation would be to model the process of generating long terminal repeats (LTRs): identical sequences of DNA that repeat hundreds or thousands of times at either end of some specific DNA sequence. Such sequences are used, for instance, by viruses to insert their genetic material into the host genomes.

In [32] the class of languages that can be defined by the iterative application of the prefix-suffix duplication to a word are investigated and the class is compared to other well studied classes of languages. It is shown that the languages of this class have a rather complicated structure even if the initial word is rather simple: they are already non-context-free as soon as the initial word contains at least two different letters. Algorithms for the membership and distance computation problems are also given. In this context we considered a weaker variant of the prefix-suffix duplications, called the bounded prefix-suffix duplication. The new model allowed us to solve some problems that remained unsolved in [32] and also the new model seems closer to the biochemical reality that inspired the definition of this operation. It seems more practical and closer to the biological reality to consider that the factor added by the prefix-suffix duplication cannot be arbitrarily long. We were able to prove that the duplication language obtained by applying bounded prefix-suffix duplication to a word, or even to a regular language is still a regular language, totally different from the case of the prefix-suffix duplication languages. Based on this result we were also able to solve the common descendent problem for the bounded prefix-suffix duplication operation. We also obtained more efficient algorithms for the membership and distance computation problems.

In the same context of prefix-suffix duplication languages we introduced the prefix-suffix square completion operation [23]. The initial motivation of studying the prefix-suffix duplication were some biological processes that essentially create repetitions at the ends of the genetic sequences; however, the formal operations defined in [32] assumed that such repetitions are created by replicating their root. In [23] we assumed a different point of view: we considered the possibility of creating squares (the simplest type of repetition) at one of the ends of the word by completing a prefix or suffix of the considered sequence to a square. In [23] we were able to solve the membership problem really efficiently and starting from that we were able to compute all primitive ancestors in linear time. In this thesis we computed all the primitive ancestors of a word with respect to the prefix-suffix operation and the bounded prefix-suffix operation, as it seems to be really natural in the context of DNA-sequence operations.

Keeping to the above mentioned operations, we found it interesting to see how our operations can be used to generate infinite words [23]. We first show that the infinite Fibonacci word, the Period-doubling word and the infinite word Thue-Morse can be generated by suffix square completion. This exhibits a property that seems interesting to us: every (infinite) word generated by suffix completion contains squares, but there are (infinite) words generated by this operation (which basically creates squares) that avoid any repetition of (rational) exponent higher than 2. In comparison, we show that the Thue-Morse infinite word cannot be generated by prefix-suffix duplication. However, we show that one can generate an infinite cube-free word by suffix-duplication. This is a weaker version of the result obtained for square completion: every (infinite) word generated by suffix duplication contains squares, but there are (infinite) words generated by this operation that avoid any repetition of integer exponent higher than 2.

We than proceed to investigate the problem of efficiently detecting the existence of repetitive structures occurring at both ends of some sequence. For instance, words that do not end or start with repetitions may model DNA sequences that went through some degenerative process that destroyed the terminal repeats, affecting their stability or functionalities. Moving away from the biological motivation, words that do not start nor end with repetitions seem to be interesting from a combinatorial point of view, as well. Indeed, repetitions-free words (i.e., words that do not contain consecutive occurrences of the same factors) are central in combinatorics on words, stringology, and their applications (see, e.g., [49, 36]); words that do not have repetitive prefixes or suffixes model a weaker, but strongly related, notion. In [24] and in this thesis we show how to efficiently compute the number of prefix-suffix-square free factors in a word, how to count them but also how to compute the longest one. In this thesis we also obtain similar results for the computations of prefix-suffix-square free ancestors in a word (or primitive ancestors).

Keeping to our biological motivation we discuss about gapped repeats and palindromes. Gapped repeats and palindromes have been investigated for a long time (see, e.g., [36, 8, 45, 43, 44, 15, 17]), with motivation coming especially from the analysis of DNA and RNA structures, where tandem repeats or hairpin structures play important roles in revealing structural and functional information of the analyzed genetic sequence (see [36, 8, 43]). More precisely, a gapped repeat (respectively, palindrome) occurring in a word $w$ is a factor $uvu$ (respectively, $u^R vu$) of $w$. The middle part $v$ of such structures is called gap, while the two factors $u$ (or the factors $u^R$ and $u$) are called left and right arms. Generally, the previous works were interested in finding all the gapped repeats and palindromes, under certain restrictions on the length of the gap or on the relation between the arm of the repeat or palindrome and the gap. In this thesis (and in [22]) we solved the problem in three new different settings.

One should note that the investigation we pursue here is not aimed to tackle real biological facts and provide solutions for them. In fact, its aim is to provide a better understanding of the structural properties of strings obtained by the specific operations discussed as well as specific tools for the manipulation of such strings.

## 1.2 Personal Contributions

In the preliminary chapter, in section 2.4, we present a few lemmas that provide information about squares and periods in a word. We show how we can compute in linear time the length of the minimum or maximum square that starts in each position of the array, or it is centered in each position of the array and other similar results. Although some of the results have already been proved [61, 46, 27] we bring a different, and from some perspectives, a more general solution. We give a new proof of these results based on a Lempel-Ziv-like factorisation of the input word and a series of combinatorial remarks on the structure of this factorisation. In order to prove these lemmas we also need to use a special case of the disjoint sets union-find problem. Thus, we firstly prove a lemma that shows the disjoint sets union-find problem works in linear time in our problem settings. These results have been published in [23, 22].

In section 3.1 we give a few results related to bounded prefix-suffix duplications. We show that every nonempty class of languages closed under union with regular languages, intersection with regular languages, and substitution with regular languages, is closed under bounded prefix-suffix duplication. We then provide an algorithm that finds a minimal finite regular language that generates a k-prefix-suffix duplication language (if any). Thus, our algorithm also decides if a language L is a finite bounded-prefix-suffix duplication language. The results in this section have been published in [25, 26].

In section 3.2 we start by showing that we can solve the membership problem in $\mathcal{O}(n \log(k))$ time in the case of bounded prefix or suffix duplications. We then prove a more general result and show that if the membership problem for a language $L$ can be decided in $\mathcal{O}(f(n))$ time, then the membership problem for $PSD_k^*(L)$ can be decided in $\mathcal{O}(nk \log(k) + n^2 f(n))$. Based on the previous two results we show that we can decide if a word $x$, with $|x| = n$, is in $PSD_k^*(w)$ in $\mathcal{O}(n \log(k))$ time if $|w| \geq k$ and in $\mathcal{O}(nk \log(k))$ time otherwise. We heavily rely on dynamic programming to solve the problems in this section. The above results have been published in [25, 26, 23].

We then switch to ancestor (or root) computations and show how to compute the number of ancestors, the shortest ancestor and the longest primitive ancestor of a word in relation to the bounded prefix-suffix duplications in $\mathcal{O}(nk\log(k))$ time. An algorithm outputting all ancestors in $\mathcal{O}(nk\log(k) + |output|)$ time is also given. We then discuss ancestors related problems in the context of the prefix-suffix square completion operation, computing the number of ancestors and the shortest one in linear time. Based on this result we give an algorithm that solves the membership problem for this operation in linear time. Our algorithms are based on the lemmas we proved in the section 2.4 combined with a pattern matching algorithm and a smart use of RMQ queries. Most of the results in this subsection have been published in [23, 25, 26, 24], while others have been published in this thesis for the first time.

Later we show how we can find a common ancestor of two words in $\mathcal{O}(nk\log(k) + n^2)$ time in the case of the bounded prefix-suffix duplication and in linear time for the prefix-suffix square completion. In order to prove the later result we use several tools from linear sorts to suffix arrays and LCP queries, and the lemmas in section 2.4. Using the same tools as before and binary search, we obtain an $\mathcal{O}(n\log(n))$ algorithm for finding the shortest common ancestor of two words with respect to the prefix-suffix square completion. Most of the results in this subsection appear hear for the first time.

In section 3.3 we start by giving a $\mathcal{O}(n\log(k))$ time algorithm for computing the bounded prefix duplication distance based on dynamic programming, deques and a series of combinatorial remarks related to runs in a word. Based on the previous result, we easily show that we can compute the bounded prefix-suffix distance between two words $x$, $w$ with $|x| \geq |w|$, in $\mathcal{O}(nk\log(k))$ if $|w| < k$ and in $\mathcal{O}(n\log(k))$ otherwise. We then give a totally different solution for computing the distance related problems for any of our operations, based on graphs, that has complexity a time complexity of $\mathcal{O}(n^3)$. We then prove that we can improve this to $\mathcal{O}(n^2)$ time for the case of the prefix-suffix square completion operation based on some analytical results. We finish the chapter with a result for computing the distance between two bounded prefix-suffix languages. The results rely on automata, dynamic programming and tries. Most of the results in this section have been published in [23, 25, 26].

In section 4.1 we show how we can generate various infinite words (like the Fibonacci word and the Thue-Morse word) using prefix-suffix square completion. We also prove that we can not generate the Thue Morse word by prefix-suffix duplications but we can generate Stewart's choral sequence using the PSD operation. Our results are mostly based on thorough analysis of the words and are partly based on computer simulations. The results in this section have been published in [23].

In section 4.2 we show how we can construct in linear time a data structure that allows us to verify in constant time if a factor of a word is *SD*, *PD*, or *PSD* primitive. We then give an $\mathcal{O}(n + |output|)$ time complexity algorithm that computes all the prefix-suffix-square free factors of a word. We continue by giving an $\mathcal{O}(n \log n)$ algorithm for counting the number of prefix-suffix square free factors that uses segment trees, besides the union find problem. A similar solution is given for finding the longest primitive ancestor of a word in relation to the prefix-suffix square completion operation. These results are followed by a theorem showing that we can find in linear time the longest prefix-suffix-square factor of a word. The section ends with a few theorems that provide various factorizations of a word. Most of the results of this section have been published in [24], but there are also a few results that have been published here for the first time.

In section 4.3 we look for factors of the form $uvu$ and $u^r vu$ and compute the longest previous factor (*LPrF*) table, and longest previous reverse factor (*LPrF*) table in three different settings. First we take into account the case where we want $v$ to be bounded by constants in two directions. We are able to compute the *LPrF* table in $\mathcal{O}(n)$ time and the LPF table in $\mathcal{O}(n \log n)$ time in this setting. In order to do so we use a mix of techniques from combinatorics on strings to the union find problem, suffix arrays and we also make use of the dictionary of basic factors. We then compute the two tables in the case where the gap is bound only by a lower bound function. We are able to compute both tables in linear time using a similar technique as the one above and also an interesting technique based on dynamic construction of trees combined with LCP queries. The last case we settle is that of long armed repeats and palindromes. Obtaining solutions of complexity $\mathcal{O}(n + |output|)$ time and $\mathcal{O}(n)$ time. The tools we use in this section are: various existing results on runs and long armed repeats and palindromes in words combined with the lemmas in section 2.4. The results in this section have been published in [22].

# Chapter 2

# Preliminaries

In this chapter we define most of the notions used in this thesis. The chapter is divided into four sections:

- The first section covers basic definitions about finite and infinite words, factors, periods and runs but also includes definitions on finite state machines and Turing machines.
- The second section introduces several operations on strings inspired from DNA biochemistry.
- The third section covers details about the various data structures used in the thesis.
- The forth and last section presents a few of the lemmas used throughout the thesis.

## 2.1 Finite and infinite words

We start with a few classical definitions on words. For further details one can check for example [21]. An *alphabet* is a finite and nonempty set of symbols. The cardinality of a finite set $V$ is written $|V|$. Any finite sequence of symbols from an alphabet $V$ is called a *word* over $V$. The set of all words over $V$ is denoted by $V^*$ and the *empty word* is denoted by $\lambda$; also $V^+$ is the set of non-empty words over $V$, $V^k$ is the set of all words over $V$ of length $k$, while $V^{\leq k}$ is the set of all words over $V$ of length at most $k$. Given a word $w$ over an alphabet $V$, we denote its length by $|w|$, while $|w|_a$ denotes the number of occurrences of the letter $a$ in $w$.

If $w = xyz$ for some $x, y, z \in V^*$, then $x, y, z$ are called *prefix, factor, suffix,* respectively, of $w$.

For a word $w$, $w[i..j]$ denotes the factor of $w$ starting at position $i$ and ending at position $j$, where $1 \leq i \leq j \leq |w|$; by convention, $w[i..j] = \lambda$ if $i > j$. If $i = j$, then $w[i..j]$ is the $i$-th letter of $w$ which is simply denoted by $w[i]$.

A *period* of a word $w$ over $V$ is a positive integer $p$ such that $w[i] = w[j]$ for all positions $i$ and $j$ with $i \equiv j \pmod{p}$. By $per(w)$ (called *the period of* $w$) we denote the smallest period of $w$. If $per(w) < |w|$ and $per(w)$ divides $|w|$, then $w$ is a *repetition*; otherwise, $w$ is called *primitive*.

A *primitively rooted square* is a word $w$ having the form $xx$ for some primitive word $x$.

A word $w$ with $per(w) \leq \frac{|w|}{2}$ is called *run*; a run $w[i..j]$ (so, $p = per(w[i..j]) < \frac{j-i+1}{2}$) is maximal iff it cannot be extended to the left or right to get a word with the same period $p$, i.e., $i = 1$ or $w[i-1] \neq w[i+p-1]$, and, $j = n$ or $w[j+1] \neq w[j-p+1]$. In [42] it is shown that the number of maximal runs of a word is linear and their list (with a run $w[i..j]$ represented as the triple $(i, j, per(w[i..j]))$ can be computed in linear time.

The *exponent* of a maximal run $w[i..j]$ occurring in $w$ is defined as $\frac{j-i+1}{per(w[i..j])}$; in [2] it is shown that the sum of the exponents of maximal runs in a word of length $n$ is upper bounded by $n$.

The dictionary of basic factors [16] of a word $w$(DBF for short) is a data structure that labels the factors $w[i..i+2^k-1]$ (called basic factors), for $k \geq 0$ and $1 \leq i \leq n-2^k+1$, such that every two identical factors of $w$ get the same label and we can retrieve the label of any basic factor in $\mathcal{O}(1)$ time. The DBF of a word of length $n$ is constructed in $\mathcal{O}(n \log n)$ time.

Note that a basic factor $w[i..i+2^k-1]$ occurs either at most twice in any factor $w[j..j+2^{k+1}-1]$ or the occurrences of $w[i..i+2^k-1]$ in $w[j..j+2^{k+1}-1]$ form a run of period $per(w[i..i+2^k-1])$ and the positions where $w[i..i+2^k-1]$ occurs in $w[j..j+2^{k+1}-1]$ for an arithmetic progression of ratio $per(w[i..i+2^k-1])$ (see [41]). Hence, the occurrences of $w[i..i+2^k-1]$ in $w[j..j+2^{k+1}-1]$ can be presented in a compact manner: either at most two positions, or the starting position of the progression and its ratio. For $c \geq 2$, the occurrences of the basic factor $w[i..i+2^k-1]$ in $w[j..j+c2^k-1]$ can be presented in a compact manner: the positions (at most $c$) where $w[i..i+2^k-1]$ occurs isolated (not inside a run) and/or at most $c$ maximal runs that contain the occurrences of $w[i..i+2^k-1]$, each run having period $per(w[i..i+2^k-1])$.

We also use the $Lempel-Ziv$-factorisation and its variants. Let $w \in \Sigma^*$ be a word. The $Lempel-Ziv$-factorisation of $w$ is defined as follows (see [50]). We factorize $w = u_1 \cdots u_r$ if the following hold for all $i \geq 1$:

- If a letter $a$ occurs in $w$ immediately after $u_1 \cdots u_{i-1}$ and $a$ did not appear in $u_1 \cdots u_{i-1}$ then $u_i = a$
- Otherwise, $u_i$ is the longest word such that $u_1 \cdots u_{i-1}u_i$ is a prefix of $w$

and $u_i$ occurs at least once in $u_1 \cdots u_{i-1}$.

Variations of the Lempel-Ziv-factorisation allow overlap of the factors. The Lempel-Ziv-factorisation of a word $w$ and its variants can be computed in linear time (see, e.g., [15]).

**Example 2.1.1** *The word* $w = abbaabbbaaabab$ *has the factorization* $S = w_1...w_8$ *where* $w_1 = a, w_2 = b, w_3 = b, w_4 = a, w_5 = abb, w_6 = baa, w_7 = ab$ *and* $w_8 = ab$.

The infinite Thue-Morse word $\mathbf{t}$ is defined as $\mathbf{t} = \lim_{n\to\infty} \phi_t^n(0)$, for the morphism $\phi_t : \Sigma_2^* \to \Sigma_2^*$ where $\phi_t(0) = 01$ and $\phi_t(1) = 10$ (see [57, 58]). It is well-known (see, for instance, [49]) that $\mathbf{t}$ does not contain any factor of the form $xyxyx$ (overlaps). Consequently, the infinite word $\mathbf{t}$ does not contain any repetition of (rational) exponent greater than 2. The first Thue-Morse words are: $0, 01, 0110, 01101001...$

The infinite Fibonacci word $\mathbf{f}$ is defined as $\mathbf{f} = \lim_{n\to\infty} f_n$, where $f_0 = 0$, $f_1 = 01$, and $f_k = f_{k-1}f_{k-2}$ for $k \geq 2$ (see, e.g., [1, 49]). The Fibonacci word contains cubes, but it does not contain repetitions of exponent 4; it is also a Sturmian word [51]. Finally, $\mathbf{f} = \lim_{n\to\infty} \phi_f^n(0)$, where $\phi_f(0) = 01$ and $\phi_f(1) = 0$. The first Fibonacci words are: $0, 01, 010, 01001, 01001010...$

The Period doubling word $\mathbf{d}$ is defined as $\mathbf{d} = \lim_{n\to\infty} \phi_d^n(0)$, for the morphism $\phi_d : \Sigma_2^* \to \Sigma_2^*$ where $\phi_d(0) = 01$ and $\phi_d(1) = 00$. This infinite word was studied both in combinatorics on words and, just like the Fibonacci word, in quasicrystal spectral theory (see, [19, 1, 18, 29] and the references therein). To generate this sequence, note that to obtain $\phi_d^n(0)$ we can concatenate $\phi_d^{n-1}(0)$ to itself and then change the last digit (1 becomes 0 and vice versa). Interestingly, this sequence is obtained by replacing the 2 letters by 0 in the square-free Hall sequence $\mathbf{h} = \lim_{\mathbf{n}\to\infty}\phi_h(2)$, with $\phi_h(2) = 210, \phi_h(1) = 20, \phi_h(0) = 1$ (see [37]). This sequence has label $A096268$ in The Online Dictionary of Integer Sequences. The first words in the Period doubling sequence are: $0, 01, 0100, 01000101...$

Stewart's choral sequence $\mathbf{s}$ is defined as $\mathbf{s} = \lim_{n\to\infty} s_n$, where $s_0 = 0$ and $s_{k+1} = s_k s_k s_k^*$ for $k \geq 0$, with $s_k^*$ being a copy of $s_k$ with the middle letter changed from 0 to 1 or vice versa. It is known that $\mathbf{s}$ does not contain cubes. For more details, see The Online Dictionary of Integer Sequences, where this sequence has the label $A116178$. The beginning of Stewarts choral sequence is: $0, 0, 1, 0, 0, 1, 0, 1, 1....$

See [49] for further details on the above definitions.

We will now define a few variants of finite state automata and grammars used in the thesis.

A finite state automata(FSA) or finite state machine is a 5-tuple $A = (Q, V, \delta, q_0, F)$, where $Q$ is a finite set of states, $V$ is the alphabet of the automaton, $\delta$ is the transition function $\delta : Q * V \to P(Q)$, $q_0$ is the start state, with $q_0 \in Q$, and $F$ is the set of final or accepting states of the automata with $F \subseteq Q$. If $\delta : Q * V \to Q$ then the automata is called a deterministic finite automata(DFA), otherwise it is called a non-deterministic finite automata(NDFA).

A prefix grammar is a type of grammar where the rules are applied such that only prefixes are rewritten. The prefix grammars are equivalent with FSA and regular grammars. Formally, a prefix grammar is a 3-tuple $G = (V, S, P)$ where:

- $V$ is the finite alphabet
- $S$ is the finite set of base words over $V$
- $P$ is a set of production rules of the form $x \to y$ where $x$ and $y$ are words over $V$

Given two words $x$ and $y$ we say that we can derive $y$ from $x$ (and write $x \Rightarrow_G y$) in one step, if there are words $u, v, w$ such that $x = vu, y = wu$, and $v \to_G w \in P$. The language of $G$, denoted $L(G)$, is the set of strings derivable from $S$ in zero or more steps: formally, the set of strings $w$ such that for some $s \in S$, $s \Rightarrow^* w$.

**Example 2.1.2** $G = (V = \{0, 1\}, S = \{01, 10\}, P = \{0 \to 010, 10 \to 100\})$ *describes the language defined by the regular expression* $01(01)^* \cup 100^*$.

In this thesis we show several results of algorithmic nature. All the time complexity bounds we give hold for the Random Access Machine with logarithmic memory-word size [35]. In the algorithmic problems we approach, we are usually given as input one or more words. These words are assumed to be over an integer alphabet; that is, if $w$ is the input word, and has length $n$, then we assume that its letters are integers from the set $\{1, \ldots, n\}$; this assumption is realistic: see the discussion in [38]. If the input of our problems is a language, then we assume that this language is specified by a procedure deciding it; for example a regular language is specified by a DFA accepting it.

## 2.2 Bio-Inspired operations on strings

In the thesis we cover a series of bio-inspired operations. In this section we focus on the formalization of these operations. All of the operations rotate around squares and duplications which we now define.

We say that the pair ${}_w(i, p)$ is a *duplication* in $w$ starting at position $i$ in $w$ if $w[i..i+p-1] = w[i+p..i+2p-1]$. Analogously, the pair $(i, p)_w$ is a *duplication* in $w$ ending at position $i$ in $w$ if $w[i-2p+1..i-p] = w[i-p+1..i]$. In both cases, $p$ is called the length of the duplication. Furthermore, the pair ${}_w(i, p)_w$ is a duplication in $w$ having the middle at position $i$ in $w$ if $w[i-p+1..i] = w[i+1...i+p]$.

Duplications are widely discussed in the thesis and in the following papers [25, 26, 24, 23], where most of the results related to duplication that appear in this thesis have been published.

The first operation covered in the thesis is the operation named prefix-suffix duplications. The operation was introduced in [32] as follows:

Given a word $x$ over an alphabet $V$, we define:

- *prefix duplication*, namely $PD(x) = \{ux \mid x = uy \text{ for some } u \in V^+\}$.
- *suffix duplication*, namely $SD(x) = \{xu \mid x = yu \text{ for some } u \in V^+\}$.
- *prefix-suffix duplication*, namely $PSD(x) = PD(x) \cup SD(x)$.

In [25, 26] we discuss a restricted variant of the prefix-suffix duplications called bounded prefix-suffix duplications or k-prefix-suffix duplications. The bounded prefix-suffix duplications are defined in a similar fashion to prefix-suffix duplications, but the length of the prefix/suffix that is to be duplicated is bounded by a constant.

Formally, given a word $x \in V^*$ and a positive integer $k$, we define:

- *the k-prefix duplication*, namely $PD_k(x) = \{ux \mid x = uy \text{ for some } u \in V^+, |u| \leq k\}$.
- *the k-suffix duplication*, namely $SD_k(x) = \{xu \mid x = yu \text{ for some } u \in V^+, |u| \leq k\}$.
- *the k-prefix-suffix duplication*, namely $PSD_k(x) = PD_k(x) \cup SD_k(x)$

The last operation that we discuss in the thesis is the prefix-suffix square completion operation defined in [24].
Given a word $w \in \Sigma^*$, we define:

- *prefix square completion*: $PSC(w) = \{xw \mid w = yxyw', \text{ with } x \in \Sigma^+, y \in \Sigma^*\}$.
- *suffix square completion*: $SSC(w) = \{wx \mid w = w'yxy, \text{ with } x \in \Sigma^+, y \in \Sigma^*\}$.
- *prefix-suffix square completion*: $PSSC(w) = PSC(w) \cup SSC(w)$.

These operations are naturally extended to languages $L$ for any $\Theta \in \{PD, SD, PSD, PD_k, SD_k, PSD_k, PSC, SSC, PSSC\}$ with $k \geq 1$, by

$$\Theta(L) = \bigcup_{x \in L} \Theta(x)$$

We further define, for $\Theta \in \{PD, SD, PSD, PD_k, SD_k, PSD_k, PSC, SSC, PSSC\}$ with $k \geq 1$, its iteration:

$$\Theta^0(x) = \{x\}, \Theta^{n+1}(x) = \Theta^n(x) \cup \Theta(\Theta^n(x)),\ \text{for } n \geq 0,\ \Theta^*(x) = \bigcup_{n\geq 0} \Theta^n(x).$$

The prefix-suffix duplication distance between two words $w$ and $x$ is defined as follows:

$$\pi(w,x) = \begin{cases} \text{the minimum number } \ell \text{ such that } w \in PSD^\ell(x) \text{ or } x \in PSD^\ell(w), \\ \infty,\ \text{if } w \notin PSD^*(x) \text{ and } x \notin PSD^*(w). \end{cases}$$

Note that if $\pi(x,w)$ is defined, then $0 \leq \pi(x,w) \leq \max\{||x| - |w||\}$.

We stress from the very beginning that the function $\pi$, applied on pairs of words, is not a distance function in the strict mathematical sense, since it does not necessarily satisfy the triangle inequality. It can be rather seen as a similarity measure between words, or, if we consider our biological motivation, as a measure that tells us how many evolution steps are needed to transform a chromosome into the other.

The bounded prefix-suffix duplication distance and prefix-suffix square completion distance between two words $w$ and $x$ are defined in a similar fashion:

$$\pi_k(w,x) = \begin{cases} \text{the minimum number } \ell \text{ such that } w \in PSD_k^\ell(x) \text{ or } x \in PSD_k^\ell(w), \\ \infty,\ \text{if } w \notin PSD_k^*(x) \text{ and } x \notin PSD_k^*(w). \end{cases}$$

$$PSSCD(w,x) = \begin{cases} \text{the minimum number } \ell \text{ such that } w \in PSSC^\ell(x) \text{ or } x \in PSSC^\ell(w), \\ \infty,\ \text{if } w \notin PSSC^*(x) \text{ and } x \notin PSSC^*(w). \end{cases}$$

Furthermore, for $\Theta \in \{PD, SD, PSD, PD_k, SD_k, PSD_k, PSC, SSC, PSSC\}$ with $k \geq 1$, we have:

$$\Theta^*(L) = \bigcup_{x \in L} \Theta^*(x) \quad .$$

Prefix-suffix duplication languages are defined as follows in [32]: a language $L \subseteq V^*$ is called a *prefix-suffix duplication language* if $L = PSD^*(x)$ for some $x \in V^*$.

Starting from the above, we define bounded prefix-suffix duplication languages and prefix-suffix square completion languages. A language $L \subseteq V^*$ is called a *bounded prefix-suffix duplication language* if $L = PSD_k^*(x)$ for some $x \in V^*$ and $k > 0$. A language $L \subseteq V^*$ is called a *prefix-suffix square completion language* if $L = PSSC^*(x)$ for some $x \in V^*$.

The previous definitions are inspired by (arbitrary) duplication languages, the case where duplications of arbitrary factors within the word are permitted, see [20] for more details about duplication languages.

We say that a word $x$ is an *ancestor* or a *root* of $y$ in relation to $\Theta$ if $y \in \Theta^*(x)$, where $\Theta \in \{PD, SD, PSD, PD_k, SD_k, PSD_k, PSC, SSC, PSSC\}$. In the same context $x$ is a *predecessor* of $y$. We have chosen to use both ancestor and root to define the same concept as the two definitions are both used in literature. We mostly use ancestor in the third chapter where we look for common ancestors of two words, and use the term root in the forth chapter where we discuss about the roots of a word.

Furthermore we say that a word $x$ is *primitive* in relation to $\Theta$ if there is no word $y$ such that $x \in \Theta(y)$.

We also combine the two previous notations and discuss of *primitive ancestors* or *roots* of a word: $x$ is a primitive ancestor (or root) of $y$ if $y \in \Theta^*(x)$ and there is no element $z \neq x$ such that $x \in \Theta(z)$.

Finally, we say that a right-infinite word $w$ is generated by the operation $\Theta$ and we write $w \in \Theta^\omega$, where $\Theta \in \{PD, SD, PSD, PD_k, SD_k, PSD_k, PSC, SSC, PSSC\}$, if there exists a sequence of finite words $(w_n)_{n \in I\!N}$ such that: $w_i$ is a prefix of $w$ and $w_{i+1} \in \Theta(w_i)$ for all $i \in I\!N$.

**Example 2.2.1** *Consider the words $w_0 = ab$, $w_{2i+1} = w_{2i}w_{2i}$ and $w_{2i+2} = w_{2i+1}b$ for $i \geq 0$. Let $w = \lim_{n\to\infty} w_n$. Clearly, $w_{2i+1} \in SD(w_{2i})$, and, as all the words $w_i$ end with $b$, $w_{2i+2} \in SD^*(w_{2i+1})$. Therefore, $w \in SD^*(w_0) \subseteq PSD^*(w_0)$.*

## 2.3 Data Structures

We continue our series of preliminaries with details about the data structures that we use inside the thesis. They range from suffix and LCP arrays to union find structures and segment trees.

The suffix array of $w$ allows us to construct in linear time a list $\mathcal{L}$ of the suffixes $w[i..n]$ of $w$ ordered lexicographically [15].

**Example 2.3.1** *For the word* banana *the sorted suffixes along with the position where they start are showed in the following table.*

| | |
|---|---|
| *a* | *6* |
| *ana* | *4* |
| *anana* | *2* |
| *banana* | *1* |
| *na* | *5* |
| *nana* | *3* |

*Then the suffix array of the word banana is $SA = \{6, 4, 2, 1, 5, 3\}$.*

We can obtain a list of both suffixes of $w$ but also of the suffixes of $w[1..i]^R$, thus the prefixes of $w$, by applying a suffix array construction algorithm on the word $w0w^R$, where $0 \notin w$. Generally, we denote by $Rank[i]$ the position of $w[i..n]$ in the ordered list of these factors $\mathcal{L}$, and by $Rank_R[i]$ the position of $w[1..i]^R$ in $\mathcal{L}$.

Starting from the suffix array we can compute in linear time for a word $u = u_1...u_n$ over an integer alphabet the LCP array [38, 36] that allows us to retrieve in constant time the length of the longest common prefix of any two suffixes $u[i..n]$ and $u[j..n]$ of $u$. We note such a query as $LCP_u(i, j)$ (the subscript $u$ is omitted when there is no danger of confusion). Similarly, we can build structures allowing us to retrieve in constant time the length of the longest common suffix of any two prefixes $u[1..i]$ and $u[1..j]$ of $u$, denoted $LCS_u(i, j)$.

**Example 2.3.2** *Given the word $w$ = abracadabra, $LCP_w(1, 4) = 1$("a"), $LCP_w(2, 9) = 3$("bra").*

**Remark 2.3.1** *Given a word $w$ of length $n$ and $\ell$ a divisor of $n$ we can use one LCP query to check in constant time whether $w = x^k$, where $|x| = \ell$ and $k = \frac{n}{\ell}$. Indeed, $w = x^k$ if and only if $LCP(1, \ell+1) = n-\ell+1$. Also, the longest prefix of $w$ that is a power of $x$ is obtained as the longest prefix $w'$ of $w$ whose length is divisible by $\ell$ and $|w'| \leq LCP(1, \ell+1) = n - \ell + 1$.*

A trie [39] is a tree based data structure used to maintain strings in an orderly fashion. A node will represent all of the letters from the node to the root and all of its descendants will have this letters as a common prefix. Insertion, search and deletion can be implemented in time $\mathcal{O}(n)$, where $n$ is the length of the word that needs to be inserted/searched/deleted.

**Example 2.3.3** *Trie example for the words "banana", "anana", "nana", "ana", "na", "a" and "and".*

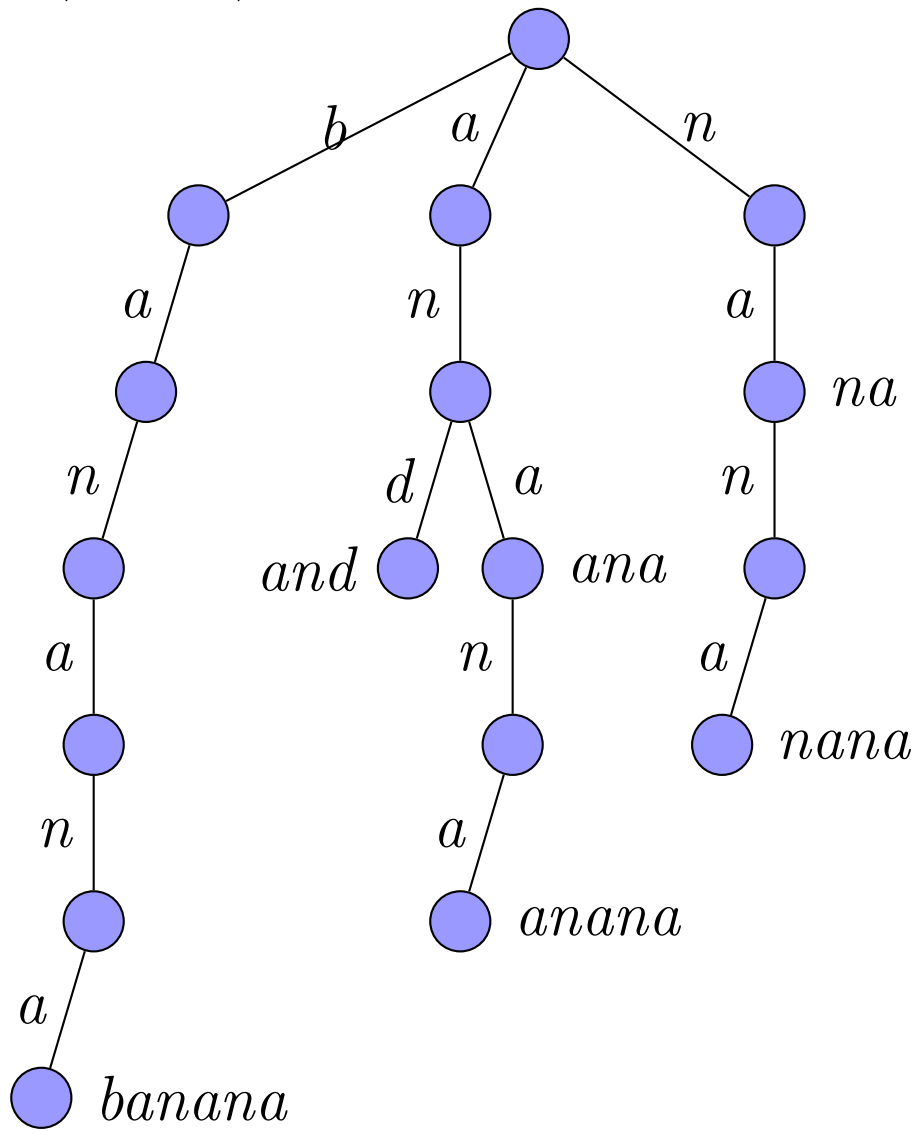

The suffix tree combines the suffix arrays with tries, resulting a data structure that has all the suffixes of a string inserted in a tree. With the help of road compression the suffix tree can be constructed in linear time [61].

**Example 2.3.4** *Suffix tree example for the word "banana".*

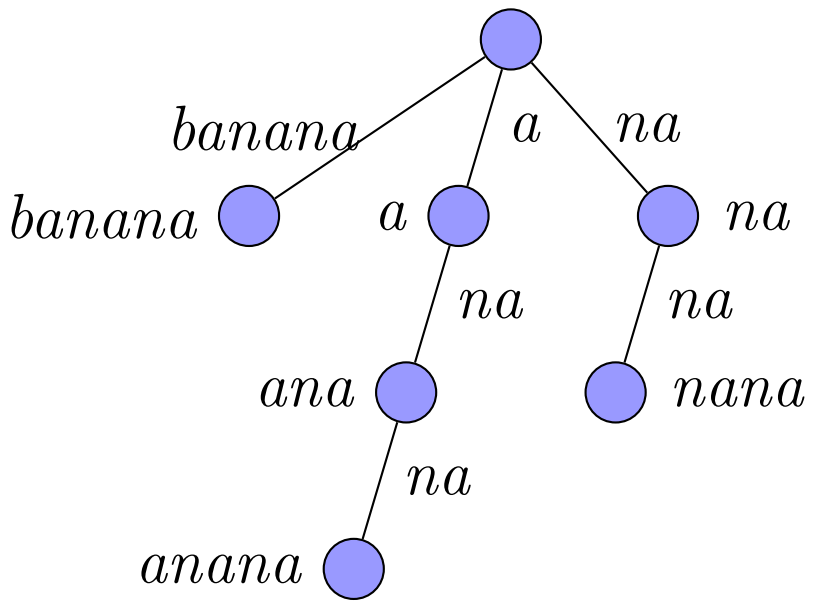

Another important tool for proving our results is an efficient solution for the *disjoint set union-find* problem. This problem asks to maintain a family consisting initially of disjoint singleton sets from the universe $U = [1, n + 1)$ (short for $\{1, \ldots, n\}$) so that given any element we can locate its current set and return the minimal (and/or the maximal) element of this set (operation called *find-query*) and we can merge two disjoint sets into one (operation called *union*). You can read more about the disjoint set union-find problem in [11, 31].

We will now shortly describe the range minimum query data structure [3]. Given an array of $n$ integers $T[]$, we can produce in $\mathcal{O}(n)$ time several data-structures for this array, allowing us to answer in constant time the queries $RMQ(i, j)$ and $posRMQ(i, j)$; asking the minimum value and, respectively, its position among T[i], T[i+1], ..., T[j]. Of course, we can obtain similar arrays for maximum range queries.

**Example 2.3.5** *Given the array* $A = \{1, 4, 9, -2, 7, 6, 14, 3, 8, 5\}$*,* $RMQ(1, 5) = -2$ *and* $posRMQ(1, 5) = 4$*.*

Another data structure that we use in the thesis is the segment tree[6].. Given an interval $U = [1, n]$, segment trees enable us to process the following operations:

- *sum(i,j)* (computes the sum $U[i]+, U[i + 1]+, ..., U[j]$ , if $i = j$ we get the value for position $i$)

- *add(i,j,val)* (with this operation we can increase the value for each element in the interval [i...j] by a value *val*)

in $\log(n)$ time per operation.

**Example 2.3.6** *Given the array* $A = \{1, 4, 9, -2, 7, 6\}$*, after* $add(1, 4, 3)$ *we obtain* $A = \{4, 7, 12, 1, 7, 6\}$*. In this setting* $sum(2, 5)$ *will return the value 27.*

## 2.4 Lemmas

In this section we present a few lemmas that will be used all throughout the thesis. The lemmas are closely related and are a common basis for many of the main results. We show how we can compute in linear time and space a lot of information related to squares such as the length of the minimum or maximum square that starts in each position of the array. Computing the minimum square was settled in [61], but only for inputs over constant size alphabets. In [33] the result was shown for integer alphabets, using a proof based on a data-structure (see [31]) designed to maintain efficiently a family of disjoint sets under union and find operations (however, the proof was not published in the conference version of the paper [33]). Here, we give a new proof of this result based on a Lempel-Ziv-like factorisation of the input word and a series of combinatorial remarks on the structure of this factorisation. The lemmas in this section are from [23, 22].

Our lemmas rely on an efficient solution for the *disjoint set union-find* problem. In our problems we assume to know from the beginning the pairs of elements whose corresponding sets can be joined (that is, we know a scheme of the unions that we are allowed to make). A data-structure fulfilling the above requirements in our setting can be constructed in $\mathcal{O}(d)$ time, where $d$ is the number of singleton sets to be united, such that both operations (find-query and union) can be implemented in amortized constant time [31] (that is, constructing the structure and performing a sequence of $m$ operations takes $\mathcal{O}(d+m)$ time). Using this data structure, we can show the following lemma, proved in [23].

**Lemma 2.4.1** *Let $U=[1,n+1)$. We are given $k$ intervals $I_1, I_2, \ldots, I_k$, with $I_j=[a_j,b_j)$ where $1 \leq a_j \leq n$ and $1 \leq b_j \leq n+1$, for all $1 \leq j \leq k$. Also, for each $j \leq k$ we are given an integer $g_j \leq n$, the weight of the interval $I_j$. We can compute in $\mathcal{O}(n+k)$ time the values $H[i]=\max\{g_j \mid i \in I_j\}$ (or, alternatively, $h[i]=\min\{g_j \mid i \in I_j\}$) for all $i \leq n$.*

*Proof.* Let us first show how the array $H[\cdot]$ is computed.

We first sort the intervals $I_1, I_2, \ldots, I_k$ with respect to their starting positions $a_j$, for $1 \leq j \leq k$. Then, we produce for each $i=1$ to $n$ the list of the intervals $I_j$ that have $g_j=i$ (again, sorted by their starting positions). Using radix-sort we can achieve this in time $\mathcal{O}(n+k)$. Further, we set up a disjoint set union-find data structure for the universe $U=[1,n]$. Initially, the sets in our structure are the singletons $\{1\}, \ldots, \{n\}$; the only unions that we are allowed to make are between the set containing $i$ and the set containing $i+1$, for all $1 \leq i \leq n-1$. Therefore, we can think that our structure only contains intervals $[a,b)$; initially, $\{i\}=[i,i+1)$. Now, we process the intervals that have weight $i$, for each $i$ from $n$ to 1 (so, considered in decreasing order). So, assume that we process the intervals of weight $i$. At this moment, $U$ is partitioned in some intervals (some of them just singletons). Let $I = [a, b)$ be an interval of weight $i$. Let now $j = a$ (in the following, $j$ will take different values between $a$ and $b$). We locate the interval $[c_1, c_2)$ where $j$ is located; if this is a singleton and $H[j]$ was not already set, then we set $H[j] = i$. Further, unless $j$ is $a$, we merge the interval containing $j$ to that containing $j - 1$ and set $j = c_2$; if $j = a$ we just set $j = c_2$. In both cases, we repeat the procedure while $j < b$. We process in this manner all the intervals $I_j$ with $g_j = i$, in the order given by their starting positions, and then continue by processing the intervals with weight $i - 1$, and so on.

The algorithm computes $H[\cdot]$ correctly. Indeed, we set $H[j] = i$ when we process an interval $I$ of weight $i$ that contains $j$, and no other interval of greater weight contained $j$. To see the complexity of the algorithm we need to count the number of union and find operations we make. First, we count the number of union operations. For this, it is enough to note that for each element $j$ we might make at most 2 unions: one that unites the singleton interval $[j, j + 1)$ to the interval of $j - 1$, which has the form $[a, j)$ for some $a$, and another one that unites the interval of $j + 1$ to the one of $j$. So, this means that we make at most $\mathcal{O}(n)$ union operations. For the find operations, we just have to note that when an interval $I$ is processed the total number of finds is $\mathcal{O}(|I \setminus U| + 2)$, where $U$ is the union of the intervals that were processed before $I$. This shows that the total number of find operations is $\mathcal{O}(k + | \cup_{i=1,k} I_i|) = \mathcal{O}(n + k)$.

By the results of [31], as we know from the beginning the structure (the tree) of the unions that we are allowed to make, our algorithm runs in $\mathcal{O}(n+k)$ time.

The computation of $h[\cdot]$ is similar. The only difference is that we consider the intervals in increasing order of their weight. □

We will first show how to find the longest square centred on each position and will give a brief proof.

**Lemma 2.4.2** *Given a word $w$ of length $n$ we can compute in $\mathcal{O}(n)$ time the values $SC[i] = \max\{|u| \mid u$ is both a suffix of $w[1..i-1]$ and a prefix of $w[i..n]\}$.*

*Proof.* Note that each square $u^2$ occurring in a word $w$ is part of a maximal run $w[i'..j'] = p^\alpha p'$, where $p$ is primitive and $p'$ is a prefix of $p$, and $u = q^\ell$, where $q$ is a cyclic shift of $p$ (i.e., $q$ is a factor of $p^2$ of length $|p|$) and $\ell \leq \frac{\alpha}{2}$.

So, if we consider a maximal run $r = p^\alpha p'$ and some $\ell \leq \frac{\alpha}{2}$, we can easily detect the possible centre positions of the squares having the form $(q^\ell)^2$ contained in this run, with $q$ a cyclic shift of $p$. These positions occur consecutively in the word $w$: the first is the $(|p|^\ell + 1)^{th}$ position of the run, and the last is the one where the suffix of length $|p|^\ell$ of the run starts. So they form an interval $I_{r,\ell}$ and we associate to this interval the weight $g_{r,\ell} = |p|^\ell$ (i.e., the length of an arm of the square). In this way, we define $\mathcal{O}(n)$ intervals (as their number is upper bounded by the sum of the exponents of the maximal runs of $w$ [42]), all contained in $[1, n+1)$, and each interval having a weight between 1 and $n$. By Lemma 2.4.1, we can process these intervals so that we can determine for each $i \in [1, n+1)$ the interval of maximum weight containing $i$, or, in other words, the maximum length $SC[i]$ of a square centred on $i$. This procedure runs in $\mathcal{O}(n)$ time. □

We follow with a similar result and a more detailed proof.

**Lemma 2.4.3** *Given a word $w$ of length $n$ we can compute in $\mathcal{O}(n)$ time the values*

$$MinRightEnd[i] = \min\left\{ j \mid \exists\ w[\ell..j] \text{ a square, such that } \ell \leq i < \ell + \frac{j-\ell+1}{2} \right\}.$$

*Proof.* To show the result, we note that each square $u^2$ occurring in a word $w$ is part of a maximal run $w[i'..j'] = p^\alpha p'$, where $p$ is primitive and $|p|$ is the period of that run. So, if we consider a maximal run $r = p^\alpha p', r = w[\ell..\ell']$, we can observe that for each position $k$ with $\ell + p - 1 \leq k \leq \ell' - p$, there is a square with root length $|p|$, that ends exactly on position $k + p$, and there is no other square contained in this run that has $k$ in its first half and ends on a smaller position (we call such squares that contain $i$ square of type 1). Also, for each position $k$ with $\ell \leq k \leq \ell + p - 1$ there is a square finishing on $\ell + 2p - 1$ with $k$ in its first half, and no other square in this run finishes before $\ell + 2p - 1$ (squares of type 2).

We treat separately these two cases and obtain, for each $i$ and each run containing it, the square containing $i$ in its first half that ends on the smallest position, and is contained, at its turn, in the respective run. Then we just have to return, for each $i$, the minimum ending position for such a square from the ones obtained for all the runs.

In the first case we can treat each run $w[\ell..\ell']$, with period $p$, as an interval $[\ell + p - 1, \ell' - p]$ with weight $p$. This means, that for each position $i$ of this interval there is a square of the type we look for finishing on position $i + p$ letters to the right of that position. We than use Lemma 2.4.1 to process these $\mathcal{O}(n)$ intervals and obtain for each position $i$ from 1 to $n$, the interval of minimum weight that contains it. If this weight is $p_i$, then there is a square containing $i$ in its first half whose ending position is $i + p_i$.

But we only processed the squares of the first type. For the squares of the second type, we also compute an interval for each run $w[\ell..\ell']$. In this case, we make an interval $[\ell, \ell + p - 1]$ with the value $\ell + 2p - 1$; this means for each element $i$ of this interval there is a square containing $i$ in its first half and ending on position $\ell + 2p - 1$. We again use Lemma 2.4.1 for each position $i$ from 1 to $n$, the interval of minimum weight that contains it. If this weight is $p'_i$, then there is a square containing $i$ in its first half and ending on position $\ell + 2p_i - 1$.

After computing these values for each $i$, we get that $MinRightEnd[i] = \min\{p_i, p'_i\}$.

In each of the applications of Lemma 2.4.1, we only processed $\mathcal{O}(n)$ intervals, with weights at most $n$. This means that the total time needed to compute the values $p_i$ and $p'_i$ for all $i$ is $\mathcal{O}(n)$. The conclusion of the lemma follows now immediately. □

It is worth noting that using the same strategy as in the proof of lemmas 2.4.2 and 2.4.3, one can detect the minimum length of a square centred at each position of a word in $\mathcal{O}(n)$ time. This leads to an alternative solution to the problem of computing the local periods of a word, solved in [27]. Compared to the solution from [27], ours uses a relatively involved data structures machinery (disjoint sets union-find structures), but it is much shorter and it seems conceptually simpler as it does not require a very long and detailed combinatorial analysis of the properties of the input word. The same strategy allows solving the problem of computing in linear time, for integer alphabets, the length of the shortest (or longest) square ending , (or starting) at each position of a given word; this improves the results from [46, 61], where such a result was only shown for constant size alphabets.

# Chapter 3

# Language theoretical and algorithmic aspects

In this section we thoroughly discuss about the prefix-suffix duplication, the bounded prefix-suffix duplication and the prefix-suffix completion operations from different points of view.

Prefix-suffix duplications were first considered in [32] as a model inspired by the case of DNA. In the paper the class of languages that can be defined by the iterative application of the prefix-suffix duplication to a word was examined and compared to other well studied classes of languages. It was shown that the languages of this class have a rather complicated structure even if the initial word is rather simple. More precisely, they are already non-context-free even for binary alphabets (regular only for unary alphabets). On the other hand, all the languages of this class have been proven to have a linear Parikh image[53] and that they belong to **NL**, hence are polynomially recognizable. Algorithmical results were also provided such as a $\mathcal{O}(n^2 \log n)$ time and $\mathcal{O}(n^2)$ space complexity recognition algorithm that verifies if a word $x$ with $|x| = n$ is in $PSD^*(w)$, with $|w| = m$. The prefix-suffix duplication distance between two given words was also computed and two algorithms were proposed: a cubic time one which uses a quadratic memory and a more efficient one, namely $\mathcal{O}(n^2 \log n)$ time complexity, but with some extra memory consumption, that is $\mathcal{O}(n^2 \log n)$ space complexity.

Several problems remained unsolved in the aforementioned paper, thus inspiring us to consider a weaker variant of the prefix-suffix duplication: the bounded prefix-suffix duplication. Using this model we are able to solve some of the problem that remained unsolved in [32]. Another reason for considering bounded prefix-suffix duplications is related to the biochemical reality that inspired the definition of this operation. It seems more practical and closer to the biological reality to consider that the factor added by the prefix-suffix duplication cannot be arbitrarily long. One should note that the investigation we pursue here is not aimed to tackle real biological facts and provide solutions for them. In fact, its aim is to provide a better understanding of the structural properties of strings obtained by prefix-suffix duplication as well as specific tools for the manipulation of such strings.

We also present a new operation on words that we introduced in [24], the operation is closely related to the prefix-suffix duplication. Instead of replicating a prefix or suffix to create a square, we assume a different point of view: we consider the possibility of creating squares (the simplest type of repetition) at one of the ends of the word by completing a prefix or suffix of the considered sequence to a square. More precisely, by *suffix square completion*, we derive from a word $w$ a word $wx$ if $w$ has a suffix $yxy$; note that the suffix we complete to a square must contain the root of the square (here, $yx$), and that suffix duplication is obtained by restricting square suffix to the case when $y$ is the empty word. The *prefix square completion* is symmetrical: from a word $w$ we derive $xw$ if $w$ has a prefix $yxy$. The *prefix-suffix square completion* just combines the previous two operations: from a word $w$ we derive $w'$ if $w'$ can be obtained either by prefix square completion or by suffix square completion from $w$.

Most of the language theoretic results obtained for the prefix, suffix, and prefix-suffix duplication operations seem to also hold for the newly defined square completion operations. Therefore, our investigation of the prefix-suffix square completion operations is aimed in the following two directions: membership/distance problems and infinite word generation(which we will cover in the following chapter) .

In this chapter we focus on finite words, and try to see whether a word can be generated by prefix-suffix square completion from one of its factors. We give an algorithm that identifies in linear time, for a given word $w$, all its prefixes from which $w$ can be generated by iterated suffix square completion. This algorithm is an essential step in showing that we can identify in linear time (and produce a compact representation of) all the factors of a word from which it can be generated by iterated prefix-suffix square completion. This leads immediately to a linear-time algorithm deciding, for two given words, whether the longer word can be generated from the shorter word by prefix-suffix square completion. This algorithm is much faster than the corresponding one from the case of prefix-suffix duplication. We will also give algorithms to solve other distance and membership problems in the case of prefix-suffix square completion.

This chapter is split into three sections:

- In the first section we give sufficient conditions for a family of languages to be closed under bounded prefix-suffix duplication. Consequently, we show that every language generated by applying iteratively the bounded prefix-suffix duplication to a word is regular. We also propose an algorithm for deciding whether there exists a finite set of words generating a given regular language w.r.t. bounded-prefix-suffix duplication. We also show that the prefix-suffix duplication and prefix-suffix square completion operations generate languages outside of Context Free even for binary alphabets.

- In the second section we discuss membership problems for the three different operations, algorithms range from $\mathcal{O}(n)$ time and space for prefix-suffix completion to $\mathcal{O}(n^2 \log n)$ time and $\mathcal{O}(n^2)$ space for prefix-suffix duplications. For the bounded prefix-suffix duplication operation we take into account two separate cases, given a word x, if $|x| \geq k$: we obtain $\mathcal{O}(n \log(k))$ time complexity, otherwise we obtain a time complexity of $\mathcal{O}(nk \log(k))$. We also give a linear time algorithm for finding a $PSD, BPSD$ primitive root. Furthermore, we discuss a few problems about predecessors and ancestors and we give algorithms for some of this problems. A lot of interesting problems remain open in this area.

- We then discuss distance related problems and give algorithms to compute the distance between two words for all of our operations. We then extend the distance to languages and we propose an algorithm for efficiently computing the $k$-prefix-suffix duplication distance between two bounded prefix-suffix duplication languages.

One should note that not all three operations have been researched in all of these directions and we will discuss in each section why we have focused on specific aspects in our research.

## 3.1 Duplication languages

In this section we focus on bounded prefix-suffix duplications as research has already been developed on different classes of duplication languages including prefix-suffix duplications. To our best knowledge no research has been made in this direction related to the prefix-suffix square completion operation, we limit ourselves to giving one result that links this operation to prefix-suffix duplication.

We start by recalling some basic language theoretical properties of the class of duplication languages. Putting together the results on arbitrary duplication languages from [7] and [28] (rediscovered in [20] and [60]), and those on prefix-suffix duplication languages from [32] we obtain the following.

**Theorem 3.1.1** *An arbitrary duplication language is regular if and only if it is a language over an alphabet with at most two symbols.*

**Theorem 3.1.2** *A prefix-suffix duplication language is context-free if and only if it is a language over the unary alphabet.*

An example of a prefix-suffix duplication language, generated starting from a binary word, that is not context free follows[32]:

**Example 3.1.1** $PSD^*(ab) \cap ab^+ab^+ab^+ = \{ab^mab^nab^p \mid n, m, p \geq 1, m \leq \min(n, p)$ *and* $n \leq m + p\} \notin CF$.

Note that $PSSC^*(ab) \cap ab^+ab^+ab^+$ also equals $\{ab^mab^nab^p \mid n, m, p \geq 1, m \leq \min(n, p)$ and $n \leq m + p\} \notin CF$, thus PSSC is not in context free for binary alphabets(and of course for larger alphabets too).

Whether or not every arbitrary duplication language is recognizable in polynomial time is open while every prefix-suffix duplication language is in **NL**.

We now turn our attention to results related to the bounded prefix-suffix duplication languages, which have been studied in [25, 26].

We say that a class $\mathcal{L}$ of languages is *closed under bounded prefix-suffix duplication* if $PSD_k^*(L)$ $\mathcal{L}$ for any $L \in \mathcal{L}$ and $k \geq 1$.

**Theorem 3.1.3** *Every nonempty class of languages closed under union with regular languages, intersection with regular languages, and substitution with regular languages, is closed under bounded prefix-suffix duplication.*

*Proof.* Let $\mathcal{L}$ be a family of languages having all the required closure properties. By [34], $\mathcal{L}$ is closed under inverse morphism. Let $L \subseteq V^*$, with $|V| = m$, be a language from $\mathcal{L}$, and $k$ be a positive integer. We define the alphabet

$$U = V \cup \{p_1, p_2, \ldots, p_{m^k}\} \cup \{s_1, s_2, \ldots, s_{m^k}\},$$

and the morphism $h : U^* \longrightarrow V^*$ defined by $h(a) = a$ for any $a \in V$ and $h(p_i) = h(s_i) =$ the $i^{th}$ word of length $k$ over $V$ in the lexicographic order, for all $1 \leq i \leq m^k$. Further, let $F$ be the finite language defined by $F = \{x \in L \mid |x| \leq 2k-1\}$ and

$$E = (L \cup PSD_k^{2k}(F)) \cap \{x \in V^+ \mid |x| \geq 2k\}.$$

As $PSD_k^{2k}(F)$ is a finite language and $\mathcal{L}$ is closed under union with regular languages and intersection with regular languages, it follows that $E$ is still in $\mathcal{L}$. The following relation is immediate:

$$PSD_k^*(L) = PSD_k^*(E) \cup PSD_k^{2k}(F).$$

It is rather easy to prove that

$$PSD_k^*(E) = \sigma(h^{-1}(E) \cap \{p_1, p_2, \ldots, p_{m^k}\}V^*\{s_1, s_2, \ldots, s_{m^k}\}),$$

where $\sigma$ is a substitution defined by $\sigma(p_i) = PD_k^*(x_i)$ and $\sigma(s_i) = SD_k^*(x_i)$, where $x_i$ is the $i^{\text{th}}$ word of length $k$ over $V$ in the lexicographic order.

Each language $PD_k^*(x_i)$ can be generated by a prefix grammar [30], hence it is regular. Analogously, each language $SD_k^*(x_i)$ is regular. Consequently, $\sigma$ is a substitution with regular languages. By the closure properties of $\mathcal{L}$, $PSD_k^*(E)$ belongs to $\mathcal{L}$, hence $PSD_k^*(L)$ is also in $\mathcal{L}$. □

According to theorem 3.1.3, we immediately obtain now a result which is very different from those described in theorems 3.1.1 and 3.1.2:

**Corollary 3.1.1** *Every bounded prefix-suffix duplication language is regular.*

A language $L$ is said to be a *multiple k-prefix-suffix duplication language* if there exists a language $E$ such that $L = PSD_k^*(E)$. If $E$ is finite, then $L$ is said to be a *finite k-prefix-suffix duplication language*. Note that given a regular language $L$ and a positive integer $k$, a necessary condition such that $L = PSD_k^*(E)$ holds, for some set $E$, is $L = PSD_k^*(L)$. By the proof of theorem 3.1.3 a finite automaton accepting $PSD_k^*(L)$ can effectively be constructed and so the above equality can be algorithmically checked. However, if the equality holds, we cannot infer anything about the finiteness of $E$. The problem is completely solved by the next theorem.

**Theorem 3.1.4** *Let $L$ be a regular language which is a multiple $k$-prefix-suffix duplication language for some positive integer $k$. There exists a unique minimal (with respect to inclusion) regular language $E$, which can be algorithmically computed, such that $L = PSD_k^*(E)$. In particular, one can algorithmically decide whether $L$ is a finite $k$-prefix-suffix duplication language.*

*Proof.* Let $L \subseteq V^*$ be a multiple $k$-prefix-suffix duplication language accepted by the deterministic finite automaton $A = (Q, V, f, q, F)$. We define the language

$$M_k(L) = \{x \in L \mid \text{ there is no } y \in L \text{ such that } x \in PSD_k(y)\}.$$

As $L = PSD_k^*(L)$, it follows that

$$M_k(L) = \{x \in L \mid \text{ there is no } y \in L, y \neq x \text{ such that } x \in PSD_k^*(y)\}.$$

**Claim.** *If* $PSD_k^*(E) = L$ *for some* $E \subseteq L$*, then the following statements hold:*
*(i)* $M_k(L) \subseteq E$*, and*
*(ii)* $PSD_k^*(M_k(L)) = L$*.*

*Proof of the claim.* (i) Let $x \in M_k(L) \subseteq L$; there exists $y \in E$ such that $x \in PSD_k^*(y)$. By the definition of $M_k(L)$, it follows that $x = y$.

(ii) Clearly, $PSD_k^*(M_k(L)) \subseteq L$. Let $y \in L$; there exists $x \in L$ such that $y \in PSD_k^*(x)$. We may choose $x$ such that $x \in PSD_k(z)$ for no $z \in L$. Thus, $x \in M_k(L)$, and $y \in PSD_k^*(M_k(L))$, which concludes the proof of the claim.

Clearly, $M_k(L) = L \setminus PSD_k(L)$; hence $M_k(L)$ is regular and can effectively be constructed.

In order to check whether $L$ is a finite $k$-prefix-suffix duplication language, we first compute $M_k(L)$. Then we check whether $M_k(L)$ is finite. Finally, by the proof of theorem 3.1.3, the language $PSD_k^*(M_k(L))$ is regular and can be effectively computed, therefore the equality $PSD_k^*(M_k(L)) = L$ can be algorithmically checked. □

## 3.2 Membership problems

In this section we discuss membership problems focusing on the bounded prefix-suffix duplication and prefix-suffix square completion operations as several membership problems about prefix-suffix operations have already been investigated in [32]. We begin with a few lemmas needed to prove our main results of the chapter which will be computing efficiently if a word $x$, with $|x| = n$, is in $PSD_k^*(w)$ and in $PSSC^*(w)$. After that we discuss unpublished results related to common ancestors and predecessors of words. Formally finding a/the shortest predecessor of two words $x$ and $y$ is finding a/the shortest word $z$ such that $z \in A^*(x) \bigcap A^*(y)$ for any $A \in \{PSD, BPSD, PSSC\}$. Conversely finding an ancestor of two words $x$ and $y$ is finding a word $z$ such that $x \in A^*(z)$ and $y \in A^*(z)$ for any $A \in \{PSD, BPSD, PSSC\}$.

Coming back to the membership problem, an algorithm in $\mathcal{O}(n^2 \log(n))$ time and space is given in [32] that tests if a word $x$ is in $PSD^*(w)$. In this section we show how to solve the membership problem in $\mathcal{O}(nk \log(k))$ for $BPSD$ and in linear time for $PSSC$.

#### Membership problem for the bounded prefix-suffix operation

We will start with some preliminaries needed to solve the prefix-suffix duplication membership problem, after that we obtain a generalization of the result (testing if $w \in PSD_k^*(L)$), and then narrow it down to verifying the membership problem for a single word.

We will make use of the following classical result from [13]. It is known that the number of primitively rooted square factors of length at most $2k$ that occur at some position of a word $w$, of length $n$, is $\mathcal{O}(\log(k))$. Moreover, one can construct the list of primitively rooted squares of length at most $2k$ occurring in $w$ in $\mathcal{O}(n \log(k))$ time. Each square is represented in the list by the starting position and the length of their root, and the list is ordered in increasing order with respect to the starting position of the squares; when more squares share the same starting position they are ordered with respect to the length of the root. Moreover, one can store an array of $n$ pointers, where the $i^{th}$ such pointer indicates the memory location of the list of the primitively rooted squares occurring at position $i$. A similar list, where the squares are ordered by their ending position, can be computed in the same time. Further, we present our main algorithmic tool needed to solve $PSD_k$ membership.

**Lemma 3.2.1** *Given $w \in V^*$, of length $n$, and an integer $k \leq n$, we can identify all prefixes $w[1..i]$ of $w$ such that $w \in SD_k^*(w[1..i])$ in $\mathcal{O}(n \log(k))$ time.*

*Proof.* We propose an algorithm that computes an array $S[\cdot]$, defined by $S[i] = 1$ if $w \in SD_k^*(w[1..i])$, and $S[i] = 0$, otherwise. The algorithm has a preprocessing phase, in which all the primitively rooted squares with root of length at most $k$ occurring in $w$ are computed. This preprocessing takes $\mathcal{O}(n \log(k))$ time.

Now, we describe the computation of the array $S$. Initially, all the positions of this array are initialized to 0, except $S[n]$, which is set to 1. Clearly, this is correct, as $w \in SD_k^*(w[1..n]) = SD_k^*(w)$. Further, we update the values in the array $S$ using a dynamic programming approach. That is, for $i$ from $n$ to 1, if $S[i] = 1$, then we go through all the primitively rooted squares $(w[j+1..i])^2$, $|i-j| \leq k$, that end at position $i$ in $w$. For each such factor $w[j+1..i]$ we set $S[j] = 1$. Indeed, $w[1..i]$ can be obtained from $w[1..j]$ by appending $w[j+1..i]$ (which is known to be a suffix of $w[1..j]$); as we already know that $w$ can be obtained by suffix duplication from $w[1..i]$, it follows that $w$ can be obtained by suffix duplication from $w[1..j]$. The processing for each $i$ takes $\mathcal{O}(\log(k))$ time.

It is not hard to see that our algorithm works correctly. Assume that $w \in SD_k^*(w[1..j])$ for some $j < n$. Let us consider the longest sequence of suffix duplication steps (or, for short, derivation) that produces $w$ from $w[1..j]$. Say that this derivation has $s \geq 2$ steps, so it can be described by a sequence of indices $j_1 = j < j_2 < \ldots < j_s = n$ such that $w[1..j_{i+1}] \in SD_k(w[1..j_i])$ for $1 \leq i \leq s-1$. We can show that $w[j_i+1..j_{i+1}]$ is primitive for all $i$. Otherwise, $w[j_i+1..j_{i+1}] = t^\ell$ for some word $t$ and $\ell \geq 2$, so we can replace in the original derivation the duplication that produces $w[1..j_{i+1}]$ from $w[1..j_i]$ by other $\ell$ duplication steps in which $t$ factors are added to $w[1..j_i]$. This leads to a sequence with more than $s$ duplications steps producing $w$ from $w[1..j]$, a contradiction. In our algorithm, $S[j_s]$ is set to 1 in the first step. Assuming that for some $i$ we already have $S[j_{i+1}] = 1$, when considering the value $j_{i+1}$ in the main loop of our algorithm, as $w[j_i+1..j_{i+1}]^2$ is a primitively rooted square ending on position $j_{i+1}$, we will set $S[j_i] = 1$. In the end, we will also have $S[j] = S[j_1] = 1$. Moreover, it is not hard to see that if $w \notin SD_k^*(w[1..i])$ for some $i < n$ then $S[i]$ remains equal to 0 during the entire execution of our algorithm. Hence, our algorithm works properly. □

**Lemma 3.2.2** *Given $w \in V^*$, of length $n$, we can identify all suffixes $w[j..n]$ of $w$ such that $w \in PD_k^*(w[j..n])$ in $\mathcal{O}(n \log(k))$ time.*

The proof is similar to the one of Lemma 3.2.1. The output of the algorithm we construct in this proof is an array $P[\cdot]$, defined by $P[j] = 1$ if $w \in PD_k^*(w[j..n])$, and $P[j] = 0$, otherwise.

The next lemma shows a way to find the factors of length at least $k$ from a list of factors $F$, from which $w$ can be obtained by iterated prefix-suffix duplication.

**Lemma 3.2.3** *Given $w \in V^*$ of length $n$ and a list $F$ of factors of $w$ of length greater than or equal to $k$, given by their starting and ending position,* *ordered by their starting position, and in case of equality by their ending position, we can check whether there exists* $x \in F$ *such that* $w \in PSD_k^*(x)$ *in time* $\mathcal{O}(n\log(k) + |F|)$.

*Proof.* The main remark of this lemma is that, if $w[i..j]$ is longer than $k$, then $w \in PSD_k^*(w[i..j])$ if and only if $w[1..j] \in PD_k^*(w[i..j])$ and $w = w[1..n] \in SD_k^*(w[1..j])$. Equivalently, we have $w \in PSD_k^*(w[i..j])$ if and only if $w[1..n] \in PD_k^*(w[i..n])$ and $w[1..n] \in SD_k^*(w[1..j])$.

This remark suggests the following approach: we first identify all the suffixes $w[j..n]$ of $w$ such that $w \in PD_k^*(w[j..n])$ and all the prefixes $w[1..i]$ of $w$ such that $w \in SD_k^*(w[1..i])$; this takes $\mathcal{O}(n\log(k))$, by Lemmas 3.2.1 and 3.2.2. Now, for every factor $w[i..j]$ in list $F$, we just check whether $S[i] = P[j] = 1$ (that is, $w \in PD_k^*(w[i..n]) \cap SD_k^*(w[1..j])$); if so, we decide that $w \in PSD_k^*(w[i..j])$. □

Building on the previous lemmas, we can solve the membership problem for $PSD_k^*(L)$ languages, provided that we know how to solve the membership problem for $L$ on the RAM with logarithmic word size model.

**Theorem 3.2.1** *If the membership problem for the language* $L$ *can be decided in* $\mathcal{O}(f(n))$ *time, then the membership problem for* $PSD_k^*(L)$ *can be decided in* $\mathcal{O}(nk\log(k) + n^2 f(n))$.

*Proof.* Assume that we are given a word $w$, of length $n$; we want to test whether $w \in PSD_k^*(L)$ or not. In the setting of our theorem, $L$ is constant (i.e., its description, given as a procedure deciding $L$ in $\mathcal{O}(f(n))$ time, is not part of the input). Otherwise, if $L$ was given as part of the input, then we can use exactly the same algorithm, but one should add to the final time complexity the time needed to read the description of $L$ and the time needed to effectively construct a procedure deciding $L$ in $\mathcal{O}(f(n))$ time.

First, let us note that we can identify trivially in $\mathcal{O}(n^2 f(n))$ the factors of $w$ that are in $L$. More precisely, we can produce a list $F$ of factors of $w$ that are contained in $L$, specified by their starting and ending position, ordered by their starting position, and, in case of equality by their ending position. The list $F$ can be easily split, in $\mathcal{O}(|F|)$ time, into two lists: $F_1$, containing the factors of length at least $k$, and $F_2$, the list of factors of length less than $k$. It is worth noting that $|F| \in \mathcal{O}(n^2)$. By Lemma 3.2.3 it follows that we can decide in time $\mathcal{O}(n\log(k) + |F_1|)$ whether $w \in PSD_k^*(x)$ for some $x \in F_1$.

It remains to test whether $w \in PSD_k^*(x)$ for some $x \in F_2$. The main remark we make in this case is that there exists $x \in F_2$ such that $w \in PSD_k^*(x)$ if and only if there exists $y \in PSD_k^*(x)$ such that $k \leq |y| \leq 2k$ and $w \in PSD_k^*(y)$. Therefore, we will produce the list $F_3$ of words $z \in \cup_{x\in F_2} PSD_k^*(x)$ such that $z$ is a factor of $w$ and $k \leq |z| \leq 2k$.

In order to compute $F_3$ we can use the $\mathcal{O}(|u|^2 \log |u|)$ algorithm proposed in [32] to decide whether a word $u$ is contained in $PSD^*(v)$. In that algorithm, one first marks the factors of $u$ that are equal to $v$. Further, for each possible length $\ell$ of the factors of $u$, from 1 to $|u|$, and for each $i \leq n$ where a factor of length $\ell$ of $u$ may start, one checks whether $u[i..i+\ell-1]$ can be obtained by prefix (respectively, suffix duplication) from a shorter suffix (respectively, prefix), that was already known (i.e., marked) to be in $PSD^*(v)$, such that in the last step of duplication a primitive root $x$ of a primitively rooted square prefix $x^2$ of $u[i..i+\ell-1]$ was appended to the shorter suffix (respectively, a primitive root $x$ of a primitively rooted square suffix $x^2$ of $u[i..i+\ell-1]$ was appended to the shorter prefix). Each time we found a factor of $w$ that can be obtained in this way from one of its marked prefixes or suffixes, we marked it as part of $PSD_k^*(v)$ and continued the search with the next factor of $w$.

In our case, we can pursue the same strategy: taking $w$ in the role of $u$, and having already marked the words of $F_2$ (which are factors of $w$) just like we did with the occurrences of $v$, we run the algorithm described above, but only for $\ell \leq 2k$. Note that the roots of primitively rooted square suffixes or prefixes of factors $w[i..i+\ell-1]$ with $\ell \leq 2k$ have length at most $k$; hence, each arbitrary duplication that is made to obtain such a factor is, in fact, a $k$-prefix-suffix duplication. In this manner we obtain the factors of $w$ of length at most $2k$ that are from $PSD_k^*(F_2)$. The time needed to obtain these factors is $\mathcal{O}(nk\log(k))$. We store this set of factors in $F_3$ just like before: they are given by their starting and ending position, ordered with respect to their starting position, and, in case of equality with respect to their ending position. The set $F_3$ may have up to $2nk$ factors.

By Lemma 3.2.3, we can decide in time $\mathcal{O}(n\log(k)+|F_3|) = \mathcal{O}(nk)$ whether $w \in PSD_k^*(F_3)$. Accordingly, adding the time needed to compute $F_3$ from $F_2$, it follows that we can decide in time $\mathcal{O}(nk\log(k))$ whether $w \in PSD_k^*(F_2)$.

Hence, we can decide whether $w \in PSD_k^*(L)$ in $\mathcal{O}(nk\log(k)+n^2 f(n))$ time. □

In fact, there are classes of languages for which a better bound than the one in theorem 3.2.1 can be obtained. If $L$ is context-free (respectively, regular) the time needed to decide whether $w \in PSD_k^*(L)$ is $\mathcal{O}(n^3)$ (respectively, $\mathcal{O}(nk\log(k)+n^2)$), where $|w| = n$. Indeed, $F$ has always less than $n^2$ elements, and in the case of context-free (or regular) languages it can be obtained in $\mathcal{O}(n^3)$ time (respectively, $\mathcal{O}(n^2)$) by the Cocke-Younger-Kasami algorithm (respectively, by running a DFA accepting $L$ on all suffixes of $w$, and storing the factors accepted by the DFA).

When $L$ is a singleton, the procedure is even more efficient.

**Corollary 3.2.1** *Given two words $w$ and $x$, with $|w| = n \geq m = |x|$, we can*

*decide whether* $w \in PSD^*_k(x)$ *in time* $\mathcal{O}(nk\log(k))$. *If* $m \geq k$, *then we can decide whether* $w \in PSD^*_k(x)$ *in time* $\mathcal{O}(n\log(k))$.

*Proof.* First, note that the list $F$ of all occurrences of $x$ in $w$ can be obtained in linear time $\mathcal{O}(n+m)$ (using, e.g., the Knuth-Morris-Pratt algorithm from [40]), and $|F| \in \mathcal{O}(n)$.

For the first part, we follow the same general approach as in theorem 3.2.1. If $|x| < k$, we produce the list of all the factors longer than $k$, but of length at most $2k$, that can be derived from $x$. This list is produced in $n*(n+1)/2$ time. Therefore, the total complexity of the algorithm is $\mathcal{O}(nk\log(k))$, in this case.

The second result follows now immediately from Lemma 3.2.3, as $F$ contains only words of length at least $k$. □

We stress that the above results comes from basically intersecting the occurrences of $x$ in $w$ with the ancestors of $w$ in an efficient way. We will now discuss a few problems related to ancestors and predecessors. Note that all ancestors of a word are also factors of the world thus the two notions are closely related (though not every factor is an ancestor).

The following results are unpublished:

**Lemma 3.2.4** *Given a word* $w$ *we can compute all of its ancestors(or roots) in* $\mathcal{O}(nk\log(k) + |output|)$ .

*Proof.* First note that the number of ancestors for a word can be as high as $n*(n+1)/2$. Take for example the word $a^n$ who has $n*(n+1)/2$ ancestors.

We start by computing in $\mathcal{O}(n\log(k))$ all long factors as in lemma 3.2.3 and also compute the arrays $P$ and $S$ as in lemma 3.2.1 and lemma 3.2.2. And then using a similar approach to the one in theorem 3.2.1 we further restrain the intervals to obtain all factors $w[i...j]$ with $j - i <= k$. The second part takes $\mathcal{O}(nk\log(k))$. Note that there are at most $nk$ short factors (and ancestors) so listing them does not affect our complexity. To obtain $\mathcal{O}(nk\log(k) + |output|)$ time complexity we have to be careful how we output elements as $\mathcal{O}(n^2)$ might be greater than $\mathcal{O}(nk\log(k) + |output|)$. Having obtained the arrays $P$ and $S$ previously we create two lists with the elements from $S$ and $P$ in increasing order. For an element $x$ in the first list we output all elements from the second list that are greater than $x+k$. To do this we simply eliminate from the second list all elements $y$ such that $y < x+k$. Thus for every element in the first list we just go through the second list and print all its pairs. As every element is inserted and deleted at most once from every list the total complexity of this part of the algorithm is $\mathcal{O}(n + |output|)$, bringing the total complexity to $\mathcal{O}(nk\log(k) + |output|)$ □

A natural question is the upper bound for the number of primitive roots of a word in relation to the *BPSD* operation. While we don't have a complete answer for this problem we can show that $|output|$ can be as high as $\Theta(n^2)$ for $k \geq n/2$, based on the following result proved for the *PSD* in [24].

**Example 3.2.1** *We define a family of words $(w_n)_{n\in\mathbb{N}}$ such that $w_n$ has $\Theta(|w_n|)$ SD primitive roots. Let $w_1 = aabbab$; this word is SD primitive. Then define $w_i = w_{i-1}w_{i-1}bb$ for $i \geq 2$. The length of $w_i$ is $2^{i+2} - 2$ for all $i \geq 1$. Note that $w_2$ has the SD primitive root $w_1$ and let $R_2 = \{w_1\}$. Now, for $i \geq 3$, $w_i$ has at least the SD primitive roots $R_i = R_{i-1} \cup \{w_{i-1}r \mid r \in R_{i-1}\}$. So $w_i$ has, indeed, at least $|R_i| \geq \frac{|w_i|}{16}$ primitive roots. Further, it is not hard to see that $w_i'^R w_i$, where $w_i'$ is obtained from $w_i$ by replacing a by c and b by d, has $\Theta(|w_i'^R w_i|^2)$ PSD primitive roots. Indeed, if $r_1$ and $r_2$ are SD primitive roots of $w_i$, then $r_1'^R r_2$ is a PSD primitive root of $w_i'^R w_i$ (where $r_1'$ is obtained from $r_1$ by replacing a by c and b by d). So the words of the family $(w_n'^R w_n)_{n\in\mathbb{N}}$ are such that $w_n'^R w_n$ has $\Theta(|w_n'^R w_n|^2)$ PSD primitive roots.*

*Proof.* Before starting the proof, note that $w_n$ ends with $2n-1$ letters $b$, and $w_n$ does not contain any other factor with more than $2n-3$ letters $b$ (i.e., any factor which is not a suffix of $b$ does not contain more than $2n-3$ letters $b$). This can be easily shown by induction. Similarly, if $r \in R_n$ then either $|r| < |w_{n-1}| - |2n-3|$ or $r = w_{n-1}r'$ with $|r'| < |w_{n-1}| - |2n-3|$; this can be easily shown by induction.

First of all, it is immediate that from each $r \in R_n$ we can generate $w_{n+1}$ by iterated *SD*. We can proceed by induction. The case $n = 2$ is immediate. Further, if $r \in R_{n-1}$, we first generate $w_{n-1}$ from it (induction hypothesis), and then duplicate $w_{n-1}$ to obtain $w_{n-1}w_{n-1}$, which ends with $b$. So, we can duplicate the final $b$ twice to get $w_n = w_{n-1}w_{n-1}bb$.

Also, it can be easily shown that $|R_n| = 2^{n-2}$ for all $n \geq 2$.

Further we proceed with the main part of the proof: showing that the elements of $R_n$ are *SD* primitive.

We show by induction that following two properties (for $n \geq 2$):

1. the elements of $R_n$ are *SD* primitive.

2. if $r \in R_n$ then $b^k r$ is *SD* primitive for all $k \geq 0$.

The properties can be manually checked for the elements of $R_2$, $R_3$, and $R_4$. Let us now assume that they are true for all elements of $R_i$ with $i \leq n$ (for some $n \geq 4$), and we want to show that they are true for $R_{n+1}$.

The first item can be shown as follows. Assume that we have $r \in R_{n+1} = R_n \cup \{w_n r \mid r \in R_n\}$ and recall that $w_{n+1} = w_n w_n bb$. We assume for the sake of a contradiction that $r$ ends with a square. If $r \in R_n$ then the fact that $r$ is *SD* primitive follows from the induction hypothesis, and we get the contradiction. So, we consider the case when $r = w_n r'$ for some $r' \in R_n$. It is immediate that if $r$ ends with a square $x^2$ then $|x^2| > |r'|$ (or $r'$ would end with a square, contradiction to the induction hypothesis). Moreover, as $b^k r'$ does not end with a square, then $x^2$ cannot start in the group of letters $b$ occurring at the end of the prefix $w_n$ of $w_{n+1} = w_n w_n b^2$. So $x^2$ starts somewhere in this prefix $w_n$. As $r$ ends with $ab$, we have that $x$ should also end with $ab$. So, the first factor $x$ of the square $x^2$ ends either on the first letter $b$ of the suffix of letters $b$ of the prefix $w_n$ of $w_{n+1}$, or somewhere inside this factor $w_n$, before the suffix of letters $b$ start. In both cases, we get that the second factor $x$ in the square contains a sequence of $2n-2$ consecutive letters $b$; this sequence should occur somewhere in the first $x$, so also as a non-suffix factor of $w_n$, and this is a contradiction. So, $r$ cannot end with a square, and the first item of our statement is shown for the elements of $R_{n+1}$. To complete the proof, we also need to show the second item.

Let us take $r \in R_{n+1}$ and we want to show that $b^k r$ does not end with a square, for any $k > 0$. We assume the contrary. Clearly, the square $x^2$ occurring at the end of $b^k r$ should start in the prefix of letters $b$. Otherwise, $r$ would end with a square, and we saw that this is not possible. We may also assume that $r = w_n r'$ with $r' \in R_n$; otherwise, if $r \in R_n$ we get a contradiction from the induction hypothesis. We have that $r$ ends with $ab$, so the first $x$ of the square $x^2$ ends either on the first $b$ from the group of letters $b$ occurring at the end of the prefix $w_n$ of $w_n r'$, or somewhere to the left of this position. Now, if the second $x$ contains $b^{2n-1}$ as a proper factor (i.e., this is not a prefix, neither a suffix of $x$), then there must be at least one $a$ before this long sequence of letters $b$, and the conclusion follows just like before: it would imply that $w_n$ contains as a proper factor $b^{2n-1}$. So, $x$ should either start with $b^{2n-2}$ or with $b^{2n-1}$ (meaning that $k$ is at least $2n-2$ or, respectively, $2n-1$). We only present the first case as the other one follows identically. We get that $x = b^{2n-2} w'_n b$, where $w'_n$ is obtained from $w_n$ by cutting its suffix of length $2n-1$, containing only letters $b$. It follows that $r' = w'_n b$, a contradiction, as $r'$ should be strictly shorter than $w'_n$ by the remarks made at the beginning of the proof. So, again, in all cases we reached contradictions, and conclude that $b^k r$ cannot end with a square.

This completes the induction proof of our two statements, and shows, thus, that $w_n$ has $\Theta(|w_n|)$ *SD* primitive roots. □

The family of words described above evidently has $\Theta(n^2)$ *BPSD* primitive roots for any $k \geq n/2$.

We now give a few results related to ancestors in relation to the bounded prefix-suffix operation that we briefly proof.

**Lemma 3.2.5** *Given a word $w$ we can compute in $\mathcal{O}(nk \log(k))$ the following:*

1. *The number of ancestors*

2. *The shortest ancestor*

3. *The longest primitive ancestor*

*Proof.* 1. We can easily count the ancestors of length at most $k$ using the argument used in the previous lemma in $\mathcal{O}(nk\log(k))$. To compute the number of ancestors of length at least $k$ we can again use the argument from the above algorithm but avoid traversing the elements from the second list. We keep both lists and still remove from the second list all elements that are not greater than the current element from the first list + k, but instead of traversing the second list we simply add to the number of ancestors the number of elements still in the second list. Thus this part of the algorithm is $\mathcal{O}(n)$, obtaining a final complexity of $\mathcal{O}(nk\log(k))$. Note that if we want to compute all factors of length at least $k$ we can do so in $\mathcal{O}(n\log(k))$.

2. We first go through all ancestors of length at most $k$, and compute the minimum($\mathcal{O}(nk\log(k))$). If there is at least one such ancestor, we stop and output the minimum found. If there is no such ancestor we again compute the previous lists, obviously the minimum ancestor starting at a position $i$ with $S[i] = 1$ will be $w[i...j]$ with $P[j] = 1$ and $j \geq i + k$, and that is exactly the first element of the second list after eliminating elements from the second list as previously explained. We compute the minimum for each position and output the global minimum. Again the complexity for computing the shortest factor of length at least $k$ is $(n\log(k))$ and the total complexity is $\mathcal{O}(nk\log(k))$.

3. We recall the result that states that at a position $i$ there are at most $\log(k)$ primitive squares of size at most $k$. Based on this idea we can check for all at most $\mathcal{O}(nk)$ ancestors of $w$ of size at most $k$ if they are primitive in $\mathcal{O}(nk\log(k))$. For the factors of size greater than $k$ we have to adopt a different strategy:

   We compute for each position the length of the shortest square that starts in that position, and place this values in an array $MinSq(O(n))$ (section 2.4). We then compute the arrays $S$ and $P$ and introduce all elements in two sorted arrays $S'$ and $P'$. For each element in $S'$ we do a binary search in $P'$ to find the largest element $j$ such $S'[i] + k <= P'[j] <= S'[i] + MinSq[i]$. We then compute the longest pair from all this pairs. The time complexity is $\mathcal{O}(n\log(n))$ for this part of the algorithm while the overall time complexity is $\mathcal{O}(nk\log(k))$.

□

Note that the third results introduces the primitive ancestor concept that will also discuss in the following chapter of the thesis.

We will now give a result that shows how we can compute a primitive ancestor in relation to the $PD, SD$ and $PSD$ operations in linear time, and we show how this can be used to compute a primitive ancestor in relation to $BPSD$ as well.

**Theorem 3.2.2** *Given a word $w$ of length $n$, we can find one PSD primitive root of $w$ in $\mathcal{O}(n)$ time.*

*Proof.* Let $w_0 = w$. We construct a sequence $(w_i)_{i\geq 0}$ such that $w_i \in PSD(w_{i+1})$. Assume that $w_i$ has a square suffix and let $t^2$ be the shortest square suffix of $w_i$. Then let $w_{i+1} = w_i[1..|w_i| - |t|]$. If $w_i$ does not end with a square, but it has a square prefix, let $t^2$ the shortest square prefix of $w_i$. Then let $w_{i+1} = w_i[|t| + 1..|w_i|]$. If $w_i$ is prefix-suffix-square free, then we stop: we reached a primitive root of $w$. Using the arrays *left* and *right* defined in the last section, we can implement this strategy in linear time. □

**Remark.** *Given a word $w$ of length $n$, we can find an SD primitive (respectively, PD primitive) root of $w$ in $\mathcal{O}(n)$ time. We just use the same procedure, but only extend our sequence as long as the current word is not suffix-square free (respectively, prefix-square free).*

**Remark.** *Given a word $w$ of length $n$, we can find an BPSD primitive root of $w$ in $\mathcal{O}(n)$ time by using the same procedure with the difference that we simply ignore the shortest square of a word $w_i$ if its length is longer than $k$.*

We will come back to solving problems related to ancestors and predecessor for the bounded prefix-suffix duplication later on in this chapter. We feel that it's interesting to solve these problems in parallel for all of the operations we cover in this thesis, thus before we do that we need to gain more knowledge about the membership problem in the context of the prefix-suffix square completion operation.

### Membership problem for the prefix-suffix square completion operation

The following results have been published in [23].

**Lemma 3.2.6** *Let $x, y, z \in \Sigma^*$ such that $x$ is a prefix of $y$ and $y$ a prefix of $z$. If $z \in SSC^*(x)$ then $z \in SSC^*(y)$.*

*Proof.* If $x = z$ the conclusion follows immediately. Otherwise, there exists a sequence of words $x_0, x_1, \ldots, x_n$, with $n \geq 1$, such that $x_0 = x, x_n = z$, and $x_i \in SSC(x_{i-1})$ for $1 \leq i \leq n$. As $x_0$ is a prefix of $y$ and $y$ is a prefix of $x_n$, there exists $1 \leq k \leq n$ such $x_{k-1}$ is a prefix of $y$ and $y$ is a prefix of $x_k$. So, let $y = x_{k-1}u$. As $x_k \in SSC(x_{k-1})$ we get that $x_{k-1} = wvv'v$ and $x_k = wvv'vv'$; it follows that $u$ is a prefix of $v'$ and $v' = uv''$. So, $x_k = wvuv''vuv'' = yv''$. Therefore, $x_k$ can be obtained by suffix square completion from $y$ by appending $v''$ to $y$. It follows that $x_j \in SSC^*(y)$ for all $j \geq k$, so $z = x_n \in SSC^*(y)$. □

A similar result does not hold for $PSD^*$ as $abaabaa \in PSD^*(aba)$ but $abaabaa \notin PSD^*(abaab)$.

We need the following immediate extension of lemma 3.2.6.

**Lemma 3.2.7** *Let $w \in \Sigma^*$ be a word, and consider two factors of this word $w[i_1..j_1]$ and $w[i_2..j_2]$, such that $i_1 \leq i_2 \leq j_2 \leq j_1$. If $w \in SSC^*(w[i_2..j_2])$ then $w \in SSC^*(w[i_1..j_1])$.*

We now recall a result we proved in the lemmas section of our preliminaries: lemma 2.4.3.The lemma states that we can construct efficiently a data structure providing insight in the structure of the squares occurring inside a word, thus obtaining the *MinRightEnd* and *MaxLeftEnd* arrays.

Note that *MinRightEnd*$[i]$ denotes the minimum position $j > i$ such that there exists a square ending on position $j$, whose first half (that is, the square's left root) contains position $i$.

Alternatively, we can compute in linear time an array

$$MaxLeftEnd[i] = \max\{\ell \mid \exists\ w[\ell..j] \text{ a square, such that } \ell + \frac{j-\ell+1}{2} \leq i \leq j\}.$$

*MaxLeftEnd*$[i]$ is the maximum position $\ell < i$ such that there exists a square starting on position $\ell$, whose second half (i.e., right root) contains position $i$.

These two arrays will be used in the following way. The array *MinRightEnd* (respectively, *MaxLeftEnd*) tells us that from a factor $w[i+1..j]$ with $j \geq$ *MinRightEnd*$[i]$ (respectively, a factor $w[j..i-1]$ with $j \leq$ *MaxLeftEnd*$[i]$) we can generate $w[i'..j]$ (respectively, $w[j..i']$) for some $i' \leq i$ (respectively, $i' \geq i$); moreover, from a factor $w[i+1..j']$ with $j' <$ *MinRightEnd*$[i]$ we cannot generate any factor $w[i'..j']$ (respectively, from $w[j'..i-1]$ with $j' >$ *MaxLeftEnd*$[i]$ we cannot generate $w[j'..i']$) with $i' \leq i$ (respectively, $i' \geq i$).

We can now move on to the main results of this section.

**Theorem 3.2.3** *Given a word $w$ of length $n$ we can identify the minimum $i \leq n$ such that $w \in SSC^*(w[1..i])$.*

*Proof.* For each $1 \leq i \leq n$, let $L[i]$ be 1 if $w \in SSC^*(w[1..i])$ and 0 otherwise. We show how to compute the values of the array $L[\cdot]$.

We first compute all the runs of the input word in linear time. We sort the runs with respect to their ending position; as these ending positions are between 1 and $n$, and we have $\mathcal{O}(n)$ runs, we can clearly do this sorting in linear time, using, e.g., count sort.

Now, we observe that if there is a square $w[i..i+2\ell-1]$ then, if $L[i+2\ell-1] = 1$, we have $L[j] = 1$, for all $j$ such that $i+\ell-1 \leq j \leq i+2\ell-2$, as well. Indeed, from $w[1..j]$ we can obtain in one suffix square completion step $w[1..i+2\ell-1]$, if $i+\ell-1 \leq j \leq i+2\ell-2$, and then from $w[1..i+2\ell-1]$ we can obtain $w$. Generally, if $w[i..j]$ is a maximal run of period $\ell$, and $L[j] = 1$ then, for all $i+\ell-1 \leq k \leq j-1$, we have $L[j] = 1$. On the other hand, if $L[j] = 1$, then $j$ must necessarily fall inside a run of $w$.

Based on the previous observations, and on Lemma 3.2.6, we get that the sequence of 1 values in the array $L$ forms, in fact, a contiguous nonempty suffix of this array. It is, thus, enough to know the starting point of this suffix.

We are now ready to present the general approach we use in our algorithm. As a first step in our algorithm, we set $L[n] = 1$ as $w \in SSC^*(w)$ clearly holds. Also, during the computation, we maintain the value $\ell$ of the leftmost position we discovered so far such that $L[\ell] = 1$; initially $\ell = n$. We now go through all positions of $w$ from $n$ to 1 in decreasing order. When we reach some position $j$ of the word $w$, for every maximal run $r = w[i..j]$ of period $p$ we test whether $i+p-1 < \ell$; if so, we update $\ell$ and set $\ell = i+p-1$.

After this traversal of the word, we make $L[i] = 1$ for all $i \geq \ell$. It is clear that our algorithm runs in linear time. □

We recall definitions about ancestors and put them in the context of $PSSC$. A factor $w[i..j]$ of a word $w$ is its ancestor if $w \in PSSC^*(w[i...j])$. An ancestor is primitive if it cannot be derived from any of its factors.

**Observation 3.2.1** *A factor $w[i...j]$ is primitive if there is no square that starts in $i$ or finishes in $j$ that is completely inside the factor $w[i...j]$.*

The observation is clear from the definition of the operation. Note that we can compute in linear time the shortest square starting in each position and the shortest square ending in each position (see section 2.4). Thus after linear computations we can answer in constant time if a factor is primitive.

There are words (for example $a^n$) which can be derived by prefix-suffix completion from all their factors. However, by Lemma 3.2.7, given a word $w$, for each position $i$ of $w$ where a factor generating $w$ starts, there is a minimum $j_i \geq i$ such that $w \in PSC^*(w[i..j_i])$; then, for all $j \geq j_i$ we have that $w \in PSC^*(w[i..j])$, thus any such factor of $w$ is one of its ancestors. Therefore, we are interested in computing for each position of a word, the shortest factors starting there, from which $w$ can be derived.

**Theorem 3.2.4** *Given a word $w$ of length $n$, for the prefix-suffix square completion operation we can identify in $\mathcal{O}(n)$ time:*

1. *The number of ancestors of $w$*

2. *The shortest ancestor of $w$*

*Proof.* We start by computing the arrays *MinRightEnd* and *MaxLeftEnd* and preprocess them so that we can answer range maximum and, respectively, minimum queries on their ranges in constant time. For the reminder of the demonstration we will write $RMQ(i,k)$ for the minimum value among $MaxLeftEnd[i..k] = \{MaxLeftEnd[i], MaxLeftEnd[i+1], \ldots, MaxLeftEnd[k]\}$ and $posRMQ(i,k)$ for respective position of that minimum value (if there are more positions with the same value we will consider the rightmost).

Our approach will be to find for each $i$ the value of $j_i$ defined as above. By Lemma 3.2.7 it follows that $j_1 \leq j_2 \leq \ldots \leq j_{n-1} \leq j_n$.

We start with computing $j_1$, using the algorithm in theorem 3.2.3. We then explain how to compute $j_2$. Note that a sufficient condition for $w \in w[2..j_1]$ is that $MinRightEnd[1] \leq j_1$. Indeed, if there is a square that starts on position 1 and finishes before or on $j_1$, then we can derive $w[1..j_1]$ from $w[2..j_1]$ in one prefix square completion step, and, thus, we can derive $w$ from $w[2..j_1]$. This means that we can set, in that case, $j_2 = j_1$. The condition is not necessary, as there could be several suffix square completion operations doable starting from $w[2 \ldots j_1]$ that would expand this factor to $w[2 \ldots p]$ with $p \geq MinRightEnd[1]$; so we could start by applying these completions first, and then derive $w[1..p]$, and then do the completions that allow us to construct $w$. However, we can easily test whether this is the case. More precisely, if $RMQ(j_1, MinRightEnd[1]) \geq 2$ then each position $j$ with $j_1 \leq j \leq MinRightEnd[1]$ is contained in a second half of a square that starts at least on position 2. Thus, by successive suffixes completions we can obtain, in order, longer and longer factors $w[2..m]$ that contain position $j$, for each $j_1 \leq j \leq MinRightEnd[1]$. In the end, we get a factor $w[2..m]$ that contains $MinRightEnd[1]$; we then derive $w[1..m]$ and from this one we can derive the entire $w$. So, also in this case, we can simply set $j_2 = j_1$. If this is not the case, then we set $j_2 = posRMQ(j_1, MinRightEnd[1])$. Indeed, we could not derive a factor that contains 1 from any factor that does not contain $MinRightEnd[1]$, and, in order to produce $MinRightEnd[1]$ from a factor that starts on 2, we must be able to derive factors containing all the positions $j$ between $j_1$ and $MinRightEnd[1]$ with $MinRightEnd[j] = 1$ (otherwise, we could not derive any factor covering these positions from smaller factors of our word).

Further, we consider the case where we know $j_1, \ldots j_{i-1}$ and want to compute $j_i$. We first verify if $MinRightEnd[i-1] \leq j_{i-1}$. Just like above, this means that we could derive $w[j..j_{i-1}]$ with $j \leq i-1$ from $w[i..j_{i-1}]$; then we could continue and derive the whole $w$. In this case we simply set $j_i = j_{i-1}$ and continue with the computation of $j_{i+1}$.

Secondly, we test whether $RMQ(j_{i-1}, MinRightEnd[i-1]) \geq i$. If this is the case, we set $j_i = j_{i-1}$, just like in the case when $i = 2$. Indeed, we can first obtain $w[i..p]$ for some $p \geq MinRightEnd[i-1]$ from $w[i..j_{i-1}]$; then we obtain $w[i-1..p]$ in one step, and from this factor we can derive $w$ in multiple steps.

On the other hand, if $RMQ(j_{i-1}, MinRightEnd[i]) < i$ then we cannot proceed as above. Let $\ell = posRMQ(j_{i-1}, MinRightEnd[i])$. We keep updating $\ell = posRMQ(\ell, MinRightEnd[i])$ while $RMQ(\ell, MinRightEnd[i]) < i$. At the end of this while cycle, $\ell$ will give us the rightmost position $j$ between $j_{i-1}$ and $MinRightEnd[i]]$ such that any square that contains $j$ in the second half starts on a position before $i$. So, basically, it is impossible to derive a factor that contains $\ell$ from a factor starting on $i$: to obtain $\ell$ we need a factor that contains (at least) $i-1$, while to obtain $i-1$ we need a factor that ends after $\ell$. Thus, we set $j_i = \ell$. It is immediate that $w$ can be derived from $w[i..j_i]$.

Repeating this process for all $i$, we compute correctly all the values $j_i$ as defined in the statement.

Having computed $j_i$ for each $i$ it is to compute the two values:

1. The number of ancestors is

$$\sum_{i=1}^{n} n - j_i + 1.$$

   This can be easily computed in linear time.

2. The shortest ancestor will be of course one of factors $w[i...j_i]$ as other factors that start with $i$ will be longer. So the shortest ancestor will be the factor $w[i...j_i]$ such that $j_i - i <= j_k - k$ for all $1 \leq k \leq n$. Of course it might not be unique.

To compute the complexity of this algorithm, it is enough to evaluate the time needed to compute $j_i$. Clearly, when computing $j_i$ we execute $\mathcal{O}(j_i - j_{i-1})$ *RMQ* and *posRMQ* queries and a constant number of other comparisons. Hence, our algorithm runs in linear time. □

Computing the longest primitive ancestor is a bit more complicated and we were only able to find a $\mathcal{O}(n\log(n))$ solution, that we will present in the next chapter.

**Open Problem 3.2.1** *Can the longest primitive ancestor of a word in respect to the PSSC operation be determined in linear time?*

A direct consequence of the previous theorem is that we can decide in linear time whether a shorter word generates a longer word by iterated prefix-suffix square completion.

**Theorem 3.2.5** *Given two words $w$ and $x$, with $|x| < |w| = n$, we can decide in $\mathcal{O}(n)$ time whether $w \in PSSC^*(x)$.*

*Proof.* We first run a linear time pattern matching algorithm (e.g., the Knuth-Morris-Pratt algorithm [40]) to locate all the occurrences of $x$ in $w$. Then we run the algorithm from the proof of theorem 3.2.4 to find, for each position $i$ of $w$, the shortest factor $w[i..j_i]$ such that $w \in PSSC^*(w[i..j_i])$. Finally, we just have to check whether one of these factors is contained in an occurrence of $x$ starting at the same position. □

We leave as an open problem the problem of computing the maximum number of primitive roots for the bounded prefix-suffix duplication and prefix-suffix square completion operation.

### Common Ancestor problem

All of the results in this subsection appear here for the first time.

We now focus on the following problem:

**Problem 3.2.1** *Given two words $x$ and $y$, with $|x| = n \geq m = |y|$. Find a word $z$ such that $x \in \Theta^*(z)$ and $y \in \Theta^*(z)$, where $\Theta \in \{PSD, PSD_k, PSSC\}$.*

Again there are multiple variants of the problem to be taken into account: any ancestor, the shortest or longest ancestor or the number of ancestors.

All of these problems can be solved in $\mathcal{O}(n^2 \log(n))$ in the case of the $PSD$ operation. As proven in [32] we can find all $PSD$-ancestors of a word in $\mathcal{O}(n^2 \log(n))$. After this we can add all ancestors of one of them in a trie in quadratic time and than query all of the ancestors from the second word in the trie. In order to obtain quadratic time for the intersection we insert all ancestors that start from a position $i$ in linear time. While doing the intersection it is easy to compute all of the variants of the problem.

Thus, we will focus on the other two operations starting with the bounded prefix-suffix duplications.

The following theorem will be useful throughout this section:

**Theorem 3.2.6** *For every ancestor $z$ of $x$ which is not primitive there exists a primitive ancestor $z_p$ of $x$ such that $z \in PSD_k^*(z_p)$. Furthermore, the shortest ancestor of a word is primitive.*

*Proof.* Let $z$ be the an ancestor of $x$ that is not primitive. As $z$ is not primitive there exists a word $z_0$ such that $z \in PSD_k^*(z_0)$. Evidently $z_0$ is a factor of $z$ and of $x$. If $z_0$ is primitive our proof is finished. If not we can repeat the same logic starting with $z_0$, and obtain other words $z_1, ..., z_p$, as x is a finite word and $z_i > z_{i+1}$ the process will finish and we will obtain a primitive word in at most $|x| - 1$ steps. Of course, we have $z_p \in z_{p-1} \in ...z_1 \in z_0 \in x$. This concludes the first part of our proof.

Let us presume the shortest ancestor of $x$ is $z$ which is not primitive. This is evidently false as $z$ will have a primitive factor that is its own and $x$'s ancestor. □

The above theorem holds for $PSSC$ operation as well.

The following result is a direct consequence of the above one:

**Remark.** *Given two words $w_1$ and $w_2$, if they have a common ancestor, they also have a primitive common ancestor and their shortest ancestor is primitive.*

Now we know that in order to find an ancestor of two words and their shortest ancestor we only have to search in the primitive ancestors of the words.

**Theorem 3.2.7** *Problem 3.2.1 can be solved in $\mathcal{O}(nk\log(k) + n^2)$ in the case of the bounded prefix-suffix duplication operation. Furthermore, we can find the shortest common ancestor of two words in the same time complexity.*

*Proof.* We can simply determine all the ancestors of the two words and intersect them using tries. Obtaining all ancestors costs $\mathcal{O}(nk\log(k)) + |output|$, which can be at most $\mathcal{O}(n^2)$. The intersection costs $\mathcal{O}(n^2)$ time and space, thus we obtain a total time complexity of $\mathcal{O}(nk\log(k) + n^2)$, and $\mathcal{O}(n^2)$ space. □

**Open Problem 3.2.2** *Can the common ancestor problem in respect to the bounded prefix-suffix duplication operation be solved in $\mathcal{O}(nk\log(k))$ ?*

We now discus the problem in the case of the prefix-suffix square competition.

**Theorem 3.2.8** *Problem 3.2.1 can be solved in $\mathcal{O}(n)$ time and $\mathcal{O}(n)$ memory for the prefix-suffix square completion operation. Furthermore, we can find the longest common ancestor of two words in the same time complexity and the shortest common ancestor in $\mathcal{O}(n\log(n))$ time, with linear memory.*

*Proof.* We start by computing the $j$ array as in theorem 3.2.4 for both words. We also compute the Suffix Array and LCP Array for the word $x0y$ where 0 is a character that does not appear in $w_1$ and $w_2$. Our strategy will be to manipulate the appearances of the prefixes of the first word in the SA so that we can efficiently detect a common ancestor. The solution is based on the idea of passing through the SA and doing the following two operations:

If the element from the SA is a prefix in the first word then we do DELETE and UPDATE in our data structures.

If the element from the SA is a prefix in the second word then we do DELETE and QUERY in our data structures.

Before going into details we need to prove the following claims:

1. Given two positions in the suffix array $x$ and $y$ where $x$ is the position of a prefix $i_1$ from $w_1$ and $y$ is the position of a prefix $i_2$ from $w_2$ then $w_1$ and $w_2$ have a common ancestor iff:

$$LCP(x, y) \geq max(j_1[i_1], j_2[i_2])$$

2. Given three position in the suffix array $x_1 \leq x_2 \leq y$ there $x_1, x_2$ are position of prefixes $i_1, i_2$ from $w_1$ and $y$ is the position of a prefix $i_3$ from $w_2$, then if $j_1[i_2] \geq j_1[i_1]$ and there is an ancestor of $w_1, w_2$ starting at position $i_1$ in $w_1$ and at position $i_3$ in $w_2$ then there is a common ancestor of the two words that starts at positions $i_2$ in $w_1$ and at position $i_3$ in $w_2$.

*Proof.*

1. It is easy to see that if the above condition is satisfied then $w_1[i_1...i_1 + max(j_1[i_1], j_2[i_2])]$ is a common ancestor as it is equal to $w_2[i_2...i_2 + max(j_1[i_1], j_2[_2])]$. Furthermore, all factors $w_1[i_1...i_1 + max(j_1[i_1], j_2[i_2])], ..., w_1[i_1...i_1 + LCP(x, y)]$ are common ancestors of $w_1, w_2$. Conversely if $LCP(x, y) < max(j_1[i_1], j_2[i_2])$ then the common prefix is shorter than the shortest ancestor starting at that position for one of the two words, thus there is no common ancestor starting at those positions.

   Note that based on the previous claim and the fact that we can compute LCP's in $\mathcal{O}(1)$ time we could obtain a $\mathcal{O}(n^2)$ algorithm, because there are $\mathcal{O}(n^2)$ pairs to test. Thus we obtain a quadratic time and linear space algorithm (a space improvement compared to the previous solution using tries that works for PSSC). We will further refine the idea.

2. Based on the previous claim, if the two words have a common ancestor at $i_1, i_3$ then $LCP(x_1, y) \geq max(j_1[i_1], j_2[i_3])$ but $LCP(x_2, y) \geq LCP(x_1, y)$ and $max(j_1[i_2], j_2[i_3]) > max(j_1[i_1], j_2[i_3])$, thus $LCP(x_2, y) \geq max(j_1[i_2], j_2[i_3])$, so we have a common ancestor of $w_1, w_2$ that starts at position $i_1, i_3$.

□

The above claims suggest that if we look for one solution and even the shortest solution for the common ancestor problem, then we can ignore a position in the SA if we find any position after it that has a smaller or equal $j_1$ value, this will ensure that our elements will be sorted increasingly by both $j_1$ and $LCP$ distance to any of the following elements. The first claim also suggests that we should throw out from our data structure all elements $x$ for which $LCP(i_1, x) < j_1(x)$.

Basically the DELETE operation will ensure that that we don't have any elements that have a $j_1$ value bigger than the LCP between their position and the current position. As our stack is increasing we simply pop all elements that do not have this property.

When a new element of the first word comes into play we do a DELETE operation followed by an UPDATE in our stack. The UPDATE operation will have two steps, in the first we pop all elements from the stack that have a greater $j_1$ and then we insert the new element in the stack.

When a new element of the second word in the next element in our SA we do a DELETE operation and then do a Query operation.

For the QUERY operation we have to find if there is any element $i$ in our stack that has $j_1[i] \leq LCP(i, pos)$ where $pos$ is the current position. Of course, if $j_2[pos] > LCP(pos, pos-1)$ we have no solution for our query(the LCP of pos and all elements behind him will be less than $j_2[pos]$ the shortest length of an ancestor starting in $pos$). We have already eliminated in the DELETE operation any elements that have $j_1[i] < LCP(i, pos)$, thus, if there is any element left in the stack after the DELETE operation, the top element will have $j_1[i] \leq LCP(i, pos)$ giving us a solution for the common ancestor problem: $w1[i....i + LCP(i, pos) - 1]$ $=w2[pos...pos + LCP(i, pos) - 1]$. Furthermore, as all other elements $k$ left in the stack have $LCP(k, pos) \leq LCP(i, pos)$ this is the longest common ancestor that starts at position $pos$ in $w2$ (in respect to prefixes that are smaller than pos in the SA, the converse will be discussed at the end of the proof). We continue the same logic for all elements in the suffix array and keep in mind the maximum solution of all solution found, thus, obtaining the maximum common ancestor (and of course a common ancestor if any exists).

In order to compute the minimum common ancestor, we consider the stack to be implemented as an array, then use binary search to find the leftmost element $i$ such that $LCP(i, pos) \geq j_2[pos]$(if the stack is not empty we will find such a position), $j_1[i] \leq LCP(i, pos)$ as it would have been eliminated at DELETE otherwise. We have found a common ancestor of the two words: $w1[i....max(j_1[i], j_2[pos])] = w2[pos...max(j_1[i], j_2[pos])]$. Furthermore, as all elements $k$ to the right of $i$ have $j_1[k] \geq j_1[i]$, we have obtained the shortest common ancestor in respect to position $pos$ in $w2$ (in respect to prefixes that are smaller than pos in the SA, the converse will be discussed at the end of the proof). This takes $\mathcal{O}(\log(n))$ per minimum ancestor query.

In order to solve the common ancestor and longest common ancestor query we needed to compute SA and LCP arrays that take linear time. Furthermore, each element is inserted and extracted only once from the stack and we do one LCP operation per Query thus we obtain a linear time for this two problems. In the case of shortest common ancestor we need to do $n$ binary searches, thus we obtain a final time complexity of $\mathcal{O}(n\log(n))$. We use $\mathcal{O}(n)$ memory in all of the cases.

□

## 3.3 Distance related problems

We now focus on different problems related to the distance between two words. We recall that the prefix-suffix duplication distance between two words $w$ and $x$ is defined as follows:

$$\pi(w,x) = \begin{cases} \text{the minimum number } \ell \text{ such that } w \in PSD^{\ell}(x) \text{ or } x \in PSD^{\ell}(w), \\ \infty, \text{ if } w \notin PSD^{*}(x) \text{ and } x \notin PSD^{*}(w). \end{cases}$$

Computing the PSD distance has been studied in [32] and the authors found two algorithms one in $\mathcal{O}(n^3)$ time and $\mathcal{O}(n^2)$ space, and a second one in $\mathcal{O}(n^2\log(n))$ time and space. This leaves as with an interesting algorithm question.

**Open Problem 3.3.1** *Is there an $\mathcal{O}(n^2\log(n))$ time, and $\mathcal{O}(n^2)$ space algorithm for computing the PSD distance?*

In the following, we show how the $k$-prefix-suffix-duplication distance between two words can be efficiently computed.

For a word $w$ of length $n$, we define the array $SDD_k[\cdot]$ with $n$ integer elements. For $1 \leq i \leq n$, the value $SDD_k[i]$ is defined as the minimum number of $k$-suffix duplication needed to obtain $w$ from $w[1\dots i]$ (or, $\infty$ if $w$ cannot be obtained from $w[1\dots i]$). That is, $SDD_k[i]$ is the $k$-suffix duplication distance between $w[1..i]$ and $w$. Similarly, we can define the array $PDD_k[\cdot]$, where the value $PDD_k[i]$ equals the minimum number of $k$-prefix duplication needed to obtain $w$ from $w[i\dots n]$ (or $\infty$ if $w$ cannot be obtained from $w[i\dots n]$). In this case, $PDD_k[i]$ is the $k$-prefix duplication distance between $w[i..n]$ and $w$.

The algorithm we devise for computing the distance $\delta_k(x,w)$ between two words $x$ and $w$ is based on the following lemmas.

**Lemma 3.3.1** *Given a number $k \geq 1$ and a word $w$ of length $n$, with $n \geq k$, we can compute all the values $SDD_k[i]$ for $1 \leq i \leq n$ in $\mathcal{O}(n\log(k))$ time.*

*Proof.* Let us begin the proof by giving a high-level description of our algorithm. We use dynamic programming to compute the value $SDD_k[i]$ for each $i$ from $n$ to 1 in decreasing order. Basically, when we have to compute $SDD_k[i]$ for some $i$ we know already all the values $SDD_k[j]$ for $j > i$. Also, we note that if $w \in SD_k^*(w[1..i])$ then there must be some $j = i+\ell$ such that $w[1..j] \in SD_k(w[1..i])$ and $w[i-\ell+1..i] = w[i+1..i+\ell]$. This means that there exists a suffix of $w[1..i]$ that has the form $t^d$ for some primitive word $t$ such that $w[i+1..j] = t^s$, and $d \geq s \geq 1$. So, we analyze all the possible repetitive factors $t^p$ of $w$ such that $t$ is primitive, $w[1..i]$ ends with a power of $t$ and at least one $t$ occurs at position $i+1$. For each such factor, we see what possible choices we have for $j$, and then select the one $j$ such that $SDD_k[j]$ is minimum; $SDD_k[i]$ is then set to be equal to $SDD_k[j]+1$.

There are two steps in the approach described above that need a special attention. Firstly, we should clarify how to find efficiently for each $i \leq n$ the factors of the form $t^p$ such that $t$ is primitive, $w[1..i]$ ends with a power of $t$, and $w[i+1..n]$ begins with a power of $t$. Secondly, we should clarify how to identify efficiently, for each primitive word $t$ determined in the previous step, the number $j > i$ such that $w[1..j] \in SD_k(w[1..i])$, $w[i+1..j]$ is a power of $t$, and $SDD_k[j]$ is minimum. We explain how these steps are implemented in the following:

To begin with, in a first phase, we identify all factors $w[i..j]$ of $w$ that can be written as $t^s$, for a primitive word $t$ with $|t| \leq k$, but with $w[i-|t|..i-1] \neq t \neq w[j+1..j+|t|]$; in other words, these factors have the form $t^s$ but cannot be extended to a factor $t^{s+1}$ neither to the left nor to the right. We call such factors maximal repetitions (with respect to the possibility of being extended by their primitive root). As each such repetition starts with a primitively rooted square, and the length of the root is at most $k$, then we can obtain all of them (i.e., their starting position and the length of their primitive root) in $\mathcal{O}(n\log(k))$. Assume that we store these repetitions in a linear data structure, sorted by their starting position and, in case of equality, by the length of the root; let $q$ be the number of these maximal repetitions. Now, for each $i \leq n$ we define a list $L_i$, initially empty. Next, if $w[i..j] = t^s$ is the $r^{th}$ maximal repetition in the list of such repetitions, then we append $(i, |t|, r)$ to the lists $L_i$, $L_{i+|t|}$, $L_{i+2|t|}, \ldots, L_{j-|t|+1}$. This takes $\mathcal{O}(n\log(k))$ time in total, as well. Indeed, $t^2$ is a primitively rooted square occurring at positions $i$, $i+|t|, \ldots, j-2|t|+1$. So, when considering all possible choices for $t$ we will make as many insertions in the lists as primitively rooted squares we have in a word, plus once more the number of maximal repetitions (coming from the fact that each $t$ as above is added to the list $L_{j-|t|+1}$). As a consequence, the sum of the number of elements that these lists contain is also bounded by $\mathcal{O}(n\log(k))$.

At this moment we achieved the first of our goals. For each $i$, the list $L_i$

provides us the factors of the form $t^p$ such that $t$ is primitive, $w[1..i]$ ends with a power of $t$, and $w[i+1..n]$ begins with a power of $t$.

When computing $SDD_k[i]$ we start by computing, for each $(i, |t|, r) \in L_i$, the value $m_{i,t}$ which equals the minimum number of $k$-suffix duplication steps needed to derive $w$ from $w[1..i]$, such that in the first such duplication step we append a power of $t$ to $w[1..i]$. Now, if $w[1..j]$ is the result of a duplication in which a power of $t$ is appended to $w[1..i]$, then $L_i$ contains a triple $(i', |t|, r)$, $w[i'..i] = t^s$ and $w[i+1..j] = t^\ell$ for some $\ell \leq \min\{s, \frac{k}{|t|}\}$. We are interested in finding efficiently the value of $j$ that fulfills the conditions above and $SDD_k[j]$ is minimum. To do this, we note that if $j_1$ and $j_2$ are such that $i < j_1 < j_2 \leq \min\{i+k, i+|t|s\}$, $w[i+1..j_1], w[i+1..j_2] \in \{t^*\}$, and $SDD_k[j_1] \leq SDD_k[j_2]$ then $j_2$ should not be analyzed when computing the value of $j$ that fulfills the aforementioned conditions and has $SDD_k[j]$ minimum. Moreover, $j_2$ should not be considered when computing the values of $m_{r,|t|}$ for any $r \in \{i', i'+|t|, \ldots, i\}$, as $j_1$ would be in any of these cases a better choice for the index $j$ we are looking for.

This suggests that we should maintain, for each maximal repetition $w[i'..j'] = t^p$, that is ranked $r^{th}$ in the ordered list of the maximal repetitions, a list $D_r$ of the pairs $(j, SDD_k[j])$. The list $D_r$ is ordered increasingly with respect to the first component and fulfills the following conditions. First, $w[1..j]$ should end with a prefix $t^\sigma$ of the maximal repetition. Then, when computing $m_{i,|t|}$, where $i$ is such that $(i', |t|, r) \in L_i$, each component $(j, SDD_k[j])$ of the list should have $i < j \leq \min\{i+k, i+s'|t|\}$. Further, the list should be ordered decreasingly with respect to the second components of the pairs, and maximal with respect to inclusion. By the latter, we mean that adding to $D_r$ in the right position according to the order of the first components to $D_r$, when $m_{i,|t|}$ is computed, a new pair $(j', SDD_k[j'])$ such that $w[1..j']$ ends with a factor $t^\sigma$ and $j' > i$ would violate either the restriction that the list is ordered decreasingly according to the second component or the aforementioned upper bound on the first component of the pairs. Clearly, once $D_r$ is constructed, we have that $m_{i,|t|} = SDD_k[j] + 1$ where $(j, SDD_k[j])$ is the last element in the list (i.e., the one with the greatest first component and smallest second component). Then, $SDD_k[i]$ is the minimum of the values $m_{i,t}$ obtained for each triple $(i', |t|, r) \in L_i$.

Accordingly, before computing effectively the values of the array $SDD_k[\cdot]$, we need to introduce some more data structures helpful to store the lists described above. For $1 \leq r \leq q$ we define the (initially empty) deque $D_r$. These deques are supposed to implement the lists described above. The time needed to initialize these structures is $\mathcal{O}(n \log(k))$.

We now start computing the values of the array $SDD_k[\cdot]$ by setting $SDD_k[n] = 0$. For each maximal repetition $w[i..n]$, if it is the $r^{th}$ in the list of maximal repetitions, we add $(n, 0)$ to $D_r$.

As already explained, we compute $SDD_k[i]$ for each $i$ from $n$ to 1 in decreasing order. Let us fix such an $i$, and explain how $SDD_k[i]$ is computed efficiently using the data structures we defined above and the values $SDD_k[j]$ for $j > i$.

Let us assume that before computing $SDD_k[i]$, each deque $D_r$ for which the triple $(i', d, r)$ is in $L_i$ fulfills the conditions stated in the definitions of these lists corresponding to the moment before $SDD_k[i]$ should be computed. This holds canonically after the first step, in which the deques are initialized and the pair $(n, 0)$ was added in some of these deques.

We go through the list $L_i$. For each triple $(i', d, k)$ from $L_i$ we do the following. Let $t = w[i + 1..i + d]$. We compute the largest power $s$ such that $t^s$ is a suffix of $w[1..i]$. We delete from the back of the deque $D_k$ all the pairs $(j, SDD_k[j])$ with $j > i+sd$. Now, the deque $D_r$ has exactly the form we needed in order to compute easily the value of $m_{i,t}$. Let $(j_{i,d}, SDD_k[j_{i,d}])$ be the element found at the back of the deque $D_k$. We set $m_{i,t} = SDD_k[j_{i,d}]+1$. After we went through all the elements of $L_i$, we set $SDD_k[i] = \min\{m_{i,t} \mid (i, |t|, k) \in L_i\}$. We then delete from the front of the deque $D_k$ all the pairs $(j, SDD_k[j])$ such that $SDD_k[i] \leq SDD_k[j]$. Finally, we insert $(i, SDD_k[i])$ into the deque $D_k$, and move on to compute $SDD_k[i-1]$. Now all the lists $D_r$ for which the triple $(i', d, r)$ is in $L_i$ (as well as the deques that were not involved in the computation of $SDD_k[i]$) are prepared to be used in the computation of the values $SDD_k[i'']$ for all $i'' < i$.

The correctness of the procedure follows immediately from the discussions above. The time complexity is $\mathcal{O}(n \log(k))$. This holds as the time needed to maintain the deques $D_r$, for $1 \leq r \leq q$ is proportional to the number of insertions and deletions we made in these structures. But the number of insertions in the deques is equal to the sum of the number of elements contained by the lists $L_i$ for all $i \leq n$. Indeed, for each element $(i, d, r) \in L_i$ we compute $m_{i,t}$ for $t = w[i..i + d - 1]$, then take the minimum of these values, save it as $SDD_k[i]$, and, finally, insert $(i, SDD_k[i])$ in every deque $D_r$. As the total number of elements in the lists $L_i$ for $i \leq n$ is $\mathcal{O}(n \log(k))$, the total number of insertions in the deques is also $\mathcal{O}(n \log(k))$. Then, each pair is deleted exactly once from each deque it belongs to. By the definition of the deques, the insertion and deletion operations are only made at their back or front, so we never have to go through their elements. In conclusion, the overall time used to maintain these data structures is $\mathcal{O}(n \log(k))$. The first phase of our algorithm, in which the maximal repetitions are detected and the lists $L_i$ computed, also runs in $\mathcal{O}(n \log(k))$ time. In conclusion, the overall complexity is upper bounded by $\mathcal{O}(n \log(k))$. The space complexity is also $\mathcal{O}(n \log(k))$. □

The following lemma follows in an identical fashion.

**Lemma 3.3.2** *Given a number $k \geq 1$ and a word $w$ of length $n$, with $n \geq k$, we can compute all the values $PDD_k[i]$ for $1 \leq i \leq n$ in $\mathcal{O}(n \log(k))$ time.*

Related to the previous results, it seems interesting to note that the strategy employed in [32] to compute the suffix (or prefix) duplication between two words cannot be immediately used to compute the $k$-suffix (respectively, prefix) duplication distance in the time complexity stated above. In the algorithm presented in [32] there is no mechanism by which the length of the duplicated factor can be bounded, and we see no direct way to implement such a mechanism without making use of a strategy similar to the one already used above. The next lemma easily follows now:

**Lemma 3.3.3** *Given $k \geq 1$, and two words $x$ and $w$ of respective length $m$ and $n$, with $n \geq m \geq k$, $\delta_k(x, w)$ can be computed in $\mathcal{O}(n \log(k))$.*

*Proof.* Just like in the proof of Lemma 3.2.3, if $w$ is obtained from $x$ in a certain number of steps, then it can be obtained in exactly the same number of steps, but executing first only suffix duplications, and then only prefix duplications. Thus, to compute the distance between $x$ and $w$ we can proceed as follows. We first compute the arrays $SDD_k$ and $PDD_k$ in $\mathcal{O}(n \log(k))$ time, by Lemmas 3.3.1 and 3.3.2.

We then find all occurrences of $x$ in $w$; this takes linear time $\mathcal{O}(n + m)$, using, e.g., the Knuth-Morris-Pratt algorithm [40]. We now consider each of these occurrences, in order. Let $w[i..j]$ be one of these occurrences of $x$ in $w$. We can obtain $w$ from $w[i..j]$ in $SDD_k[j] + PDD_k[i]$ $k$-prefix-suffix duplication steps. Thus, we compute these values for all occurrences of $x$ in $w$, and then $\delta_k(w[i..j], w)$ will be the shortest of them. This part of the algorithm takes $\mathcal{O}(n)$ time, as we can have at most $n - k + 1$ occurrences of $x$ in $w$. The total time complexity of our algorithm is $\mathcal{O}(n \log(k))$. □

Consequently, we get the following theorem:

**Theorem 3.3.1** *Given $k \geq 1$, let $x$ and $w$ be two words of respective length $m$ and $n$, $n > m$. If $m \geq k$, then $\delta_k(x, w)$ can be computed in $\mathcal{O}(n \log(k))$. If $m < k$, then $\delta_k(x, w)$ can be computed in $\mathcal{O}(nk \log(k))$.*

*Proof.* We distinguish two cases. If $m \geq k$, then by Lemma 3.3.3 the value of $\delta_k(x, w)$ is computed in $\mathcal{O}(n \log(k))$ time. Consequently, we further focus on the case $m < k$.

In this case, we follow closely the approach in Theorem 3.2.1. First, the technique developed in Section 6.2 of [25] allow us to compute in $\mathcal{O}(nk \log(k))$ time the set $F = \{(i, j, d) \mid \delta_k(x, w[i, j]) = d \text{ and } k \leq j - i + 1 \leq 2k\}$. This set has $\mathcal{O}(nk)$ elements. Now, by a reasoning similar to Lemma 3.3.3, by computing $SDD_k[j] + PDD_k[i] + d$ for each $w[i..j]$ with $(i, j, d) \in F$ we get the minimum number of $k$-prefix-suffix duplication steps needed to obtain $w$ from $x$, provided that we first derive $w[i..j]$ from $x$ and then $w$ from $w[i..j]$. Then, we just have to note that due to the fact that in each duplication step at most $k$ symbols are added to the current word to obtain a new one, and that $|x| \geq k$, then at least one of the intermediate words obtained in a derivation of $w$ from $x$ should have length between $k$ and $2k$. Thus, we can get the $k$-prefix-suffix duplication distance between $x$ and $w$ by taking the minimum of the values $SDD_k[j] + PDD_k[i] + d$ with $(i, j, d) \in F$. So the time needed to compute $\delta_k(x, w)$, once $F$ is constructed is $\mathcal{O}(nk)$. The overall time complexity of computing this distance is, thus, $\mathcal{O}(nk \log(k))$. □

Though we have managed to obtain a complexity of $\mathcal{O}(n \log(k))$ case when $|x|, |w| \geq k$ we still have a $\mathcal{O}(nk \log(k))$ complexity for some cases, thus, over quadratic time complexity. We now proceed to computing the prefix-suffix square completion distance, we first show a quadratic solution for the problem and then show a different solution for the easier suffix square completion distance that works in linear time, as a first step to potentially solving the main problem in linear time.

Although this problem can be solved using dynamic programming as well, we will model this problem a bit differently. The model could be useful in future research so we consider it worth mentioning. The following is unpublished material.

We can consider each factor $w[i...j]$ of a word $w$ a node in a directed graph, with edges connecting two nodes if we can obtain the destination node from the source node with one operation (be it $PSD, PSD_k, PSSC$). To find whether a factor is an ancestor of the word or not, we can verify if there is a path from the factor to the word in the above mentioned graph. We can use a breadth first search algorithm [12] that has complexity $\mathcal{O}(|V| + |E|)$. Unfortunately, if we do not make any limitations our graph that has $\mathcal{O}(n^2)$ nodes can have up to $\mathcal{O}(n^3)$ edges. Thus even if our algorithm is general it is not optimal.

Note that we can extend this algorithm to generate the distance from a word to another quite easily, we first find all appearances of the small word in the second word using $KMP$ in linear time, and then we use a multiple source breadth first algorithm that also runs in $\mathcal{O}(|V| + |E|)$.

We start with a lemma that strongly reassembles the lemma 3.2.7.

**Lemma 3.3.4** *For any two factors $w[i_1..j_1]$ and $w[i_2..j_2]$ of a word $w$, if $i_1 \leq i_2 \leq j_2 \leq j_1$, then $PSSCD(w[i_2..j_2], w) \geq PSSCD(w[i_1...j_1], w)$.*

*Proof.* Let $PSSCD(w[i_1...j_1], w) = d_1$, $PSSCD(w[i_2..j_2], w) = d_2$ and $w_0 = w[i_2...j_2]$. No matter how we choose $w_1, w_2, ..., w_{d_2}$ factors of $w$ such that $w_l \in PSSC(w_{l-1})$ for all $1 \leq l \leq d_2$, there will be a first factor $w_p = w[i_p...j_p]$ such that $i_p < i_1$ or $j_p > j_1$, that is, $w_p$ is the first factor from $w_0, w_1, w_2, ..., w_{d_2} \notin$ $w[i_1...j_1]$. If $w_p$ is the first such factor than $w_{p-1}$ is included in $w[i_1...j_1]$. To obtain $w_p$ from $w_{p-1}$ we complete a square $sq$ this has at least a half in $w_{p-1}$, but, as $w_{p-1} \in w[i_1...j_1]$ than we can complete the same square $sq$ to obtain $w_p$ from $w[i_1...j_1]$. Naturally we can mirror the same square completion operation from there on to obtain $w$ from $w[i_1...j_1]$ in $d_2 - p + 1$ steps. So $d_2 < d_1$ only if $p = 0$, but we have chosen $w_p \notin w[i_1...j_1]$ and $w_0 = w[i_2...j_2] \in w[i_1...j_1]$ so $p \geq 1$, thus $d_2 \geq d_1$. □

**Theorem 3.3.2** *Given two words $x$ and $w$ of respective length $m$ and $n$, with $n > m$, then the prefix-suffix square completion distance can be computed in $\mathcal{O}(n^2)$.*

*Proof.* Our solution is based on the general description given above, that we manage to improve in this particular case. We again start by finding all appearances of $w$ in $x$ and then we do a multiple source breadth first. Based on the previous lemma we are able to modify the graph. We will create an edge in our graph from a factor $w[i...j]$ to only two other factors $w[i - MaxPrefSq(i,j)...j]$ and $w[i...j + MaxSufSq(i,j)]$, where $MaxPrefSq(i,j)$ is the longest square that has the middle in $i$ and finishes before $j$, $MaxSufSq$ is defined similarly. Thus, from every node we will only have two outgoing edges, resulting in $\mathcal{O}(n^2)$ edges. Thus, our multi source breadth first algorithm works in $\mathcal{O}(n^2)$ time. In order to prove the total complexity of the algorithm is quadratic we have to show how we compute the two values($MaxPrefSq(i,j), MaxSufSq(i,j)$) for every pair $(i,j)$. $MaxPrefSq(i,j)$ will either be equal to $MaxPrefSq(i,j-1)$ or with $j-i+1$ if $w[i-j+1,j]$ is a square. Testing if a factor is a square can be done in $\mathcal{O}(1)$ using an LCP query: $LCP(RANK(i-j+1), RANK(i)) \geq j-i+1$, thus $\mathcal{O}(1)$ for each pair $i,j$ and $\mathcal{O}(n^2)$ in total. □

**Lemma 3.3.5** *Given a word $w$ of length $n$, we can obtain an array $SSCD[]$ computing the suffix square completion distance from each prefix to the word in linear time.*

*Proof.* We start computing the maximum square ending on each position in the word $MaxSq$. We can do this in linear time as showed in section 2.4. We will also use an array $Limit[]$, where $Limit[i]$ will be the leftmost position $k$ in $w$ such that $SSCD[k] = i$. Of course $SSCD[n] = 0$ as we need zero duplications to obtain $w$ from itself, and we also set $limit[0] = n$. It is also clear that we can obtain $w$ by one operation from all prefixes $w[1...n - MaxSq[n] + 1]$ as we complete exactly the square $w[n - maxSq * 2 - 1, n]$, so we set $Limit[1] = n - maxsq[n] + 1$. We then compute $Limit[2] = mini - MaxSq[i] + 1 | Limit[1] \leq i \leq Limit[0]$, as for any element $k \geq Limit[2]$ we can generate $w$ through two suffix square completions. Similarly $Limit[i] = min\{i - MaxSq[i] + 1 | Limit[i-1] \leq i \leq Limit[i-2]\}$. After computing the $Limit$ array we can easily set $SSCD[i] = j$ for any $i$ such that $Limit[j] \geq i > Limit[j-1]$. Basically we have two passes through the array one to compute the limit values and the second to write the values in the $SSCD$ array. As all of our operations are constant the complexity of our algorithm is linear. □

We can of course compute a similar array to compute the prefix square completion distance. Unfortunately, we cannot directly say something like $PSSCD(w[i...j], w) = SSCD[j] + PSCD[i]$ as $w$ might not even be in $PSSC^*(w[i...j])$ as in the following example:

**Example 3.3.1** *$w = abacabac$, though $SSCD[7] = 1$, and $PSCD[5] = 1$, $w \notin PSSC^*(w[5...7])$.*

Given the above results we consider t there might exist a linear solution or at least a sub quadratic solution for the prefix-suffix square competition distance problem.

**Open Problem 3.3.2** *Is there a sub quadratic solution for the prefix-suffix square competition distance problem?*

We finish this section with an interesting result computing the distance between two BPSD languages. We have been able to obtain such a result only for the case of the BPSD operation.

The $k$-prefix-suffix duplication distance between two words can be extended to the $k$-prefix-suffix duplication distance between a word $x$ and a language $L$ by $\delta_k(x, L) = \min\{\delta_k(x, y) \mid y \in L\}$. Moreover, one can canonically define the distance between languages: for two languages $L_1, L_2$ and a positive integer $k$, we set $\delta_k(L_1, L_2) = \min\{\delta_k(x, y) \mid x \in L_1, y \in L_2\}$.

Before describing the main result of this section, we need some data structures prerequisites. In the algorithm described in the following proof, we are given two finite automata $A_1 = (Q, V, \delta', q_0, Q_f)$ and $A_2 = (S, V, \delta'', s_0, S_f)$ and a positive integer $k \geq 1$. We work with 5-tuples $(q, s_1, s_2, w_1, w_2)$ and 4-tuples $(s_1, s_2, w_1, w_2)$, where $w_1, w_2 \in V^*, |w_1| = |w_2| \leq k$ and $q \in Q$, $s_1, s_2 \in S$; moreover, whenever $|w_1| < k$ then $w_1 = w_2$.

A set $T$ of 5-tuples as above is implemented as a 3-dimensional array $M_T$, where $M_T[q][s_1][s_2]$ contains a representation of the set $\{(w_1, w_2) \in V^* \times V^* \mid (q, s_1, s_2, w_1, w_2) \in T\}$ which is implemented using a trie data structure essentially storing all possible words of length $k$, augmented with suffix links.

More precisely, for each set $M_T[q][s_1][s_2]$ we construct a trie of all the words $w \in V^k$. Such a trie is, in fact, a tree whose all leaves have depth $k$ and where each internal node has exactly $|V|$ children; the $|V|$ edges connecting each node to its the children are labeled with the letters of $V$, respectively. Basically, in such a tree there is a bijective relation between the paths connecting the root to other nodes (internal or leaves) and the words of $V^{\leq k}$. Assuming that the letters of $V$ are ordered, then the children of each node in the trie are also ordered, and there is an induced order on the nodes of depth $j$ (or, alternatively, on the paths of length $j$), for all $j \leq k$. We can also assume that the nodes of each trie we use are labeled with integers from 1 to $\frac{|V|^{k+1}-1}{|V|-1}$, in the order they are encountered in a level traversal of the tree, the nodes on the same level being considered in the induced order. This labeling is the same in each of the tries, so the internal node associated to a word $w$ has always the same label, denoted $\#(w)$; the labels can be seen as pointers to the real nodes of the trie. Finally, it is immediate that in this setting we can retrieve in constant time the label of the node corresponding to the word $x$ when we know the label of the node corresponding to $ax$: if $x$ labels the $p^{th}$ path among the children of the node which can be reached from the root by a path labeled with $a$, then the node corresponding to $x$ in the trie is the end of the $p^{th}$ path of length $|x|$ (and we know its label from the numbering procedure). Thus, for each internal node of the trie (say corresponding to the path $ax$) we compute the *suffix link*: a link connecting it to the node corresponding to the word $x$. The time needed to construct the trie and the suffix links is $\mathcal{O}(|V|^k)$.

Now, when manipulating the sets of tuples as above, instead of working with a pair of words $(w_1, w_2)$, we shall use the pair of labels $(\#(w_1), \#(w_2))$. However, in order to keep the notation simpler, we will keep using the notation $(q, s_1, s_2, w_1, w_2)$ (instead of $(q, s_1, s_2, \#(w_1), \#(w_2))$) for the tuples we work with. We just have to keep in mind that such a notation provides us pointers to the nodes corresponding to $w_1$ and $w_2$ in the trie representing the set $M_T[q][s_1][s_2]$; in other words, we keep track of a succinct representation of these words, instead of storing the full words. Further, to maintain the sets of tuples efficiently, if $(w_1, w_1)$ is in the respective set for some $w_1 \in V^{\leq k}$, then we mark the node corresponding to $w_1$ in the trie. If $(w_1, w_2)$ is in the set for some $w_1, w_2 \in V^k$, $w_1 \neq w_2$, then we add a *connecting edge* in the trie between the leaf corresponding to $w_1$ and the leaf corresponding to $w_2$. Using this representation of the tuples we can check in constant time whether a certain pair of words is in the set or not. That is, $(w_1, w_2)$ is in the set if $w_1 = w_2$ and $\#(w_1)$ is marked, or $w_1 \neq w_2$ and there is a connecting edge between $\#(w_1)$ and $\#(w_2)$. Adding a pair of words to a set can be also done in constant time: we check whether that pair of words is already in the set or not, and if not, we either have to mark the node $\#(w_1)$, if $w_1 = w_2$, or construct a connecting edge between $\#(w_1)$ and $\#(w_2)$; in both cases, we can do that in constant time according to our assumption that whenever having a pair of words $(w_1, w_2)$ we actually work with the pair of node-labels $(\#(w_1), \#(w_2))$.

Using this representation of sets of tuples we can check in constant time whether or not a certain pair of words (given as pair of nodes of the trie we construct) is in the set. The same strategy can be used for implementing a set of 4-tuples.

Using these data structures, we can show the following result.

**Theorem 3.3.3** *Given two regular languages $L_1$ and $L_2$ over an alphabet $V$, recognized by deterministic finite automata with sets of states $Q$ and $S$, respectively, and a positive integer $k \geq 1$, the distance $\delta_k(L_1, L_2)$ can be computed in $\mathcal{O}((k+N)M^2|V|^{2k})$ time, where $M = \max\{|Q|, |S|\}$ and $N = \min\{|Q|, |S|\}$.*

*Proof.* Let us assume that both $L_1$ and $L_2$ are given by the minimal deterministic finite automata accepting them, namely $A_1$ and, respectively, $A_2$. Let $A_1 = (Q, V, \delta', q_0, Q_f)$ and $A_2 = (S, V, \delta'', s_0, S_f)$. As a rule, we denote the states of $Q$ and $S$ by $q$ and $s$, respectively, with or without indices. Further, for a word $w$ we denote by $pref_k(w)$ its longest prefix of length at most $k$; similarly, let $suf_k(w)$ be its longest suffix of length at most $k$.

The algorithm computing $\delta_k(L_1, L_2)$ has two similar main parts. In the first one, we compute the minimum value $d_1$ such that there exists a word $x \in L_2 \cap PSD_k^{d_1}(L_1)$. In the second part, we compute, using exactly the same procedure, the minimum value $d_2$ such that there exists a word $y \in L_1 \cap PSD_k^{d_2}(L_2)$. Then, we conclude that $\delta_k(L_1, L_2) = \min\{d_1, d_2\}$. Hence, it suffices to describe how the minimum value $d_1$ such that there exists a word $x \in L_2 \cap PSD_k^{d_1}(L_1)$ is computed.

Essentially, our algorithm implements the following idea: if there exists $x \in L_2 \cap PSD_k^{d_1}(L_1)$ then $x$ must contain some factor $w \in L_1$ from which it was derived. As such an $x$ must be derived by $k$-bounded prefix-suffix duplications from $w$, it is enough to guess and memorize the length-$k$ prefix $w_1$ and suffix $w_2$ of $w$ (instead of the entire $w$, which could be, just like $x$, arbitrarily long), as well as the state $s_1$ where we start reading $w$ when we use $A_2$ to accept $x$ (as we do not know $x$ and $w$, we just make a guess here), and the state $s_2 = \delta_2(s_1, w)$ (we only know the prefix and suffix of length $k$ of $w$, so again we need to make a guess here). Then, we basically check (and this can be done deterministically) whether we can obtain a word that takes us from $s_0$ to a final state of $A_2$ by applying duplications to the word $w$ whose prefix and suffix we guessed; if yes, then we conclude that $x$ actually exists. This idea is non-deterministic, as it is based on several guessing steps; to solve this, we exhaustively try all the possibilities that we may have guessed.

As a preprocessing phase of our algorithm, we compute in $\mathcal{O}(k|Q|^2|V|^k)$ time (in a naive manner), for each $q_1 \in Q$ and $w \in V^{\leq k}$ all states $q_2$ such that $\delta'(q_2, w) = q_1$ and the state $q_3 = \delta(q_1, w)$. Provided that we use the ideas described before this proof (of storing words as labels of nodes from the trie, the label of $w$ being denoted $\#(w)$), we can store this information in space $\mathcal{O}(|Q|^2|V|^k)$, so that we can obtain in constant time, for $q_1$ and $\#(w)$, the states $q_2$ and $q_3$ defined above. We then process the automaton $A_2$ in a similar manner, in time $\mathcal{O}(|S|^2|V|^k)$.

We present now the main part of our algorithm. All the sets constructed in this phase are implemented using the data structures described before the theorem.

First, we compute the set
$R_0 = \{(s_1, s_2, w_1, w_2) \mid$ there exists $w \in L_1$ such that $\delta''(s_1, w) = s_2$, $pref_k(w) = w_1, suf_k(w) = w_2\}$.
This computation is done as follows. We compute iteratively the sets $T_s^i$, $i \geq 1$, each one containing the tuples $(q, s, s_1, w_1, w_2)$ for which there exists a word $w$ of length $i$, with $pref_k(w) = w_1$, $suf_k(w) = w_2$, $\delta'(q_0, w) = q$ and $\delta''(s, w) = s_1$, but there exists no word $w'$ shorter than $w$ with the same properties. Clearly, in such a 5-tuple, $|w_1| = |w_2|$ and if $|w_1| < k$ then $w_1 = w_2$. We can implement the union (over all values of $i$) of the sets $T_s^i$ by marking in a trie storing all the words of length $k$ over $V$ the nodes corresponding to the words of this set. The sets $T_s^i$ are computed as long as they are nonempty; now, if $T_s^i$ is empty, then the sets $T_s^j$ are empty, for all $j \geq i$. On the other hand, as the number of all the tuples $(q, s, s_1, w_1, w_2)$ as above is upper bounded by $2|Q||S||V|^{2k}$, there exists $i_0$ such that $T_s^i = \emptyset$ when $i \geq i_0$ and $T_s^{i_0-1} \neq \emptyset$. It is not hard to see that $T_s^{i+1}$ can be computed in time $\mathcal{O}(k|T_s^i|)$, given the elements of $T_s^i$. Indeed, for each 5-tuple $(q, s, s_1, w_1, w_2) \in T_s^i$ and letter $a \in V$, we compute the 5-tuple $(\delta'(q, a), s, \delta''(s_1, a), pref_k(w_1 a), suf_k(w_2 a))$; note that the nodes of the trie corresponding to the words $pref_k(w_1 a)$ and $suf_k(w_2 a)$ can be obtained in $\mathcal{O}(1)$ time, by knowing the nodes corresponding to $w_1$ and $w_2$ and using their suffix links. Then, if the new tuple does not belong to $\bigcup_{i=1}^{i} T_s^i$, we add it to $T_s^{i+1}$; by maintaining another trie-structure for $\bigcup_{i=1}^{i} T_s^i$, we obtain that checking whether an element is in this set or adding an element to it, is done in $\mathcal{O}(1)$ time. To go efficiently through the elements of $T_s^i$, we store them in a linked list.

We now set $\hat{T}_s = \bigcup_{i=1}^{i_0} T_s^i$. It follows that $\hat{T}_s$ is computed in $\mathcal{O}(|Q||S||V|^{2k})$ time. Therefore, $R_0 = \{(s_1, s_2, w_1, w_2) \mid (q, s_1, s_2, w_1, w_2) \in \bigcup_{s \in S} \hat{T}_s, q \in Q_f\}$. It is not hard to see that it takes $\mathcal{O}(|Q||S|^2|V|^{2k})$ time to compute $R_0$. We now set $\hat{R}_j = \bigcup_{i=0}^{j} R_i$ and iteratively compute the sets $R_j$, $j = 1, 2, \ldots$ as follows:
– $R_{j+1} = (R_{j+1}^1 \cup R_{j+1}^2) \setminus \hat{R}_j$,
– $R_{j+1}^1 = \{(s_1, s', w_1', w_2') \mid$ there exists $(s_1, s_2, w_1, w_2) \in R_j$, and $w' \in V^*$ a suffix of $w_2$, such that $\delta''(s_2, w') = s', pref_k(w_1 w') = w_1', suf_k(w_2 w') = w_2'\}$,
– $R_{j+1}^2 = \{(s', s_2, w_1', w_2') \mid$ there exists $(s_1, s_2, w_1, w_2) \in R_j$, and $w' \in V^*$ a prefix of $w_1$, such that $\delta''(s', w') = s_1, pref_k(w' w_1) = w_1', suf_k(w' w_2) = w_2'\}$.

Actually, $(s_1, s_2, w_1, w_2) \in R_j$ if and only if there exists a word $w$ which can be obtained by applying $j$ times the $k$-prefix-suffix duplication to a word from $L_1$ such that $pref_k(w) = w_1$, $suf_k(w) = w_2$, and $\delta''(s_1, w) = s_2$; Furthermore, there is no word $w'$ that fulfills the same conditions and can be obtained by applying less than $j$ times the $k$-prefix-suffix duplication to the words of $L_1$. Further, all the elements of these sets fulfill the conditions allowing us to use again a trie implementation for the union of the sets. Using this implementation, and additionally storing each $R_j$ as a list, the time needed to compute the set $R_{j+1}$ is upper bounded by $\mathcal{O}(k|R_j|)$. Indeed, first we construct $R^2_{j+1}$: for each tuple $(s_1, s_2, w_1, w_2) \in R_j$ and prefix $x$ of $w_1$, we use the precomputed data structures to obtain the state $s$ such that $\delta'(s, x) = s_1$ and decide that $(s, s_2, pref_k(xw_1), suf_k(xw_2))$ should be added to $R_{j+1}$ (but only if it is not already in other $R_{j'}$ with $j' < j + 1$). To implement this efficiently, we consider the prefixes of $x$ in increasing order with respect to the length, and so we will get the node corresponding to $xa$ in the trie in $\mathcal{O}(1)$ time when we know the node corresponding to $x$. Then we construct $R^1_{j+1}$: for each tuple $(s_1, s_2, w_1, w_2)$ and for each suffix $x$ of $w_2$, we use the precomputed data structures to obtain $s = \delta'(s_2, x)$ and decide that $(s_1, s, pref_k(w_1x), suf_k(w_2x))$ should be added to $R_{j+1}$ (again, only if it is not in other $R_{j'}$ with $j' < j+1$). This time we consider the suffixes $x$ of $w_2$ in decreasing order with respect to their length; in this way, we get the node corresponding to $x$ from the node corresponding to $ax$ in $\mathcal{O}(1)$ time using the suffix links. The sets $R_j$ are computed until either one meets a value $j_0$ such that $(s_0, s, w_1, w_2) \in R_{j_0}$ for some $s \in S_f$ and $w_1, w_2 \in V^{\leq k}$, or $R_j = \emptyset$. As the number of all 4-tuples that may appear in all the sets $R_j$ is bounded by $\mathcal{O}(|S|^2|V|^{2k})$, the computation of the sets $R_j$ ends after at most $\mathcal{O}(k|S|^2|V|^{2k})$ steps. It is clear that if the process of computing the sets $R_j$ ends by reaching the value $j_0$ mentioned above, then we conclude that $d_1 = j_0$. Otherwise, $d_1 = \infty$ holds. The correctness of the computation of $d_1$ follows from the discussions above.

Consequently, the total time needed to compute $d_1$ is $\mathcal{O}(|V|^k + k|Q|^2|V|^k + 2k|S|^2|V|^{2k} + |Q||S|^2|V|^{2k}) = \mathcal{O}(k|Q|^2|V|^k + |Q||S|^2|V|^{2k})$. We can use the same procedure to compute $d_2$, just by changing the roles of $L_1$ and $L_2$.

We return as $\delta_k(L_1, L_2) = \min\{d_1, d_2\}$. The time needed to compute this distance is $\mathcal{O}((k + N)M^2|V|^{2k})$, where $M = \max\{|Q|, |S|\}$ and $N = \min\{|Q|, |S|\}$. □

Note that if $V$ is a constant size alphabet, then the previous result provides a cubic algorithm computing the distance between two regular languages. The following corollary follows from Theorem 3.3.3 for $L_1 = \{x\}$ and $L_2 = L$.

**Corollary 3.3.1** *Given a word $x$, a regular language $L$ accepted by a DFA with $q$ states, and a positive integer $k \geq 1$, one can algorithmically compute $\delta_k(x, L)$ in $\mathcal{O}((k + |N|)|M|^2|V|^{2k})$ time, where $M = \max\{q, |x|\}$ and $N = \min\{q, |x|\}$.*

We leave as an open problem whether one can compute the distance between two languages in the case of the more complicate PSD and PSSC operations.

We finish the chapter with a short round up of the results within it.

In the first part of the chapter we discussed about the family of languages that we generated with our operations. We concluded that languages generated by bounded prefix-suffix generations are regular, whereas languages generated through prefix-suffix duplication and prefix-suffix square completion are not even in CF.

In the second section we showed how we can verify if a word is generated by our operations from another word and we gave algorithms to check this for each of the operations. The algorithms range from $\mathcal{O}(n^2 \log(n))$ time complexity for PSD to linear time for PSSC. We then discussed several problems related to ancestors and predecessors of a word, rising a few interesting open problems.

We concluded the chapter by giving algorithms for computing the distance between two words in the case of all our operations and the distance between two languages in the case of the bounded prefix-suffix duplication operation.

# Chapter 4

# Combinatorial aspects

This chapter is split into three sections. The first section covers generation of infinite words, the second provides results about suffix-square free words and prefix-suffix-square free factors, while the third covers gapped repeats and palindromes within a word. A paragraph or two explaining each in detail will follow. Most of the work presented here was published in [24, 22, 23].

- We first tackle the problem of generating finite and infinite words using the prefix-suffix square completion. We investigate how square completion operations can be used to generate infinite words. We say that a right-infinite word $\mathbf{w}$ can be generated by one of the square completion operations if there is an infinite sequence of finite prefixes of $\mathbf{w}$, namely $w_0, w_1, \ldots$, such that $w_i$ is obtained by the respective operation from $w_{i-1}$. For instance, the infinite Fibonacci word or the Period-doubling word can be generated by suffix square completion. furthermore,,, we show that the infinite Thue-Morse word can also be generated by suffix square completion. This exhibits a property that seems interesting to us: every (infinite) word generated by suffix completion contains squares, but there are (infinite) words generated by this operation (which basically creates squares) that avoid any repetition of (rational) exponent higher than 2. In comparison, we show that the Thue-Morse infinite word cannot be generated by prefix-suffix duplication. However, we show that one can generate an infinite cube-free word by suffix-duplication. This is a weaker version of the result obtained for square completion: every (infinite) word generated by suffix duplication contains squares, but there are (infinite) words generated by this operation that avoid any repetition of integer exponent higher than 2. These results have appeared in [24]. We also extend some of the above results to the bounded prefix-suffix duplication operations.

- In this section we tackle the problem of efficiently detecting the existence of repetitive structures occurring at both ends of some sequence. For instance, words that do not end or start with repetitions may model DNA sequences that went through some degenerative process that destroyed the terminal repeats, affecting their stability or functionalities. Moving away from the biological motivation, words that do not start nor end with repetitions seem to be interesting from a combinatorial point of view, as well. Indeed, repetitions-free words (i.e., words that do not contain consecutive occurrences of the same factors) are central in combinatorics on words, stringology and their applications (see, e.g., [49, 36]); words that do not have repetitive prefixes or suffixes model a weaker, but strongly related, notion.

- After focusing mostly on duplications and squares that happen at the start or end of a word we focus on a closely related topic, namely repeats inside a word and discuss about gapped repeats and palindromes within a word. The two have been investigated for a long time(see, e.g., [36, 8, 45, 43, 44, 15, 17] and the references therein), with motivation coming especially from the analysis of DNA and RNA structures, where tandem repeats or hairpin structures play important roles in revealing structural and functional information of the analyzed genetic sequence (see [36, 8, 43] and the references therein). More precisely, a gapped repeat (respectively, palindrome) occurring in a word $w$ is a factor $uvu$ (respectively, $u^R vu$) of $w$. The middle part $v$ of such structures is called *gap*, while the two factors $u$ (or the factors $u^R$ and $u$) are called left and right arms. Generally, the previous works were interested in finding all the gapped repeats and palindromes, under certain restrictions on the length of the gap or on the relation between the arm of the repeat or palindrome and the gap.

  In this section that contains material published in [22], we propose an alternative point of view in the study of gapped repeats and palindromes. The longest previous factor table (LPF) was introduced and considered in the context of efficiently computing Lempel-Ziv-like factorisations of words (see [14, 15]). Such a table provides for each position $i$ of the word the longest factor occurring both at position $i$ and once again on a position $j < i$. Several variants of this table were also considered in [15]: the longest previous reverse factor (*LPrF*), where we look for the longest factor occurring at position $i$ and whose mirror image occurs in the prefix of $w$ of length $i-1$, or the longest previous non-overlapping factor, where we look for the longest factor occurring both at position $i$ and somewhere inside the prefix of length $i-1$ of $w$. Such tables may be seen as providing a comprehensive image of the long repeats and symmetries occurring in the analyzed word. In our work we approach the construction of longest previous gapped repeat or palindrome tables: for each position $i$ of the word we want to compute the longest factor occurring both at position $i$ and once again on a position $j < i$ (or, respectively, whose mirror image occurs in the prefix of length $i-1$ of $w$) such that there is a gap (subject to various restrictions) between $i$ and the previous occurrence of the respective factor (mirrored factor). Similar to the original setting, this should give us a good image of the long gapped repeats and symmetries of a word.

## 4.1 Generating Infinite Words

We now discuss the generative power of our operation focusing on the square completion operation. Most results in this section have been published in [24]. The focus on the square completion operation is natural as it has the biggest power of generation, as the next result is naturally derived from the definitions of the operations.

**Lemma 4.1.1** *For any word $w$ and any $k \in N$ we have $PSD_k^*(w) \subseteq PSD^*(w) \subseteq PSSC^*(w)$.*

Given the above natural result we have focused on generating infinite words using $PSSC$.

The next three propositions show how the suffix square completion can be used to generate three important infinite words.

**Proposition 4.1.1** *The Fibonacci word* $\mathbf{f}$ *is in* $SSC^\omega$.

*Proof.* By Lemma 3.2.6, it is enough to show that for all $n \geq 4$, there exists $f'_n$ such that $f'_n \in SSC^*(f_n)$, $f_{n+1}$ is a prefix of $f'_n$, and $f''_n$ is a prefix of $f_{n+2}$. Indeed, it follows that we can derive $f'_{n+1}$ from $f'_n$ (because $f'_{n+1}$ can be derived from the prefix $f_{n+1}$ of $f'_n$), and so on. For short, we would be able to derive, starting with $f_4$, an infinite sequence of prefixes of $\mathbf{f}$. Now, $f_n = f_{n-1}f_{n-2}$; we can derive from $f_n$ the word $f_{n-1}f_{n-2}f_{n-2}$, and then $f_{n-1}f_{n-2}f_{n-2}f_{n-2} = f_{n-1}f_{n-2}f_{n-2}f_{n-3}f_{n-4}$; but, $f_{n-1}f_{n-2}f_{n-2}f_{n-3}f_{n-4} = f_nf_{n-1}f_{n-4} = f_{n+1}f_{n-4}$. So, taking $f'_n = f_{n+1}f_{n-4}$ we get exactly what we wanted: $f'_n \in SSC^*(f_n)$, $f_{n+1}$ is a prefix of $f'_n$, and $f''_n$ is a prefix of $f_{n+2}$. □

**Proposition 4.1.2** *The Period-doubling word* $\mathbf{d}$ *is in* $SSC^\omega$.

*Proof.* Let $d_n = \phi_d^n(0)$, for $n \geq 0$. It is not hard to note that if $d_{n+1} \in SSC^*(d_n)$, then $d_{n+2} \in SSC^*(d_{n+1})$; indeed, if in the $i^{th}$ step of the derivation of $d_{n+1}$ from $d_n$ we derived the word $wxyxy$ from the word $wxyx$, then in the $i^{th}$ step of the derivation of $d_{n+2}$ from $d_{n+1}$ we derive $\phi_d(wxyxy)$ from the word $\phi_d(wxyx)$. Therefore, it is enough to show that $d_4$ can be derived from $d_3 = 01000101$. Now, in the first step we derive $010001\underline{01}(01)$ (we place the root of the square between parentheses, and the suffix which we completed is underlined). Then, from 0100010101 we derive $0100010\underline{(10)(1}0)$. From 01000101010 we derive in two steps 0100010101000 (by duplicating twice the 0 letter occurring at the end of the word). Now, from 0100010101000 we derive $01000101\underline{0100(0}100) = d_4$. This concludes our proof. □

**Proposition 4.1.3** *The Thue-Morse word* $\mathbf{t}$ *is in* $SSC^\omega$*.*

*Proof.* First we fix some notations. Let $\bar{t}_n$ be $t_n$ in which we change all the 1 letters into 0 and all the 0 letters into 1. It is well known that $t_{n+1} = t_n\bar{t}_n$. Also, let $t'_n$ be the word obtained from $t_n$ by deleting its last letter. Let us first note that $t_6$ ends with 10110 and $\bar{t}_6$ starts with 1; so, from $t'_6$, which has the suffix 1011, we can derive in one $SSC$-step the word $t_6 1$, which is a prefix of $t_7$. Moreover, for all $n \geq 7$ we have that $t_n$ ends with $\phi_t^{n-6}(101)\phi_t^{n-6}(10)$, so from $t'_n$ we can generate $t_n\phi_t^{n-6}(1)$, which is a prefix of $t_{n+1}$. We now note that $t_{2n} = t_{2n-1}\bar{t}_{2n-2}t_{2n-3}\bar{t}_{2n-4}\ldots t_1\bar{t}_0 t_0$. As $t_{2i-1}$ ends with $\bar{t}_{2i-2}$ for all $i$, it is immediate that $t'_{2n} \in SSC^*(t_{2n-1})$. Similarly, $t_{2n+1} = t_{2n}\bar{t}_{2n-1}t_{2n-2}\bar{t}_{2n-3}\ldots\bar{t}_1 t_0\bar{t}_0$. Now, $t_{2i}$ ends with $\bar{t}_{2i-1}$ for all $i$, so $t'_{2n+1} \in SSC^*(t_{2n})$.

Now we have all the ingredients to show that $\mathbf{t} \in SSC^\omega$. We start with $t_5$ and derive from it, in multiple steps, $t'_6$. From $t'_6$ we derive in one step $t_6 1$. From $t_6 1$ we can derive $t'_7$ by Lemma 3.2.6, because we can derive $t'_7$ from the prefix $t_6$ of $t_6 1$. Then we continue this process. Generally, at some point of our derivation we obtained $t'_n$. From this we derive $t_n\phi_t^{n-6}(1)$, and further we derive in multiple steps $t'_{n+1}$ (again, we can do this according to Lemma 3.2.6, because we can derive $t'_{n+1}$ from the prefix $t_n$ of $t_n\phi_t^{n-6}(1)$). This concludes our proof. □

The previous results shows the existence of infinite words which can be generated by iterated suffix square completion, which avoid any power of rational exponent strictly greater than 2. However, $\mathbf{t}$ cannot be generated by the prefix-suffix duplication, so we cannot get in the same way a similar result for this operation.

**Proposition 4.1.4** *The Thue-Morse word* $\mathbf{t}$ *is not in* $PSD^\omega$*.*

*Proof.* To begin with, let $\bar{t}_n$ be $t_n$ in which we change all the 1 letters into 0 and all the 0 letters into 1, and let $\bar{\mathbf{t}} = \lim_{n\to\infty}\bar{t}_n$.

Our proof is based on a series of claims regarding the occurrences of squares inside the Thue-Morse word.

*Claim 1.* For all $n \geq 0$, $t_n$ does not start with a square. Same holds for $\bar{t}_n$.

*Proof of the claim:* It is easy to show that none of $t_1, \ldots, t_6$ start with a square. Now, assume that $t_n$ has no square prefix, and we show that $t_{n+1}$ has no square prefix. If $t_{n+1}$ starts with a square $xx$ with $x$ of even length, then $x = \phi_t(y)$ for some prefix $y$ of $t_n$. thus,, $t_n$ also starts with $yy$, a contradiction. If $t_{n+1}$ starts with a square $xx$ with $x$ of odd length it is immediate that $|x| > 1$. Then $x$ starts with 0; as the last letter of the first $x$ and first letter of the second $x$ are the image of some letter from $t_n$, we get that $x$ ends with 1. This means that the second $x$ also ends with 1; its last two letters are the image of a letter from $x$, so $x$ ends with 01. Now, the first $x$ ends with 01, and the letters $x[|x|-2]x[|x|-1]$ are the image of a letter from $x$; thus,, they should be 01. So $x$ ends with 101. By repeating this reasoning, we get that every odd suffix of $x$ has the form $(10)^k1$, a contradiction, as the first letter of $x$ is 0. This concludes the proof. □

*Claim 2.* For all $n \geq 0$, $t_n$ does not end with a square. Same holds for $\bar{t}_n$.
*Proof of the claim:* Similar to the proof of Claim 1. □

*Claim 3.* For all $n \geq 0$, if $t_n[i..j]$ is a square such that $i \leq 2^{n-1}$ and $j > 2^{n-1}$ (i.e., this square goes over the centre of $t_n$) then $i > 2^{n-2}$ and $j \leq 2^{n-1} + 2^{n-2}$ (i.e., the square is completely contained in $t_n[2^{n-2}+1..2^{n-1}+2^{n-2}] = \bar{t}_{n-1}\bar{t}_{n-1}$). Same holds for $\bar{t}_n$.

*Proof of the claim:* Again, we show this by induction. The claim clearly holds for $n \leq 6$. Assume it holds for $t_k$ with $k \leq n$, and we show it holds for $t_{n+1}$. Suppose for the sake of a contradiction that $t_n[i..j]$ is a square such that $j > 2^{n-1}$ and $i \leq 2^{n-2}$ (same argument can be used in the case when $j > 2^{n-1} + 2^{n-2}$).

If this square is of even length and starts on an odd position, then it is the image of a square of $t_n$ which is not contained in the factor $\bar{t}_{n-1}\bar{t}_{n-1}$ whose centre coincides with the centre of $t_n$. This is a contradiction. If this square is of even length and starts on an even position, then we can shift it one position to the left and get a square of the same length, so we get an overlap in $\mathbf{t}$. Again, a contradiction.

If the square is of odd length, then the length of its root is strictly greater than 5 (actually, it is much longer, but we do not use that). Say that our square is $xx$. Let us assume that this square starts on an odd position, and it starts with 0. This 0 should be followed by an 1. thus, the first $x$ starts with 01. So, the second $x$ starts with a 0 (on an even position), followed by an 1 on an odd position; this 1 should be followed by an 0. So, the first $x$ starts with 010; clearly, the following letter is an 1. We repeat this procedure one more time to get that $x$ starts with 01010, a contradiction to the fact that $\mathbf{t}$ is overlap free. A very similar argument works for the case when the square is of odd length and starts on an even position. □

*Claim 4.* For all $n \geq 0$, if $t_n[i..j]$ is a square such that $i \leq 2^{n-1}$ and $j > 2^{n-1}$ (i.e., this square goes over the centre of $t_n$) and either $i \neq 2^{n-2}+1$ or $j \neq 2^{n-1} + 2^{n-2}$, then $i > 2^{n-2} + 2^{n-3}$ and $j \leq 2^{n-1} + 2^{n-3}$ (i.e., the square is completely contained in $t_n[2^{n-2} + 2^{n-3} + 1..2^{n-1} + 2^{n-3}]$). Same holds for $\bar{t}_n$.

*Proof of the claim:* The proof is pretty similar to the one above. Just like there, we do induction. The property holds for $t_n$ with $n \leq 6$. Then, when trying to show it for $t_{n+1}$, we check first squares with root of even length that start on odd positions; these are images of squares that have similar positions with respect to $t_n$, so the existence of such a square would lead to a contradiction. When analyzing squares with root of even length that start on even positions, we realize that the existence of such a squares means that we have overlaps in $\mathbf{t}$. Finally, our squares should be fairly long, so such squares of odd length lead again to the existence of overlaps in $\mathbf{t}$. □

We now move on to the main proof.

According to Claim 1, it is enough to show that $\mathbf{t} \notin SD^\omega$. Indeed, if $\mathbf{t} \in PSD^\omega \setminus SD^\omega$, then $\mathbf{t}$ has a prefix $x$ obtained by prefix duplication. Clearly, $x$ has a square prefix, so $\mathbf{t}$ has a square prefix, as well; this is a contradiction.

We now show by induction that it is impossible to derive from a prefix of $t_n$ (respectively, a prefix of $\overline{t}_n$) a prefix of $\mathbf{t}$ (respectively, of $\overline{\mathbf{t}}$) longer than $|t_{n+1}| + \frac{|t_n|}{2} = 2^{n+1} + 2^{n-1}$. This property can be manually checked for $n \leq 6$.

Let us assume the above property holds for all $t_k$ and $\overline{t}_k$ with $k \leq n-1$ and show that it holds for $t_n$. Clearly, we have that:

$t_{n+2} = t_{n-2}\overline{t}_{n-2}\overline{t}_{n-2}t_{n-2}\overline{t}_{n-2}t_{n-2}t_{n-2}\overline{t}_{n-2}\overline{t}_{n-2}t_{n-2}t_{n-2}\overline{t}_{n-2}t_{n-2}\overline{t}_{n-2}\overline{t}_{n-2}t_{n-2}$.

Let $\tau_i = t_{n+2}[(i-1)2^{n-2}+1]$, for $1 \leq i \leq 16$. Basically, the factors $\tau_i$ are the factors $t_{n-2}$ or $\overline{t}_{n-2}$ emphasized in the decomposition of $t_{n+2}$ from above. That is: $\tau_1 = t_{n-2}$, $\tau_2 = \overline{t}_{n-2}$, and so on. Clearly, $\tau_1 \cdots \tau_4 = t_n$, $\tau_5 \cdots \tau_8 = \overline{t}_n$, $\tau_7 \cdots \tau_{10} = t_n$, and $\tau_9 \cdots \tau_{16} = \overline{t}_{n+1}$.

We assume, for the sake of a contradiction, that we can derive from a prefix of $\tau_1 \cdots \tau_4$ a word longer than $\tau_1 \ldots \tau_{10}$. Such a derivation starts with the initial prefix of $\mathbf{t}$, and reaches in several steps a word $y$ that is shorter than $\tau_1 \ldots \tau_{10}$ but in the next step we produce a word $yz$, which is longer than $\tau_1 \ldots \tau_{10}$. So $zz$ has the centre somewhere in $\tau_1 \ldots \tau_{10}$. We want to see where it may begin and where it centre might be.

If $yz$ ends inside $t_{n+2}$, we note that, according to Claim 3, applied to $t_{n+2} = \tau_1 \cdots \tau_{16}$, the factor $zz$ cannot begin inside $\tau_1 \ldots \tau_4$; similarly if $yz$ ends inside $t_{n+3}$, then we apply Claim 3 to $t_{n+3}$. It also cannot start at the same position as $\tau_5$: in that case, $yz$ must end inside $t_{n+2}$, so $zz$ could only be the square $\tau_5 \cdots \tau_{12} = \overline{t}_n\overline{t}_n$; this would mean that $y = \tau_1 \cdots \tau_8 = t_{n+1}$ and this word can be obtained by suffix duplication. Moreover, $zz$ cannot start somewhere else inside $\tau_5\tau_6$: such a square (centred in $\tau_1 \cdots \tau_{10}$) would end in $\tau_1 \cdots \tau_{16} = t_{n+2}$, and this leads to a contradiction with Claim 4 ($zz$ would not be contained in $t_{n+2}[2^n + 2^{n-1} + 1..2^{n-1} + 2^{n-3}]$, as it should). So, the first $z$ factor of the square $zz$ is completely contained in $\tau_7 \ldots \tau_{10}$. By Claim 4 (applied to $t_{n+2}$), there is no square that starts in $\tau_7\tau_8$ and ends strictly after the ending position of $\tau_1 0$. In conclusion, both the starting position and the centre of the square $zz$ occur in $\tau_9\tau_{10}$. Moreover, as $zz$ ends after the ending position of $\tau_{10}$, its centre cannot occur in the first half of $\tau_9$.

So, $y$ ends inside $\tau_9\tau_{10}$, but not in the first half of $\tau_9$. We repeat the reasoning above for $y$. This word was obtained from some $y'$ by appending $v$ to it. That is, $y = y'v$ and $y'$ ends with $v$. Just like before, $vv$ cannot start in $\tau_1 \cdots \tau_6$, so it must start in $\tau_7\tau_8$; also, $vv$ cannot be $\tau_8\tau_9$ (again, this would mean that $t_{n+1}$ ends with a square). Moreover, $vv$ cannot start in $\tau_7$, by Claim 3 applied to $\tau_7\tau_8\tau_9\tau_{10} = t_{n-1}$. If $y'$ ends in $\tau_7\tau_8$ it is fine; if not, we repeat the procedure and consider the string from which $y'$ was obtained in the role of $y'$. Generally, we repeat this reduction until we reach an intermediate word obtained in the derivation of $y$ from $x$ which ends inside $\tau_7\tau_8$. So, suppose $y'$ ended inside $\tau_7\tau_8$.

Assume first that $y'$ does not end inside $\tau_7$ (so it ends inside $\tau_8$). Say that $y'$ was derived from some $y''$ by duplicating the factor $u$ (i.e., $y' = y''u$ and $u$ is a suffix of $y''$). Then, by applying Claim 3 to $\bar{t}_n = \tau_5 \cdots \tau_8$, we get that the square $uu$ cannot start in $\tau_5\tau_6$. This means $y''$ ends inside $\tau_7\tau_8$, just like $y'$. We repeat the process with $y''$ in the role of $y'$ until we reach a word that ends inside $\tau_7$.

So, let us assume that $y'$ ends inside $\tau_7$. Take $w_1$ be the suffix of $y'$ contained in $\tau_7$ and $w_2$ the suffix of $y$ contained in $\tau_7 \cdots \tau_{10}$. It is immediate that the duplication steps that were used to produce $y$ from the current $y'$ can be used to produce $w_2$ from $w_1$. But $w_1$ is a prefix of $\bar{t}_{n-2}$ and $w_1$ is a prefix of $\bar{\mathbf{t}}$ of strictly longer than $|t_{n-1}| + \frac{|t_{n-2}|}{2}$ (as $y$ after the first half of $\tau_9$). This is a contradiction with our induction hypothesis.

Clearly our claim also holds for $\bar{t}_n$, so our induction proof is complete.

Hence, it is impossible to derive from a prefix of $t_n$ (respectively, a prefix of $\bar{t}_n$) a prefix of $\mathbf{t}$ (respectively, of $\mathbf{t}$) longer than $|t_{n+1}| + \frac{|t_n|}{2} = 2^{n+1} + 2^{n-1}$, for all $n$. This means that we cannot construct by suffix duplication an infinite sequence of finite prefixes of $\mathbf{t}$ whose limit is $\mathbf{t}$. □

Despite the negative result of the previous Proposition, we show that, in fact, the duplication operation can still be used to generate infinite words that do not contain repetitions of large exponent. More precisely, while suffix square completion was enough to generate words that avoid rational powers greater than 2, the suffix duplication is enough to generate words that avoid integer powers greater than 2 (so, automatically, rational powers greater or equal to 3).

**Proposition 4.1.5** *Stewart's choral sequence* $\mathbf{s}$ *is in* $SD^\omega$.

*Proof.* It is enough to show that for all $n \geq 2$ we have $s_{n+1} \in SD^*(s_n)$. To show this, note first that, because $s_{n+1} = s_n s_n s_n^*$, we have $s_{n+1}^* = s_n s_n^* s_n^*$. Now, from $s_n$ we derive $s_n s_n = (s_{n-1}s_{n-1}s_{n-1}^*)(s_{n-1}s_{n-1}s_{n-1}^*)$, in one duplication step. Then we duplicate the suffix $s_{n-1}s_{n-1}^*$ of $s^2n$, and get $s_n s_n s_{n-1} s_{n-1}^*$. We now duplicate the suffix $s_{n-1}^*$ of the word we obtained, and derive $s_n s_n s_{n-1} s_{n-1}^* s_{n-1}^* = s_n s_n s_n^* = s_{n+1}$. This concludes our proof: we derived, starting with $s_w$, an infinite sequence of prefixes of $\mathbf{s}$. □

Research has yet to be conducted on the generative power of $PSD_k^*$, and results of a similar manner to the one above would be interesting. While it is quite probable a word like Stewart's choral sequence is not in $PSD_k^*$ for any k, infinite words that construct by concatenation like the Fibonacci word could be generated through $PSD_k^*$. Following this direction it would be interesting to determine for given infinite words what is the smallest $k$ such that the word is in $PSD_k^*$.

As a start in this direction we give a conjuncture which contains two similar results.

**Conjecture 4.1.1**

- $F_{2*n} \in PSD_k^*(F_{2*p})$, *for any* $n \geq p \geq 1, k \geq 3$
- $F_{2*n+1} \in PSD_k^*(F_{2*p+1})$, *for any* $n \geq p \geq 1, k \geq 5$

The above conjecture states that we can generate the infinite Fibonacci word by prefix-suffix duplication if we use $k \geq 3$, if we want to actually generate exact words from the Fibonacci sequence than we will only be able to generate odd indexed words from odd indexed words, and even indexed words from even indexed words. We will never be able to generate an odd indexed word from an even indexed word or vice versa as they finish in different characters 0, respectively 1, and any *PSD* operation keeps the last character constant.

## 4.2 On Prefix/Suffix-Square Free Words

Results that appear in this section have been published in [24].

In this section we address a series or questions related to the simplest repetitive structure that may occur in a word: squares. We say that a word is suffix-square free (respectively, prefix-square free or prefix-suffix-square free) if that word does not end (respectively, start, or both start and end) with two consecutive occurrences of a factor. Alternatively, these are the words that cannot be obtained from shorter factors by the *SD*−operation (respectively, *PD*, or *PSD*). In comparison in section 3.2 we discussed about ancestors and primitive ancestors of a word and not about all its factors.

We show how the following tasks can be performed efficiently, for an input word $w$ of length $n$. First, we show how $w$ can be processed in linear time so that we can answer in constant time queries asking whether the factors of $w$ are suffix-square free, prefix-square free, or prefix-suffix-square free. Then, we give an algorithm that outputs in $\mathcal{O}(n + |S|)$ the set $S$ of all factors of $w$ which are prefix-suffix-square free (respectively, suffix-square free or prefix-square free); computing the size of $S$ (without enumerating all its elements) can be done in $\mathcal{O}(n \log(n))$. Note that, for square-free words (see [49]) of length $n$ we have $|S| \in \Theta(n^2)$, so there are indeed cases in which we can compute $|S|$ faster than just going through its elements. Finally, the longest prefix-suffix square free factor of a word can be obtained in $\mathcal{O}(n)$ time.

As described in the previous chapter in [32] the authors gave algorithms deciding in $\mathcal{O}(n \log(n))$ and, respectively, $\mathcal{O}(n^2 \log(n))$ time, for two given words, whether the longer one (whose length was $n$) can be generated from the shorter one by iterated *SD*, respectively, *PSD*. This algorithms can be easily modified to compute all the *SD* primitive roots(or ancestors) of a word or all its PSD primitive roots(ancestors). These are suffix-square free (respectively, prefix-suffix-square free) factors of a word that can generate it by iterated *SD*, respectively, *PSD* operations. Intuitively, these are the factors of a word containing its core information: on one hand, the rest of the word consists only in repetitions of parts of these factors, and on the other hand, they cannot be further reduced by eliminating repeated information from their ends, as they do not end with any repetition. We have showed in the previous chapter that there exist words of length $n$ having $\Theta(n)$ *SD* primitive roots and, respectively, words having $\Theta(n^2)$ *PSD* primitive roots; so our algorithms finding all these roots are only a $\log(n)$-factor slower than what one could expect in the worst case.

We conclude the section by addressing a series of related problems. Essentially, detecting a *SD* primitive root of some word $w$ is equivalent to asking whether there is a (non-trivial) factorisation of that word into $k > 1$ factors $w = s_1 \cdots s_k$ such that $s_i$ is a suffix of $s_1 \cdots s_{i-1}$ for $i > 1$; that is, for all $i > 1$, we have the square $s_i^2$ centred on the position that follows immediately after $s_1 \cdots s_{i-1}$, and $s_1$ is suffix-square free. This factorisation, as well as the similar factorisations derived from the detection of $PD$ or $PSD-$roots of a word, seems to be strongly connected to the square-structure of $w$. So, we further discuss other factorisations also strongly related to the way squares and, more general, repetitions occur in the factored words. Namely, we address the problem of factoring a word into squares, or into periodicities of exponent at least 2, or in such a way that the number of square factors of the factorisation is maximum, compared to any other possible factorisation of that word.

It seems important to stress that, in fact, all our results in this section are about understanding the way squares occur in a word. Either we discuss about prefix or suffix-square free factors of a word or about various factorisations of a word, in fact we gather information on the occurrences of squares inside the given word: sometimes we are interested in avoiding these occurrences (i.e., finding words that do not start or end with such an occurrence), while sometimes we want to use them to factor that word. We provide here a different algorithm than the one we provided in the lemma section of our introductory chapter, our goal being to provide a different technique for proving this results. Our technique uses a similar approach as the one used in [50, 42] where it was used to find efficiently all the periodicities of a word in linear time , and in [27] where it was used to compute, also in linear time, all local periods of a word; the details of this algorithm are, however, quite different.

The following lemma will be used extensively during this section:

For a word $w$ of length $n$ we define the array $left[\cdot]$ as follows: for $1 \leq i \leq n$, $left[i] = \max\{j \mid w[j..i] \text{ is a square}\}$.

**Lemma 4.2.1** *Given a word $w$ we can compute in linear time the array $left[\cdot]$.*

*Proof.* In fact, on $left[j]$ starts the shortest square ending at position $j$ of $w$.

Before describing our algorithm, we note that if $x^2$ and $y^2$ are two squares ending on position $j$ of $w$, and $\frac{3\ell}{2} \geq |y| > |x| \geq \ell$ for some $\ell > 0$, then there exists a square $u^2$ with $|u| \leq \frac{\ell}{2}$ also ending at position $j$. Indeed, $y = ux$ for some word $u$ with $|u| \leq \frac{\ell}{2}$. As $xx$ is a suffix of $yy$ and $2|x| \geq |y|$, we get that $u$ is a suffix of (the first) $x$. It follows that $u$ is also a suffix of $y$. Consequently, $uy = uux$ is a suffix of $y^2$. However, as $|u| \leq \frac{\ell}{2}$, we get that $x^2$ is longer than $uy = uux$, so $uu$ is also a suffix of (both occurrences of) $x$. This shows our claim. Moving now to the algorithm, we first compute the Lempel-Ziv-factorisation of the input word $w = u_1 \cdots u_k$, using the tools from [15]. During this computation of the $s$-factorisation we also get for each $i \leq k$ the position $\ell_i$ where $u_i$ occurs in $u_1 \cdots u_{i-1}$; let also $k_i = |u_1 \cdots u_{i-1}| + 1$ be the starting position of $u_i$, for all $i$.

We compute separately, for each $i$ from 1 to $k$, considered in increasing order, the values $left[j]$ for each position $j$ of the factor $u_i$. In the following, we explain how these values are computed for some fixed $i$; our approach ensures that when considering $u_i$ we already know *left*$[j']$ for every position $j'$ of $u_1 \cdots u_{i-1}$.

First, note that if $x^2$ is a square ending on position $j$ of $u_i$ then the centre of this square occurs inside $u_{i-1}u_i$. Otherwise, $u_{i-1}$ was not correctly chosen: in that case, a longer factor starting on position $k_{i-1}$ would be $w[k_{i-1}..j]$, a suffix of the second $x$ factor of the square. Hence, $|x| \leq |u_{i-1}u_i|$ and there are three cases to be analyzed: the shortest square ending on position $j$ might be completely contained in $u_i$, centred in $u_i$ but starting in $u_{i-1}$, or centred in $u_{i-1}$.

We begin with the simplest case: if the shortest square ending on position $j$ of $u_i$ is completely contained in $u_i$ then it should be equal to the shortest square ending at position $\ell_i + (j - k_i)$ (that is, the shortest factor ending on the position corresponding to $j$ from the previous occurrence of $u_i$ inside $u_1 \cdots u_{i-1}$). So, in the first step of our algorithm, we just check for every position $j$ of $u_i$ if the shortest square ending on $\ell_i + (j - k_i)$ is short enough to be contained in $u_i$, and, if yes, we decide that the respective square occurs again as the shortest square ending at $j$.

Secondly, we detect for each position $j$ of $u_i$ the shortest square $x^2$ ending on $j$, starting in $u_{i-1}$ and whose centre is in $u_i$; clearly, $|x| \leq |u_i|$. Following the strategy of [50], we detect for each possible length $\ell$ of $x$ a range of $u_i$ where the centre of $x^2$ may occur. Basically, for each $\ell$, we compute the longest common prefix $u_i[1..b_\ell]$ of $u_i[1..\ell]$ and $u_i[\ell+1..|u_i|]$ and the longest common suffix $u_i[a_\ell..\ell]$ of $u_{i-1}$ and $u_i[1..\ell]$; the range we look for is $[a_\ell, b_\ell]$. Now, the range where a square $x^2$ as above may end is obtained by intersecting $[a_\ell + \ell - 1, b_\ell + \ell - 1]$ with $u_i$; in this way, we obtain a new range $[c_\ell, d_\ell]$ where the squares of length $\ell$ may end. As $b_\ell \leq \ell$, we have $d_\ell \leq b_\ell + \ell - 1 < 2\ell$.

So, for each possible length $\ell$ of $x$ (i.e., $1 \leq \ell \leq |x_i|$) we now have a range $[c_\ell, d_\ell]$ where a square $x^2$ may end inside $u_i$; each range was computed using a constant number of longest common prefix queries, so in constant time. Now we just have to report for each position $j$ of $u_i$ which is the minimum $\ell$ such that $j$ is in the range $[c_\ell, d_\ell]$. To do this, we report for each $k \geq 0$ the positions where a square $x^2$ with $(3/2)^k \leq |x| \leq (3/2)^{k+1}$ ends, and we did not find so far a shorter square ending at that position. Clearly, for $\ell$ with $(3/2)^k \leq \ell \leq (3/2)^{k+1}$ we have $d_\ell \leq k_i + 2 \cdot (3/2)^{k+1}$. So, we can sort the ends of the ranges $[c_\ell, d_\ell]$ with $(3/2)^k \leq \ell \leq (3/2)^{k+1}$ in $\mathcal{O}((3/2)^{k+1})$ time. This allows us to find in $\mathcal{O}((3/2)^{k+1})$ time the positions of $u_i$ contained in exactly one range $[c_\ell, d_\ell]$ (all these positions are in the prefix of length $2 \cdot (3/2)^{k+1}$ of $u_i$). For each such position $j$ we store the value $\ell$ such that $j \in [c_\ell, d_\ell]$; we conclude that a square $xx$ with $|x| = \ell$ ends on $j$, and, if we did not already find a shorter square ending on that position, we store this square as the shortest one we found so far that ends on $j$. By the claim shown at the beginning of this proof, if a position is contained in at least two ranges $[c_\ell, d_\ell]$ and $[c_{\ell'}, d_{\ell'}]$ with $(3/2)^k \leq \ell \leq \ell' \leq (3/2)^{k+1}$, then there exists a square with root of length at most $\frac{(3/2)^k}{2}$ ending at that position, so we do not have to worry about it: we should have already found the shorter square occurring at there (note that this square might be completely contained in $u_i$).

Consequently, processing all the ranges $[c_\ell, d_\ell]$ with $(3/2)^k \leq \ell \leq (3/2)^{k+1}$ takes $\mathcal{O}((3/2)^{k+1})$ time. Now, we iterate this process for all $k \leq \log_{3/2} |u_i|$, and, alongside the analysis of the first case, obtain for each position $j$ the shortest square ending there that either starts in $u_{i+1}$ but it is centred in $u_i$ or is completely contained in $u_i$. The total time is $\mathcal{O}\left(\sum_{k \leq \log_{3/2}|u_i|} (3/2)^{k+1}\right) = \mathcal{O}(|u_i|)$.

Finally, we look for the shortest squares ending in $u_i$ that are centred in $u_{i-1}$; the length of such squares may go up to $|u_{i-1}u_i|$. The analysis is very similar to the one of the second case, when we searched for squares centred in $u_i$, starting in $u_{i-1}$. In $\mathcal{O}(|u_{i-1}u_i|)$ time we find, for all $\ell \leq |u_{i-1}u_i|$ the ranges where the centers of squares $x^2$ ending in $u_i$, with $|x| = \ell$, occur in $u_{i-1}$. This gives the ranges of $u_i$ where such squares end. Now, using the same ideas as above, we detect, in $\mathcal{O}(|u_{i-1}|+|u_i|)$ time, the shortest square ending at each position where we did not already find a shortest square. Putting together the results of the three analyzed cases, we get for each position $j$ of $u_i$ the value *left*$[j]$.

In conclusion, the time needed to compute for all $i$ from 1 to $k$ the values *left*$[j]$ for every position $j$ of the factors $u_i$ is $\mathcal{O}(\sum_{1 \leq i \leq k} |u_{i-1}u_i|) = \mathcal{O}(n)$. □

Clearly, the same algorithm can be used on the mirror image of $w$ to obtain *right*$[\cdot]$ defined as follows: for $1 \leq i \leq n$, *right*$[i] = \min\{j \mid w[i..j]$ is a square$\}$.

The previous lemma allows us to answer efficiently queries asking whether the factors of a given word are or not prefix-square free, or suffix-square free, or prefix-suffix-square free.

**Theorem 4.2.1** *Given a word $w$ of length $n$, we can construct in $\mathcal{O}(n)$ data structures allowing us to answer in constant time the following three types of queries (for all $1 \leq i \leq j \leq n$): $q_p(i,j)$: "is $w[i..j]$ prefix-square free?"; $q_s(i,j)$: "is $w[i..j]$ suffix-square free?"; $q_{ps}(i,j)$: "is $w[i..j]$ prefix-suffix-square free?".*

*Proof.* We construct the arrays *left* and *right*. Then we can answer a query $q_p(i,j)$ positively if *right*$[i] > j$ (the shortest square that starts at $i$ ends after $j$); otherwise, we answer it negatively. Similarly, a query $q_s(i,j)$ returns true if and only if *left*$[j] < i$ (the shortest square that ends at $j$ starts before $i$). Finally, $q_{ps}(i,j)$ is answered positively if and only if both $q_s(i,j)$ and $q_p(i,j)$ are true. □

However, this lemma does not enable us to enumerate the prefix-suffix-square free factors efficiently. The following theorem shows how this can be done.

**Theorem 4.2.2** *Given a word $w$ of length $n$, we can find the set $S$ of prefix-suffix-square free factors of $w$ in $\mathcal{O}(n + |S|)$ time.*

*Proof.* The idea is to find separately for each $i \leq n$, all the prefix-suffix-square free factors $w[i..j]$. This can be done as follows: we first construct the arrays $\mathit{left}[\cdot]$ and $\mathit{right}[\cdot]$ for the word $w$, and define data structures allowing us to answer in constant time range minimum queries for $\mathit{left}[\cdot]$. Let $RMQ(j_1, j_2)$ denote the minimum value occurring in the range between $j_1$ and $j_2$; this is, in fact, the rightmost starting position of a square $x^2$ that ends on a position between $j_1$ and $j_2$. Let $posRMQ(j_1, j_2)$ denote the position where $RMQ(j_1, j_2)$ occurs (in case of equality, we take the rightmost such position); this position is the ending position of the aforementioned square $x^2$.

Let us now fix some $i \leq n$. We note that if $w[i..j]$ is prefix-square free then $j < \mathit{right}[i]$ (or $w[i..\mathit{right}[i]]$ would be a square prefix). So, we only consider the values between $i + 1$ and $\mathit{right}[i] - 1$ as possible ending positions of prefix-suffix-square free words starting on $i$. The procedure detecting these words works recursively. In a call of the procedure, we find the prefix-suffix-square free words starting on $i$ and ending somewhere between two positions $j_1$ and $j_2$; initially, $j_1 = i + 1$ and $j_2 = \mathit{right}[i] - 1$, and $j_1$ and $j_2$ are always between $i + 1$ and $\mathit{right}[i] - 1$.

So, let us explain how our search is conducted for a pair of positions $(j_1, j_2)$, where $j_1 < j_2$. We compute $RMQ(j_1, j_2)$ and $posRMQ(j_1, j_2)$. If $RMQ(j_1, j_2) < i$ and $posRMQ(j_1, j_2) = j$ then clearly the shortest square ending on $j$ starts before $i$, and the shortest square starting on $i$ ends after $j$, so $w[i..j]$ is prefix-suffix-square free; in this case, we run the procedure for the two new pairs $(j_1, posRMQ(j_1, j_2) - 1)$ and $(posRMQ(j_1, j_2) + 1, j_2)$. The other case is when $RMQ(j_1, j_2) \geq i$. Then the shortest square ending on any position $j$ between $j_1$ and $j_2$ ends after $i$, so none of the words $w[i..j]$ is suffix-square free; thus, we stop the procedure. A call of the recursive procedure for a pair $(j_1, j_2)$ where $j_1$ is not strictly smaller than $j_2$ does not do anything.

It is straightforward that, for some $i \leq n$, we obtain in this manner all the positions $j < \mathit{right}[i]$ such that $w[i..j]$ is suffix-square free. But these are, in fact, all the positions $j \leq n$ such that $w[i..j]$ is prefix-suffix-square free. We iterate this algorithm for all $i$, and obtain all prefix-suffix-square free factors of $w$.

To evaluate the total time complexity of this algorithm, let us again fix $i$ and see how much time we spend to detect the prefix-suffix-square free words starting on $i$. Each call of the recursive procedure either returns a valid position $j$ and calls the procedure for two new pairs (defining disjoint ranges), or stops the search in the range defined by the pairs for which it was called. So, the calls of this procedure can be pictured as a binary tree with $|S_i|$ internal nodes, where $S_i = \{j \mid w[i..j] \text{ is prefix-suffix-square free}\}$. thus,, we have, in total $\mathcal{O}(|S_i|)$ calls. With the help of the data structures we constructed, each call of the procedure can be executed in constant time. thus,, the time needed to

compute $S_i$ is $\mathcal{O}(|S_i|)$. Now, adding this up for all possible $i$, we get that the set $S$ of prefix-suffix-square free factors of $w$ can be computed in $\mathcal{O}(n + |S|)$ time (including the time used to construct *left*, *right* and *RMQ*-structures for *left*). □

**Remark.** *The suffix-square-free (resp., prefix-square-free) factors of a word can be detected easier. That is, $w[i..j]$ is suffix-square (resp., prefix-square) free iff $i > left[j]$ (resp., $j < right[i]$). So, we can output the suffix-square-free (resp., prefix-square-free) factors of $w$ in $\mathcal{O}(n + |S|)$ time, where $S$ is the solution set.*

*Proof.* Let us assume that we want to report the suffix-square-free factors of $w$. For each $j \leq n$ we report all the factors $w[i..j]$ with $i > left[j]$; they are clearly suffix-square-free. The complexity is clearly $\mathcal{O}(n + |S_s|)$ where $S_s$ is the set of such factors. When we want to report the prefix-square-free factors, for each $i \leq n$ we output all the factors $w[i..j]$ with $j < right[i]$; they are clearly prefix-square-free. The complexity is again $\mathcal{O}(n + |S_p|)$ where $S_p$ is the set of square-prefix-free factors. □

It is not hard to construct words with a high number of prefix-suffix-square-free factors. E.g., take any finite prefix $w$ of a right-infinite square-free word $\mathbf{w}$ (see [49]). Then $w$ does not contain squares, so all its factors are prefix-suffix-square free and the size of the set of prefix-suffix-square free factors of $w$ is $\Theta(n^2)$. However, the size of $S$ can be computed without enumerating all its elements. The main idea is that, for a position $j$ of the input word $w$, we need to report as many squares $w[i..j]$ as there are positions $i$ greater than $left[j]$ and less or equal to $j$ such that $right[i] > j$. This can be efficiently done with the help of segment trees [6].

**Theorem 4.2.3** *Given a word $w$ of length $n$, we can compute the number of prefix-suffix-square free factors of $w$ in $\mathcal{O}(n \log(n))$ time.*

*Proof.* Our strategy is to compute the number of factors that start or end with a square and subtract this number from the total number of factors of the input word $w$, namely $\frac{n(n+1)}{2}$. To begin with, we construct the arrays *left* and *right*.

The number $s_1$ of factors that start with a square is $\sum_{1 \leq i \leq n}(n + 1 - right[i])$, as every factor $w[i \ldots j]$ with $j \geq right[i]$ starts with the square $w[i \ldots right[i]]$. Clearly, $s_1$ is computed in linear time.

Further, we have to count the factors that end with a square, but do not start with a square (to avoid double counting). So, for a position $i$ we only want to count the positions $j$ from 1 to $left[i]$ (so, ending with a square) that have $right[j] > i$ (that is, not starting with a square). This can be done with the help of segment trees, as we show in the following algorithm.

We maintain an array $T[\cdot]$ such that at the beginning of the $k^{th}$ step of the algorithm, $T[j] = |\{i \leq k \mid right[i] = j\}|$ for $j \leq n$; that is, $T[j]$ counts the

number of positions $i \leq k$ such that the shortest squares starting on $i$ ends exactly on $j$. Initially, $T$ contains only 0 values and we initialize in linear time the segment tree for this array.

Further, we produce for each position $i \leq n$ the set of all the positions $j \leq n$ such that $left[j] = i$; that is, for each $i$ we store the set of positions $j$ such that the shortest square ending on $j$ starts on $i$. This takes linear time.

Now, let us describe the $k^{th}$ step of our algorithm. Let us assume that $i$ is a position of $w$ such that $left[i] = k$. Then, the number of factors $w[j'..i]$ which end with a square but have no square prefix equals $\sum_{i<j\leq n} T[j]$. Indeed, this sum counts all the positions $j'$, $1 \leq j' \leq k = left[i]$, such that the shortest square starting on $j'$ ends on a position $j > i$ (that is, $right[j'] > i$). We can obtain this sum by querying the segment tree associated to $T$: $sum(i+1, n)$. We compute this sum for each $i$ such that $left[i] = k$, and store the value as $m_j$. Now, we have to update the structure so that it can be used in the $(k+1)^{th}$ step of the algorithm. We only have to increase with 1 the value of $T[right[k]]$, and this can be done by $add_{val}(right[k], 1)$.

It is not hard to see that after $n$ steps we computed the number $m_j$ of factors $w[i..j]$ that end with a square but have no square prefix, for all $j \leq n$. The total time to do this is $\mathcal{O}(n \log(n))$. Indeed, we initialize the segment tree for $T$, then perform $n$ $add_{val}$ operations on it, and, in each step $k$ we perform exactly as many *sum* queries as the number of positions $j$ with $left[j] = k$; this means that we perform in total exactly $n$ *sum* queries. Each $add_{val}$ and *sum* operation is executed in $\mathcal{O}(\log(n))$ time, so running all of them takes $\mathcal{O}(n \log(n))$ time.

Clearly, $|S| = \frac{n(n+1)}{2} - \left(s_1 + \sum_{1\leq j\leq n} m_j\right)$, and this value is computed in linear time, once we have $s_1$ and all the values $m_j$. The conclusion follows. □

Now we return to a result from the following chapter were we discussed primitive ancestors (or roots) of a word, and were we left the following problem unsolved:

**Theorem 4.2.4** *Given a word $w$ of length $n$, we can identify in $\mathcal{O}(n \log(n))$ time the longest primitive ancestor in relation to PSSC of $w$.*

Interestingly the solution to this problem highly resembles the solution to the above theorem.

*Proof.* We will maintain as in the above solution a segment tree where the leaves hold the same significance, but the internal nodes will keep the maximum position in the interval where we have a value of 0. Updates are done in a similar fashion as above, the only big difference is in the query. In this problem for each element $i$ we query the segment tree for the largest index in the segment tree with value 0 in the interval $[i, right[i]]$, let this index be $pos[i]$. We can obtain this in $\mathcal{O}(\log(n))$, thus, we obtain an $\mathcal{O}(n \log(n))$ algorithm. Clearly the longest primitive ancestor in relation to PSSC of $w$ is the factor $[i, pos[i]]$ where $pos[i] - i \geq pos[j] - j$ for all $1 \leq j \leq n$, thus, we have proved our result. □

We now show that the longest prefix-suffix-square free factor $w[i..j]$ (i.e., the longest element of the set $S$ from Theorem 4.2.2) can be computed in linear time, so without enumerating all the elements of $S$.

**Theorem 4.2.5** *Given a word $w$ of length $n$, we can find its longest prefix-suffix-square free factor in $\mathcal{O}(n)$ time.*

*Proof.* We first construct the arrays $left[\cdot]$ and $right[\cdot]$ for the input word $w$. Moreover, this time we produce a range minimum query structure for the array $left[\cdot]$. Let $RMQ(j_1, j_2)$ denote the minimum value occurring in the range between $j_1$ and $j_2$ and $posRMQ(j_1, j_2)$ denote the position where $RMQ(j_1, j_2)$ occurs (in case of equality, we take the rightmost such position).

The main idea of our algorithm is the following. We go through the positions of the word $w$, and try to maintain the rightmost position $j$ where a prefix-suffix-square free factor starting on one of the positions already considered may end. Initially, before considering the first position of the word, we have $j = 0$. Let us assume that we reached the point where we consider position $i$, and the factor $w[i'..j']$ with $i' < i$ is the prefix-suffix-square free factor that ends on the rightmost position among all the prefix-suffix-square free factors that start on positions less than $i$. We want to see if we can construct a factor $w[i..j]$ with $j > j'$ and $j < right[i]$ (so that the factor $w[i..j]$ is prefix-square free); if yes, we want to find the largest such $j$. First, such a position $j$ exists if and only if $RMQ(j'+1, right[i]-1) < i$; then $j = posRMQ(j'+1, right[i]-1)$. Indeed, if $RMQ(j'+1, right[i]-1) < i$ then there is some position $j$ such that $j' < j < right[i]$ and the shortest square ending on $j$ starts before $i$; this means that $w[i..j]$ is both prefix and suffix-square free, and it also ends after $j'$, so it is the kind of factor we were looking for. Now, we try to maximize $j$. For this we set $j' = j$ and repeat the procedure above: check if $RMQ(j'+1, right[i]-1) < i$ and, if yes, set $j = posRMQ(j'+1, right[i]-1)$. We keep repeating this procedure until if $RMQ(j'+1, right[i]-1) \geq i$ or $j'+1 = right[i]$. After the procedure cannot be repeated anymore, we have the rightmost position $j$ where a prefix-suffix-square free factor $w[i'..j]$ with $i' \leq i$ may end. Clearly, before considering the next position $i+1$ and repeating the entire process above, we have also obtained the longest prefix-suffix-square free factor starting on one of the positions less or equal to $i$: if this factor starts on $i'$ and is $w[i'..j']$ then it was exactly the factor ending on the rightmost position produced by our algorithm when the position $i'$ was considered as a potential starting position of the prefix-suffix-square free factor with the rightmost ending position. We then repeat the whole process for $i+1$, and so on, until each position of the word was considered.

By the explanations given above, it is clear that this process correctly identifies the longest prefix-suffix-square free factor of $w$. The time complexity is $\mathcal{O}(n)$. Indeed, the time used to process each $i$ is proportional to the number of *RMQ* and *posRMQ* queries we perform during this processing. However, each position of $w$ can be the answer to at most one such *posRMQ*-query (when some new *RMQ* and *posRMQ* queries are asked, they are asked for a range strictly to the right of the position that was the answer of the previous *posRMQ*-query). So, in total (for all $i$) we only ask $\mathcal{O}(n)$ queries, and each is answered in constant time. Consequently, the total running time of our algorithm is linear. □

**Factorizations**

Essentially, deciding for a word $w$ the existence of a *SD* primitive root is equivalent to asking whether we can factor $w$ in $k > 1$ factors $w = s_1 \cdots s_k$ such that $s_i$ is a suffix of $s_1 \ldots s_{i-1}$, for $1 < i \leq k$, and $s_1$ is suffix-square free. It is not hard to see that such a factorisation is deeply related to the square-structure of $w$: the square $s_i^2$ occurs centred at the border between the factors $s_{i-1}$ and $s_i$ of the factorisation, for all $i > 1$. So, it seems naturally to us also to investigate other factorisations that can be easily linked to the square-structure, or, more generally, to the repetitions-structure, of the word.

The first question we ask is how to test whether a word can be factored into squares. If such a factorisation exists, then one where every factor is a primitively rooted square also exists. As the list of $\mathcal{O}(\log(n))$ primitively rooted squares ending at each position of a word of length $n$ can be obtained in total time $\mathcal{O}(n\log(n))$, we can easily obtain within the same complexity a factorisation of $w$ in primitively rooted squares by dynamic programming.

**Theorem 4.2.6** *Given a word $w$ of length $n$, we can find (if it exists) in $\mathcal{O}(n\log(n))$ time a factorisation $w = s_1 \cdots s_k$ of $w$, such that $s_i$ is a square for all $1 \leq i \leq k$.*

*Proof.* In a preprocessing phase, we produce for each position $i$ of the input word, the list of primitively rooted squares that start on position $i$; this takes in total $\mathcal{O}(n\log(n))$ time [13].

Our algorithm works as follows. We compute by dynamic programming an array $S[\cdot]$, where, for $1 \leq i \leq n$, we have $S[i] = j$ if $w[i..j]$ is a square and $w[j+1..n]$ can be factored into squares (there might be multiple choices for $j$, we just store one of them), and $S[j] = 0$ otherwise; we also set $S[n+1] = 1$. Once this array computed, we have that $S[i] \neq 0$ if and only if $w[i..n]$ can be factored in squares; the factorisation can also be retrieved easily: if $S[i] = j$ then the factorisation into squares of $w[i..n]$ starts with $w[i..j-1]$ and continues with the factorisation of $w[j..n]$, which can be recursively constructed. So, $w$ can be factored into squares if and only if $S[1] \neq 0$ and the factorisation is obtained by looking at the actual values stored in $S[\cdot]$.

Moving to the actual details, we compute the values $S[i]$ with $i$ from $n$ to 1, in decreasing order. Going through all the $\mathcal{O}(\log(n))$ primitively rooted squares starting at position $i$, if we find a primitively rooted square $w[i..j']w[j'+1..j]$ such that $S[j+1] \neq 0$, then we set $S[i] = j$ and move on to computing $S[i-1]$. Clearly, this computation is correct. The time complexity of computing each value $S[i]$ is $\mathcal{O}(\log(n))$, so in total this adds up to $\mathcal{O}(n\log(n))$. □

While we were not able to find in linear time a factorisation of a word into squares, we can find in linear time a factorisation of $w$ that contains as many squares as possible. In fact, there exists such a factorisation where all the square factors are the shortest squares occurring at their respective position. The factorisation can be obtained in linear time by dynamic programming, using the *left*$[\cdot]$ array constructed in the previous section.

**Theorem 4.2.7** *Given a word $w$, we can find (if it exists) in linear time a factorisation $w = s_1 \cdots s_k$ of $w$, such that, for any other factorisation $= s'_1 \cdots s'_p$ we have $|\{i \mid 1 \leq i \leq k, s_i \text{ is a square}\}| \geq |\{i \mid 1 \leq i \leq p, s'_i \text{ is a square}\}|$.*

*Proof.* To solve this problem, we first make some basic remarks. Consider a factorisation $w = s_1 \cdots s_k$ of $w$ into $k$ factors. Now, if $s_i$ is a square starting at position $j$, but it is not the shortest square $t^2$ starting at that position, we can replace $s_i$ by two factors: $s'_i = t^2$ and $s''_i$, which is the suffix of $s_i$ occurring after $t^2$. Applying this procedure for all $i$, we get a factorisation with as many square factors as the original one, but each of these square factors does not have a square prefix. Similarly, each non-square factor can be split into factors of length 1: the number of squares in the factorisation stays the same.

This means that we can use a dynamic programming strategy to find a factorisation of $w = s_1 \cdots s_k$ of $w$ into $k$ factors, such that any other factorisation contains at most as many square factors as $s_1, \ldots, s_k$. We compute two arrays $S_1[\cdot]$ and $S_2[\cdot]$, defined as follows, for $i \leq n$:

- $S_1[i] = m$ if there exists a factorisation of $w[i..n]$ into $m'$ factors, out of which $m$ are squares, and no other factorisation of $w[i..n]$ has more than $m$ square factors.

- $S_2[i] = j$ if there exists a factorisation of $w[i..n]$ into $m'$ factors out of which $S_1[i]$ are squares, and the first factor in this factorisation is $w[i..j]$.

The second array $S_2$ is useful to obtain the factorisation of $w[i..n]$ with the maximum number of squares. it starts with $w[i..S_2[i]]$ followed by the factorisation of $w[S_2[i]+1..n]$, which is obtained recursively. Clearly, $S_1[1]$ gives the number of square factors in a factorisation of $w$ with a maximum number of square factors, while $S_2[\cdot]$ can be used as above to obtain this factorisation.

Our algorithm computing these two arrays has a preprocessing phase, in which the array $left[\cdot]$ from Lemma 4.2.1 is computed. Then, we compute the values stored in the two arrays for $i$ from $n$ down to 1. More precisely, assume that we want to compute $S_1[i]$ and $S_2[i]$ (and the values $S_1[j]$ and $S_2[j]$ are already computed for all $j > i$). Clearly, if there exists a factorisation of $w[i..n]$ into $m'$ factors, out of which $m$ are squares, including the factor starting on position $i$, and no other factorisation of $w[i..n]$ has more than $m$ square factors, then there exists such a factorisation where the factor starting on position $i$ is the shortest square occurring at that position. So, we compute $m_1 = S_1[left[i]+1]+1$ and $n_1 = left[i]$. But it may be the case that there exists a factorisation of $w[i..n]$ into $m'$ factors, out of which $m$ are squares,but not the factor starting on position $i$, and no other factorisation of $w[i..n]$ has more than $m$ square factors; in this case, we can safely assume that this factorisation starts with a factor of length 1. So, we also compute $m_2 = S_1[i+1]+1$ and $n_2 = i+1$. Finally, we set $S_1[i] = \max\{m_1, m_2\}$ and $S_2[i] = n_1$ if $m_1 \geq m_2$ or $S_2[i] = n_2$, otherwise.

The above procedure computes the arrays $S_1[\cdot]$ and $S_2[\cdot]$ correctly. Moreover, $S_1[i]$ and $S_2[i]$ are computed in constant time, for each $i$, so the total time needed to compute the values stored in these arrays is linear. This completely shows the statement. □

Another extension of the factorisation into squares of a word, that can still be solved in linear time, is to decide whether a given word can be factored into $k$ factors $w = s_1 \cdots s_k$, such that $per(s_i) \leq \frac{|s_i|}{2}$ for all $i$; in such a factorisation, each $s_i$ is a run (not necessarily maximal). The main idea in finding such a factorisation is that if $w[1..i]$ is a prefix of $w$ that can be factored in runs, and $w[i'..j']$ is a maximal run of period $p$ containing $i+1$ and with $i+2p \leq j'$, then all the factors $w[1..k]$ with $i+2p \leq k \leq j'$ can also be factored in runs. Using an interval union-find data-structure ([31]) to maintain the positions $i$ incrementally discovered during the execution of our algorithm such that $w[1..i]$ can be factored into runs we obtain in linear time a factorisation of $w$ in runs.

**Theorem 4.2.8** *Given a word $w$, we can find (if it exists) in linear time a factorisation $w = s_1 \cdots s_k$ of $w$, such that $per(s_i) \leq \frac{|s_i|}{2}$ for all $1 \leq i \leq k$.*

*Proof.* The first step in this algorithm is to obtain all the maximal runs of the input word $w$, and sort them by their starting position. This takes linear time (the sorting can be done by, e.g., count sort). The basic idea in our algorithm is that if $w[1..i]$ can be written as the concatenation of several runs, then we look at position $i+1$ and see all the maximal runs that contain this position. Now, for each such run $w[i'..j']$, if the period of $w[i'..j']$ is $p$ and $|w[i+1..j']| \geq 2p$, we conclude that all the prefixes $w[1..k]$ with $i+2p \leq k \leq j'$ can also be written as concatenation of runs. So, we consider in increasing order the prefixes $w[1..i]$, see if they can be factored in runs, and extend each of them as above, discovering longer prefixes that can be factored in runs. We describe in the following an efficient implementation of this general strategy.

As explained above, in our algorithm we consider each prefix $w[1..i]$ at a time and, for each such prefix that can be factored into runs, we discover several other longer prefixes $w[1..k]$ that can be factored into runs, at their turn. So, we also maintain an array $M[\cdot]$, where $M[k] = i$ if and only if we discovered that $w[1..k]$ can be factored in runs while we were trying to extend $w[1..i]$ by concatenating to it some run, and we did not already knew that $w[1..k]$ can be factored in runs.

Now we can give the details of our algorithm. Initially, the interval partition we maintain consists in the singletons $[i, i+1)$ for $1 \leq i \leq n$, and $M$ has all the positions set to 0. We first consider all the runs that start on position 1. For each run $w[1..k]$ we execute the following procedure. Assume that $per(w[1..k]) = p$. We simply make the union of all the intervals that contain the numbers between $2p$ and $k$(remember lemma 2.4.1). While executing the union operation as above, if we add to the current interval a singleton $[j, j+1)$ with $M[j] = 0$, we simply set $M[j] = 1$. Then we discard all the runs starting on position 1. We then advance in the word until we reach a new position $i$ such that $M[i] \neq 0$. For each run $w[s..f]$ with $1 < s \leq i+1$ we check whether $i + 2p \leq f$. If no, we discard the run: it cannot be used to extend $w[1..i]$. If yes, we make the union of all the intervals that contain the numbers between $i + 2p$ and $k$. While executing the union operation, if we add to the current interval a singleton $[j, j+1)$ with $M[j] = 0$, we simply set $M[j] = i$. After we considered all the runs that contain $i+1$ we discard them, and repeat the procedure: continue moving from left to right in the word and find the next position $i'$ with $M[i'] \neq 0$ and do the same processing with $i'$ in the role of $i$ and processing the runs that start between the previously considered position $i+1$ and $i'$. We stop when we went through the entire word. We decide that $w$ can be factored in runs if and only if $M[n] \neq 0$.

The correctness of our approach is immediate. The prefix $w[1..k]$ can be factored into runs if and only if there exists $i$ such that $w[1..i]$ can be factored into runs and $w[i+1..k]$ is a run. The time complexity is $\mathcal{O}(n)$. Basically, the time we use equals the time spent to perform the union operations plus the time needed to update $M$. As the union operations are done in amortized constant time, and their number equals the number of runs, and, also, each position in $M$ is updated only once, we get that the running time of our algorithm is linear.

To effectively obtain the factorisation of $w$ in runs, we use the following general remark: if $M[i] = j$ then $w[1..i] = w[1..j]w[j+1..i]$, where $w[1..j]$ can be factored in runs and $w[j+1..i]$ is a run; the factoring of $w[1..j]$ can be obtained by the same procedure, recursively. □

## 4.3 Longest Gapped Repeats and Palindromes

As it was presented in the introduction of this chapter, in this section we discuss the problem of finding the longest previous factor (LPF) table, and longest previous reverse factor (*LPrF*) table in three different settings. Note that if we restrict the gap to being lower bound by a constant, than we look for factors $uvu$ (or $u^Rvu$) with $|v| > g$ for some $g \geq 0$. The techniques of [15] can be easily adapted to compute for each position $i$ the longest prefix $u$ of $w[i..n]$ such that there exists a suffix $uv$ (respectively, $u^Rv$) of $w[1..i-1]$, forming, thus, a factor $uvu$ (respectively, $u^Rvu$) with $|v| > g$. Here we consider three other different types of restricted gaps.

We first consider the case when the length of the gap is between a lower bound $g$ and an upper bound $G$, where $g$ and $G$ are given as input (so, may depend on the input word). This extends naturally the case of lower bounded gaps.

**Problem 4.3.1** *Given $w$ of length $n$ and two integers $g$ and $G$, such that $0 \leq g < G \leq n$, construct the arrays $LPrF_{g,G}[\cdot]$ and $LPF_{g,G}[\cdot]$ defined for $1 \leq i \leq n$:*

*a.* $LPrF_{g,G}[i] = \max\{|u| \mid$ *there exists $v$ such that $u^Rv$ is a suffix of $w[1..i-1]$ and $u$ is prefix of $w[i..n]$, with $g \leq |v| < G$*$\}$.

*b.* $LPF_{g,G}[i] = \max\{|u| \mid$ *there exists $v$ such that $uv$ is a suffix of $w[1..i-1]$ and $u$ is prefix of $w[i..n]$, with $g \leq |v| < G$*$\}$.

We are able to solve Problem 4.3.1(a) in linear time $\mathcal{O}(n)$. Problem 4.3.1(b) is solved here in $\mathcal{O}(n\log(n))$ time. Intuitively, when trying to compute the longest prefix $u$ of $w[i..n]$ such that $u^Rv$ is a suffix of $w[1..i-1]$ with $g < |v| \leq G$, we just have to compute the longest common prefix between $w[i..n]$ and the words $w[1..j]^R$ with $g < i-j \leq G$. The increased difficulty in solving the problem for repeats (reflected in the increased complexity of our algorithm) seems to come from the fact that when trying to compute the longest prefix $u$ of $w[i..n]$ such that $uv$ is a suffix of $w[1..i-1]$ with $g < |v| \leq G$, it is hard to see where the $uv$ factor may start, so we have to somehow try more variants for the length of $u$. In [8], the authors give an algorithm that finds all maximal repeats (i.e., repeats whose arms cannot be extended) with gap between a lower and an upper bound, running in $\mathcal{O}(n\log(n)+z)$ time, where $z$ is the number of such repeats. It is worth noting that there are words (e.g., $(a^2b)^{n/3}$, from [8]) that may have $\Theta(nG)$ maximal repeats $uvu$ with $|v| < G$, so for $G > \log(n)$ and $g = 0$, for instance, our algorithm is faster than an approach that would first use the algorithms of [8] to get all maximal repeats, and then process them somehow to solve Problem 4.3.1(b).

In the second case, the gaps are only lower bounded; however, the bound on the gap allowed at a position is defined by a function depending on that position.

**Problem 4.3.2** *Given $w$ of length $n$ and the values $g(1), \ldots, g(n)$ of $g : \{1, \ldots, n\} \rightarrow \{1, \ldots, n\}$, construct the arrays $LPrF_g[\cdot]$ and $LPF_g[\cdot]$ defined for $1 \leq i \leq n$:*

*a.* $LPrF_g[i] = \max\{|u| \mid$ *there exists $v$ such that $u^R v$ is a suffix of $w[1..i-1]$ and $u$ is prefix of $w[i..n]$, with $g(i) \leq |v|$*$\}$.

*b.* $LPF_g[i] = \max\{|u| \mid$ *there exists $v$ such that $uv$ is a suffix of $w[1..i-1]$ and $u$ is prefix of $w[i..n]$, with $g(i) \leq |v|$*$\}$.

The setting of this problem can be seen as follows. An expert preprocesses the input word (in a way specific to the framework in which one needs the problem solved), and detects the length of the gap occurring at each position (so, computes $g(i)$ for all $i$). These values and the word are then given to us, to compute the arrays defined in our problems. We are solve both problems in linear time.

Finally, following [43, 44], we analyze gapped repeats $uvu$ and palindromes $u^R vu$ where the length of the gap $v$ is upper bounded by the length of the arm $u$; these structures are called long armed repeats and palindromes, respectively.

**Problem 4.3.3** *Given $w$ of length $n$, construct the arrays $LPal[\cdot]$ and $LRep[\cdot]$, defined for $1 \leq i \leq n$:*

*a.* $LPal[i] = \max\{|u| \mid$ *there exists $v$ such that $u^R v$ is a suffix of $w[1..i-1], u$ is a prefix of $w[i..n]$, and $|v| \leq |u|$*$\}$.

*b.* $LRep[i] = \max\{|u| \mid$ *there exists $v$ such that $uv$ is a suffix of $w[1..i-1], u$ is a prefix of $w[i..n]$, and $|v| \leq |u|$*$\}$.

In [43] one proposes an algorithm finding the set $S$ of all factors of a word of length $n$ which are maximal long armed palindromes (i.e., the arms cannot be extended to the right or to the left) in $\mathcal{O}(n + |S|)$ time; no upper bound on the possible size of the set $S$ was given in [43], but it is widely believed to be $\mathcal{O}(n)$. Using the algorithm of [43] as an initial step, we solve Problem 4.3.3(a) in $\mathcal{O}(n + |S|)$ time. In [44] the set of maximal long armed repeats with nonempty gap is shown to be of linear size and is computed in $\mathcal{O}(n)$ time. We use this algorithm and a linear time algorithm finding the longest square centred at each position of a word to solve Problem 4.3.3(b) in linear time.

We now approach Problems 4.3.1(a) and 4.3.1(b).

**Theorem 4.3.1** *Problem 4.3.1(a) can be solved in linear time.*

*Proof.* Let $\delta = G - g$; for simplicity, assume that $n$ is divisible by $\delta$. Further, for $1 \leq i \leq n$, let $u$ be the longest factor which is a prefix of $w[i..n]$ such that $u^R$ is a suffix of some $w[1..k]$ with $g < i - k \leq G$; then, $B[i]$ is the rightmost position $j$ such that $g < i - j \leq G$ and $u^R$ is a suffix of $w[1..j]$. Knowing $B[i]$ means knowing $LPrF_{g,G}[i]$: we just have to return $LPrF(w[i..n], w[1..j]^R)$.

We split the set $\{1, \ldots, n\}$ into $\frac{n}{\delta}$ ranges of consecutive numbers: $I_0 = \{1, 2, \ldots, \delta\}$, $I_1 = \{\delta + 1, \delta + 2, \ldots, 2\delta\}$, and so on. Note that for all $i$ in some range $I_k = \{k\delta + 1, k\delta + 2, \ldots, (k + 1)\delta\}$ from those defined above, there are at most three consecutive ranges where some $j$ such that $g < i - j \leq G$ may be found: the range containing $k\delta - G + 1$, the range containing $(k + 1)\delta - g - 1$, and the one in between these two (i.e., the range containing $k\delta + 1 - g$). Moreover, for some fixed $i$, we know that we have to look for $B[i]$ in the interval $\{i - G, \ldots, i - g - 1\}$, of length $\delta$; when we search $B[i + 1]$ we look at the interval $\{i + 1 - G, \ldots, i - g\}$. So, basically, when trying to find $B[i]$ for all $i \in I_k$, we move a window of length $\delta$ over the three ranges named above, and try to find for every content of the window (so for every $i$) its one element that fits the description of $B[i]$. The difference between the content of the window in two consecutive steps does not seem major: we just removed an element and introduced a new one. Also, note that at each moment the window intersects exactly two of the aforementioned three ranges. We try to use these remarks, and maintain the contents of the window such that the update can be done efficiently, and the values of $B[i]$ (and, implicitly, $LPrF_{g,G}[i]$) can be, for each $i \in I$, retrieved very fast. Intuitively, grouping the $i$'s on ranges of consecutive numbers allows us to find the possible places of the corresponding $B[i]$'s for all $i \in I_k$ in $\mathcal{O}(\delta)$ time.

We now go into more details. As we described in the preliminaries, let $\mathcal{L}$ be the lexicographically ordered list of the the suffixes of $w[i..n]$ of $w$ and of the mirror images $w[1..i]^R$ of the prefixes of $w$ (which correspond to the suffixes of $w^R$). For the list $\mathcal{L}$ we compute the arrays $Rank[\cdot]$ and $Rank_R[\cdot]$. We use the suffix array for $w0w^R$ to produce for each of the ranges $I_k$ computed above the set of suffixes of $w$ that start in the respective range (sorted lexicographically, in the order they appear in the suffix array) and the set of prefixes of $w$ ending in $I$, ordered lexicographically with respect to their mirror image.

We consider now one of the ranges of indexes $I_k = \{k\delta{+}1, k\delta{+}2, \ldots, (k{+}1)\delta\}$ from the above, and show how we can compute the values $B[i]$, for all $i \in I_k$. For some $i \in I_k$ we look for the maximum $\ell$ such that there exists a position $j$ with $w[j{-}\ell{+}1..j]^R = w[i..i{+}\ell{-}1]$, and $i{-}G \leq j \leq i{-}g$. As already explained, for the $i$'s of $I_k$ there are three consecutive ranges where the $j$'s corresponding to the $i$'s of $I_k$ may be found. Let us denote them $J_1, J_2, J_3$.

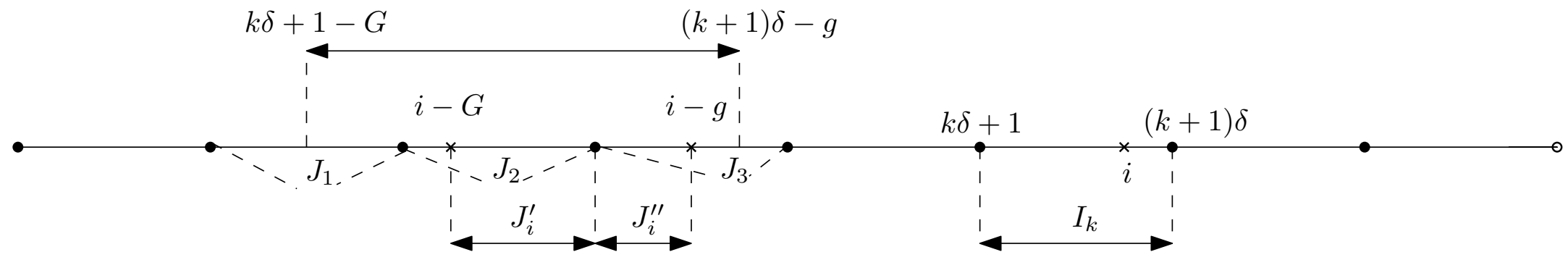

Figure 4.1: Proof of Theorem 4.3.1: Construction of the ranges $J_1, J_2, J_3$ for some range $I_k$, and of the sliding window $J'_i \cup J''_i$ for some $i \in I_k$

Now, for an $i \in I_k$ we have that $B[i]$ (for which $g < i - B[i] \leq G$) belongs to $J'_i \cup J''_i$, where $J'_i$ is an interval starting with $i - G$ and extending until the end of the range that contains $i - G$ (which is one of the ranges $J_1, J_2, J_3$) and $J''_i$ is an interval ending with $i - g - 1$, which starts at the beginning of the range that contains $i - g - 1$ (which is the range that comes next after the one containing $i - G$). Referencing back to the intuitive explanation we gave at the beginning of this proof, $J'_i \cup J''_i$ is the window we use to locate the value of $B[i]$ for an $i \in I_k$.

To compute $B[i]$ for some $i$, take $f_i \in J'_i$ such that $LCP(w[1..f_i]^R, w[i..n]) \geq LCP(w[1..j']^R, w[i..n])$ for all $j' \in J'_i$. Similarly, take $s_i \in J''_i$ such that for all $j' \in J''_i$ we have $LCP(w[1..s_i]^R, w[i..n]) \geq LCP(w[1..j']^R, w[i..n])$. Once $s_i$ and $f_i$ computed, we just have to set $B[i] = s_i$ if $LCP(w[1..s_i]^R, w[i..n]) \geq LCP(w[1..f_i]^R, w[i..n])$; we set $B[i] = f_i$, otherwise. So, in order to compute, for some $i$, the value $B[i]$, that determines $LPrF_{g,G}[i]$, we first compute $f_i$ and $s_i$.

We compute for all the indices $i \in I_k$, considered in increasing order, the values $f_i$. We consider for each $i \in I_k$ the interval $J'_i$ and note that $J'_i \setminus J'_{i+1} = \{i - G\}$, and, if $J'_i$ is not a singleton (i.e., $J'_i \neq \{i - G\}$) then $J'_{i+1} \subset J'_i$. If $J'_i$ is a singleton, than $J'_{i+1}$ is, in fact, one of the precomputed range $I_p$, namely the one which starts on position $i + 1 - G$ (so, $p = \frac{i-G}{\delta}$).

These relations suggest the following approach: we start with $i = k\delta + 1$ and consider the set of words $w[1..j]^R$ with $j \in J'_i$; this set can be easily obtained in $\mathcal{O}(\delta)$ time by finding first the range $J_1$ in which $i - G$ is contained (which takes $\mathcal{O}(1)$ time, as $J_1 = I_p$ for $p = \lfloor \frac{i-G}{\delta} \rfloor$), and then selecting from $J_1$ of the set of prefixes $w[1..d]$ of $w$, ending in $J_1$ with $d \geq i - G$ (ordered lexicographically with respect to their mirror image). The ranks corresponding to these prefixes in the ordered list $\mathcal{L}$ (i.e., the set of numbers $Rank_R[d]$) define a partition of the universe $U = [0, 2n + 1)$ in at most $\delta + 2$ disjoint intervals. So, we can maintain an interval union-find data structures like in Remark 2.4.1, where the ranks are seen as limits of the intervals in this structure. We assume that the intervals in our data structure are of the form $[a, b)$, with $a$ and $b$ equal to some $Rank_R[d_a]$ and $Rank_R[d_b]$, respectively. The first interval in the structure is of the form $[0, a)$, while the last is of the form $[b, 2n+1)$. We now find the interval to which $Rank[i]$ belongs; say that this is $[a, b)$. This means that the words $w[1..d_a]^R$ and $w[1..d_b]^R$ are the two words of $\{w[1..d]^R \mid d \in J'_i\}$ which are closest to $w[i..n]$ lexicographically ($w[1..d_a]^R$ is lexicographically smaller, $w[1..d_b]$ is greater). Clearly, $f_i = d_a$ if $LCP(w[1..d_a]^R, w[i..n]) \geq LCP(w[1..d_b]^R, w[i..n])$ and $f_i = d_b$, otherwise (in case of a tie, we take $f_i$ to be the greater of $d_a$ and $d_b$). So, to compute $f_i$ we query once the union-find data structure to find $a$ and $b$, and the corresponding $d_a$ and $d_b$, and then run two more $LCP$ queries.

When moving on to compute $f_{i+1}$, we just have to update our structure and then run the same procedure. Now, $i-G$ is no longer a valid candidate for $f_{i+1}$, and it is removed from $J'_i$. So we just delete it from the interval union-find data structure, and merge the interval ending right before $Rank_R[i-G]$ and the one starting with $Rank_R[i-G]$. This means one union operation in our interval union-find structure. Then we proceed to compute $f_{i+1}$ as in the case of $f_i$.

The process continues until $J'_i$ is a singleton, so $f_i$ equals its single element.

Now, $i-G$ is the last element of one of the ranges $J_1, J_2$, or $J_3$; assume this range is $I_p$. So far, we performed alternatively at most $\delta$ find queries and $\delta$ union operations on the union-find structure. Now, instead of updating this structure, we consider a new interval partition of $[0, 2n+1)$ induced by the ranks of the $\delta$ prefixes ending in $I_{p+1}$. When computing the values $f_i$ for $i \in I_k$ we need to consider a new partition of $U$ at most once: at the border between $J_1$ and $J_2$.

It is not hard to see from the comments made in the above presentation that our algorithm computes $f_i \in I_k$ correctly. In summary, in order to compute $f_i$, we considered the elements $i \in I_k$ from left to right, keeping track of the left part $J'_i$ of the window $J'_i \cup J''_i$ while it moved from left to right through the ranges $J_1, J_2$ and $J_3$. Now, to compute the values $s_i$ for $i \in I_k$, we proceed in a symmetric manner: we consider the values $i$ in decreasing order, moving the window from right to left, and keep track of its right part $J''_i$.

As already explained, by knowing the values $f_i$ and $s_i$ for all $i \in I_k$ and for all $k$, we immediately get $B[i]$ (and, consequently $LPrF_{g,G}[i]$) for all $i$.

We now evaluate the running time of our approach. We can compute, in $\mathcal{O}(n)$ time, from the very beginning of our algorithm the partitions of $[1, 2n-1)$ we need to process (basically, for each $I_k$ we find $J_1$, $J_2$ and $J_3$ in constant time, and we get the three initial partitions we have to process in $\mathcal{O}(\delta)$ time), and we also know the union-operations and find-queries that we will need to perform for each such partition (as we know the order in which the prefixes are taken out of the window, so the order in which the intervals are merged). In total we have $\mathcal{O}(n/\delta)$ partitions, each having initially $\delta + 2$ intervals, and on each we perform $\delta$ find-queries and $\delta$-union operations. So, by Remark 2.4.1, we can preprocess this data (once, at the beginning of the algorithm) in $\mathcal{O}(n)$ time, to be sure that the time needed to obtain the correct answers to all the find queries is $\mathcal{O}(n)$. So, the total time needed to compute the values $f_i$ for all $i \in I_k$ and for all $k$ is $\mathcal{O}(n)$. Similarly, the total time needed to compute the values $s_i$ for all $i$ is $\mathcal{O}(n)$. Then, for each $i$ we get $B[i]$ and $LPrF_{g,G}[i]$ in $\mathcal{O}(1)$ time.

Therefore, Problem 4.3.1(a) can be solved in linear time. □

We will now give a remark that involves the dictionary of basic factors that we have defined in the preliminaries. This remark will help us in solving problem 4.3.1(b).

**Remark.** *Using the DBF of $w$, for a given $\ell > 0$, we can produce in $\mathcal{O}(n\log(n))$ time a data structure answering the following type of queries in $\mathcal{O}(1)$ time: "Given i and k return the compact representation of the occurrences of $w[i..i+2^k-1]$ in $w[i-\ell-2^{k+1}..\,i-\ell-1]$". Similarly, given $\ell > 0$ and a constant $c > 0$ (e.g., $c = 10$), we can produce in $\mathcal{O}(n\log(n))$ time a data structure answering the following type of queries in $\mathcal{O}(1)$ time: "Given i and k return the compact representation of the occurrences of $w[i..i+2^k-1]$ in $w[i-\ell-c2^k..i-\ell-1]$ ".*

*Proof.* Once we construct the dictionary of basic factors of a word $w$ of length $n$ in $\mathcal{O}(n\log(n))$ time, we reorganize it such that for each distinct basic factor we have an array with all the positions where it occurs, ordered increasingly. Now, we traverse each such array, keeping track of the current occurrence $w[i..i_2^k-1]$ and a window containing its occurrences from the range between $i-\ell-c2^k$ and $i-\ell$ (and a compact representation of these occurrences); when we move in our traversal to the next occurrence of the current basic factors, we also slide the window in the array, and, looking at the content of the previous window and keeping track of the occurrences that were taken out and those that were added to its content, we can easily obtain in constant time a representation of the occurrences of the considered basic factors inside the new window. □

**Theorem 4.3.2** *Problem 4.3.1(b) can be solved in $\mathcal{O}(n\log(n))$ time.*

*Proof.* For $1 \le i \le n$, let $B[i]$ denote the value $j$ such that $w[j..j+LPF_{g,G}[i]-1] = w[i..i+LPF_{g,G}[i]-1]$ and $g < i-(j+LPF_{g,G}[i]-1) \le G$. In other words, to define $B[i]$, let $u$ be the longest factor which is both a prefix of $w[i..n]$ and a suffix of some $w[1..k]$ with $g < i-k \le G$; then, $B[i]$ is the rightmost position $j$ such that $g < i-(j+|u|-1) \le G$ and $u$ is a suffix of $w[1..j+|u|-1]$. Clearly, knowing $B[i]$ means knowing $LPrF_{g,G}[i]$.

Intuitively, computing $B[i]$ is harder in this case than it was in the case of the solution of Problem 4.3.1(a). Now we have no real information where $B[i]$ might be, we just know the range of $w$ where the longest factor that occurs both at $B[i]$ and at $i$ ends. So, to determine $B[i]$ we try different variants for the length of this factor, and see which one leads to the right answer.

Let $\delta = G - g$ and $k_0 = \lfloor \log(n) \rfloor$; assume w.l.o.g. that $n$ is divisible by $\delta$.

As already noted, the solution used in the case of gapped palindromes in the proof of Theorem 4.3.1 does not work anymore in this case: we do not know for a given $i$ a range in which $B[i]$ is found. So, we try to restrict the places where $B[i]$ can occur. We split our discussion in two cases.

In the first case, we try to find, for each $i$ and each $k \geq \frac{k_0}{2}$, the factor $uvu$ with the longest $u$, such that the second $u$ occurs at position $i$ in $w$, $2^k \leq |u| < 2^{k+1}$ and $g < |v| \leq G$. Clearly, $u$ should start with the basic factor $w[i..i+2^k-1]$, and the left arm $u$ in the factor $uvu$ we look for should have its prefix of length $2^k$ (so, a factor equal to the basic factor $w[i..i+2^k-1]$) contained in the factor $w[i-G-2^{k+1}..i-g-1]$, of length less than $10 \cdot 2^k$. So, by Remark 4.3, using the dictionary of basic factors we can retrieve a compact representation of the occurrences of $w[i..i+2^k-1]$ in $w[i-G-2^{k+1}..i-g-1]$: these consist in a constant number of isolated occurrences and a constant number of runs. For each of the isolated occurrences $w[j'..j'+2^k-1]$ we compute $LCP(j', i)$ and this gives us a possible candidate for $u$; we measure the gap between the two occurrences of $u$ (the one at $j'$ and the one at $i$) and if it is between smaller than $G$, we store this $u$ as a possible solution to our problem (we might have to cut a suffix of $u$ so that the gap is also longer than $g$). Further, each of the runs has the form $p^\alpha p'$, where $|p|$ is the period of $w[i..i+2^k-1]$ and $p'$ is a prefix of $p$ (which varies from run to run); such a run cannot be extended to the right: either the period breaks, or it would go out of the factor $w[i-G-2^{k+1}..i-g-1]$. Using a $LCP$ query we find the longest factor $p^\beta p''$ with period $|p|$, occurring at position $i$. For a run $p^\alpha p'$, the longest candidate for the first $u$ of the factor $uvu$ we look for starts with $p^\gamma p'''$ where $\gamma = \min\{\alpha, \beta\}$ and $p'''$ is the shortest of $p'$ and $p''$; if $p' = p''$, then this factor is extended with the longest factor that occurs both after the current run and after $p^\beta p''$ and does not overlaps the minimal gap (i.e., end at least $g$ symbols before $i$). This gives us a candidate for the factor $u$ we look for (provided that the gap between the two candidates for $u$ we found is not too large). In the end, we just take the longest of the candidates we identified (in case of ties, we take the one that starts on the rightmost position), and this gives the factor $uvu$ with the longest $u$, such that the second $u$ occurs at position $i$ in $w$, $2^k \leq |u| < 2^{k+1}$ and $g < |v| \leq G$.

Iterating this process for all $k$ and $i$ as above, we obtain for each $i$ the factor $uvu$ with the longest $u$, such that the second $u$ occurs at position $i$ in $w$, $2^{k_0/2} \leq |u|$ and $g < |v| \leq G$. By Remark 4.3, the time we spent in this computation for some $i$ and $k$ is constant, so the overall time used in the above computation is $\mathcal{O}(n \log(n))$. Clearly, if for some $i$ we found such a factor $u$, then this gives us both $B[i]$ and $LPF_{g,G}[i]$; if not, we continue as follows.

In the second case, we try to identify for each $i$ the factor $uvu$ with the longest $u$, such that the second $u$ factor occurs at position $i$ in $w$, $u < 2^{k_0/2}$ and $g < |v| \leq G$. Again, we consider each $k < \frac{k_0}{2} - 1$ separately and split the discussion in three cases.

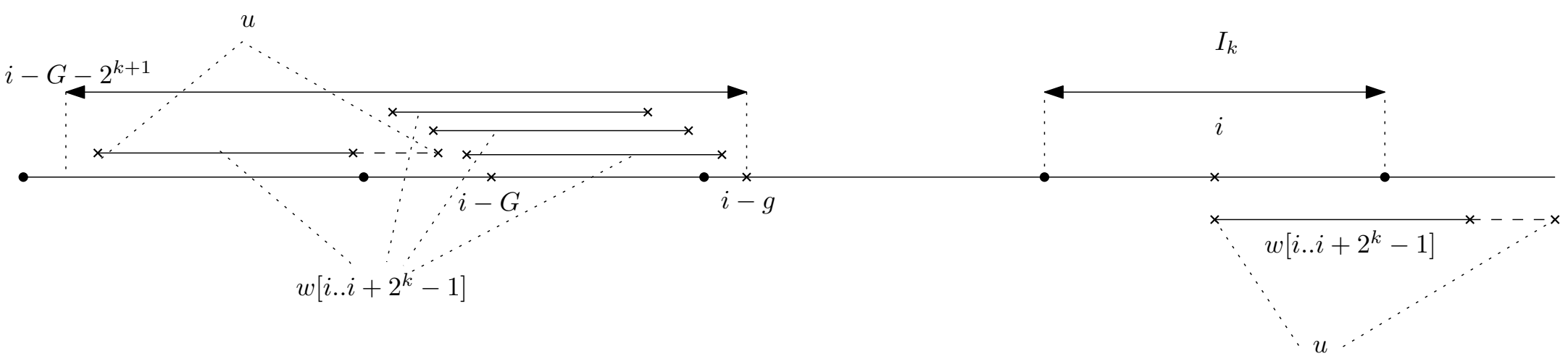

Figure 4.2: Occurrences of $w[i..i+2^k-1]$ inside $w[i-G-2^{k+1}..i-g-1]$: one separate occurrence, and a run containing three occurrences. The separate occurrence can be prolonged to produce a gapped palindrome $uvu$, which is not, however, long enough as it does not reach the range between $i-G$ and $i-g$. The rightmost occurrence in the run produces the gapped repeat $w[i..i+2^k-1]v'w[i..i+2^k-1]$, which fulfills the conditions imposed on the gap.

First, for each $i$, we find the factor $uvu$ with the longest $u$, such that the second $u$ occurs at position $i$ in $w$, $2^k \leq u < 2^{k+1}$, $g < |v| \leq G$, and the first $u$ has its prefix of length $2^k$ in the factor $w[i-G-2^{k+1}..i-G+2^k]$. This can be done similarly to the above (report all the occurrences of $w[i..i+2^k-1]$ in that range, and try to extend them to get $u$); for all $i$ and all $k$ takes $\mathcal{O}(n\log\delta)$ time.

Second, for each $i$, we find the factor $uvu$ with the longest $u$, such that the second $u$ occurs at position $i$ in $w$, $2^k \leq u < 2^{k+1}$, $g < |v| \leq G$, and the first $u$ has its prefix of length $2^k$ in the factor $w[i-g-2^{k+1}..i-g-1]$. Again, this can be done like above, and for all $i$ and all $k$ takes $\mathcal{O}(n\log\delta)$ time.

The third and more complicated subcase is when the first $u$ starts in the factor $w[i-G..i-g-2^{k+1}-1]$, of length $\delta_k = \delta-2^{k+1}-1$. Let $g_k = g+2^{k+1}+1$; obviously, $\delta_k = G-g_k$. Note that, in this case, every factor of length at most $2^{k+1}$ starting in $w[i-G..i-g-2^{k+1}-1]$ ends before $i-g$, so it is a good candidate for the $u$ we look for. In this case we can follow the algorithm from the proof of Theorem 4.3.1. We split the set $\{1,\ldots,n\}$ into ranges of consecutive numbers: $I_0 = \{1,2,\ldots,\delta_k\}$, $I_1 = \{\delta_k+1,\delta_k+2,\ldots,2\delta_k\}$, and so on. For some $I_\ell = \{\ell\delta_k+1,\ldots,(\ell+1)\delta_k\}$, considering all $i \in I_\ell$ there are three consecutive ranges from the ones defined above, where the first $u$ of the factor $uvu$ we look for may occur. The first (leftmost, with respect to its starting position) such range is the one containing $\ell\delta_k - G$, and let us denote it $J_1$; the last one (rightmost) is the one that contains $(\ell+1)\delta_k - g_k - 1$, and we denote it by $J_3$. Clearly, between the ranges containing $\ell\delta_k - G$ and $(\ell+1)\delta_k - g_k - 1$, respectively, there is exactly one complete range, call it $J_2$.

Moreover, for a precise $i \in I_\ell$ the possible starting positions of the left arm $u$ of the repeat $uvu$ of the type we are searching for (i.e., with the gap between $g$ and $G$), with the second $u$ starting on $i$, form a contiguous range $J'_i \cup J''_i$ where $J'_i$ is an interval starting with $i - G$ and extending until the end of the range $J_p$ that contains $i - G$, and $J''_i$ is an interval ending with $i - g_k - 1$, contained in $J_{p+1}$ (when $p < 3$). Like before, $J'_i \cup J''_i$ can be seen as the content of a window that slides through $J_1, J_2, J_3$ while searching for $B[i]$. We denote by $f_i$ the position of $J'_i$ such that $LCP(f_i, i)$ is maximum among all such positions; we denote by $s_i$ the position of $J''_i$ such that $LCP(s_i, i)$ is maximum among all such positions. Then we just have to check where, at $f_i$ or $s_i$, occurs a longer factor that also occurs at $i$. We just explain how to compute $f_i$.

We start with $i = \ell\delta_k + 1$ and consider the set of words $w[j..n]$ with $j \in J'_i$; this set can be easily obtained in $\mathcal{O}(\delta)$ time by finding first the range $J_1$ in which $i - G$ is contained and then selecting from $J_1$ of the set of words $w[d..n]$ of $w$ starting in $J_1$ with $d \geq i - G$ (ordered lexicographically). The ranks corresponding to these suffixes in the suffix array of $w$ (i.e., the set of numbers $Rank[d]$) define a partition of the universe $U = [0, n + 1)$ in at most $\delta_k + 2$ disjoint intervals. So, we can maintain an interval union-find data structures like in Remark 2.4.1, where the ranks are seen as limits of the intervals in this structure. We assume that the intervals in our data structure are of the form $[a, b)$, with $a$ and $b$ integers that are equal to some $Rank[d_a]$ and $Rank[d_b]$. The first interval in the structure is of the form $[0, a)$, while the last is of the form $[b, n + 1)$. Recall that we want to compute $f_i$. To this end, we just have to find the interval to which $Rank[i]$ belongs; say that this is $[a, b)$. This means that the words $w[d_a..n]$ and $w[d_b..n]$ are the two words of the set $\{w[d..n] \mid d \in J'_i\}$ which are closest to $w[i..n]$ lexicographically ($w[d_a..n]$ is lexicographically smaller, while $w[d_b..n]$ is lexicographically greater). Clearly, $f_i = d_a$ if $LCP(d_a, i) \geq LCP(d_b, i)$ and $f_i = d_b$, otherwise (in case of a tie, we take $d_a$ if $d_b < d_a$, and $d_b$ otherwise). So, to compute $f_i$ we have to query once the union-find data structure to find $a$ and $b$, and the corresponding $d_a$ and $d_b$, and then compute the answer to two $LCP$ queries.

When moving on to compute $f_{i+1}$, we just have to update our structure. In this case, $i - G$ is no longer a valid candidate for $f_{i+1}$, as it should be removed from $J'_i$. So we just delete it from the interval union-find data structure, and merge the interval ending right before $Rank[i - G]$ and the one starting with $Rank[i - G]$. This means one union operation in our interval union-find structure. Then we proceed to compute $f_{i+1}$ just in the same manner as in the case of $f_i$.

The process continues in this way until we try to compute $f_i$ for $J'_i$ being a singleton. This means that $i$ is the last element of one of the ranges $J_1$, $J_2$, or $J_3$; let us assume that this range is $I_p$ from the ranges defined above. Clearly, this time $f_i$ is the single element of $J'_i$. Until now, we performed alternatively at most $\delta_k$ find operations and $\delta_k$ union operations on the union-find data structure. Further, instead of updating the union-find data structure, we consider a new interval partition of $[0, n+1)$ induced by the ranks of the $\delta_k$ suffixes starting in $I_{p+1}$. Note that when computing the values $f_i$ for $i \in I_k$ we need to consider a new partition of $U$ once: at the border between $J_1$ and $J_2$.

Now, by the same arguments as in the proof for gapped palindromes, the process takes $\mathcal{O}(n)$ in total for all intervals $I_\ell$ (so, when we iterate $\ell$), for each $k$. In this way we find the factor $uvu$ with the longest $u$, such that the second $u$ occurs at position $i$ in $w$, $2^k \leq u < 2^{k+1}$, $g < |v| \leq G$, and the first $u$ starts in $w[i-G..i-g-2^{k+1}-1]$. Iterating for all $k$, we complete the computation of $B[i]$ and $LPF_{g,G}[i]$ for all $i$; the needed time is $\mathcal{O}(n \log \delta)$ in this case.

Considering the three cases above leads to finding for each $i$ the factor $uvu$ with the longest $u$, such that the second $u$ factor occurs at position $i$ in $w$, $u < 2^{k_0/2}$ and $g < |v| \leq G$. The complexity of this analysis is $\mathcal{O}(n \log \delta)$. After concluding this, we get the value $B[i]$ for each $i$. Therefore, the entire process of computing $B[i]$ and $LPF_{g,G}[i]$ for all $i$ takes $\mathcal{O}(n \log(n))$ time. □

**Lower bounded gap**

**Theorem 4.3.3** *Problem 4.3.2(a) can be solved in linear time.*

*Proof.* In the preprocessing step of our algorithm, we produce the suffix array of $w0w^R$ and the lexicographically ordered list $\mathcal{L}$ of the suffixes of $w[i..n]$ of $w$ and of the mirror images $w[1..i]^R$ of the prefixes of $w$ (which correspond to the suffixes of $w^R$). For the list $\mathcal{L}$ we compute the arrays $Rank[\cdot]$ and $Rank_R[\cdot]$.

We first want to find, for each $i$, the prefix $w[1..j]$, such that $i - j > g(i)$, $w[1..j]^R$ occurs before $w[i..n]$ in $\mathcal{L}$ (i.e., $Rank_R[j] < Rank[i]$), and the length of the common prefix of $w[1..j]^R$ and $w[i..n]$ is greater or equal to the length of the common prefix of $w[1..j']^R$ and $w[i..n]$ for $j'$ such that $Rank_R[j'] < Rank[i]$; for the prefix $w[1..j]$ as above, the length of the common prefix of $w[1..j]^R$ and $w[i..n]$ is denoted by $LPrF_g^<[i]$, while $B_<[i]$ denotes $j$. If $w[i..n]$ has no common prefix with any factor $w[1..j]^R$ with $i - j > g(i)$ and $Rank_R[j] < Rank[i]$, then $LPrF_g^<[i] = 0$ and $B_<[i]$ is not defined; as a convention, we set $B_<[i]$ to $-1$

Afterwards, we compute for each $i$ the prefix $w[1..j]$, such that $i - j > g(i)$, $w[1..j]^R$ occurs after $w[i..n]$ in $\mathcal{L}$ (i.e., $Rank_R[j] > Rank[i]$), and the length of the common prefix of $w[1..j]^R$ and $w[i..n]$ is greater or equal to the length of the common prefix of $w[1..j']^R$ and $w[i..n]$ for $j'$ such that $Rank_R[j'] > Rank[i]$; for the prefix $w[1..j]$ as above, the length of the common prefix of $w[1..j]^R$ and $w[i..n]$ is denoted by $LPrF_g^>[i]$, while $B_>[i]$ denotes $j$. Clearly, $LPrF_g[i] = \max\{LPrF_g^<[i], LPrF_g^>[i]\}$. If $w[i..n]$ has no common prefix with any factor $w[1..j]^R$ with $i - j > g(i)$ and $Rank_R[j] > Rank[i]$, then $LPrF_g^>[i] = 0$ and $B_>[i]$ is set to $-1$

For simplicity, we just present an algorithm computing $LPrF_g^<[\cdot]$ and $B_<[\cdot]$. The computation of $LPrF_g^>[\cdot]$ is performed in a similar way.

The main idea behind the computation of $LPrF_g^<[i]$, for some $1 \leq i \leq n$, is that if $w[1..j_1]$ and $w[1..j_2]$ are such that $Rank_R[j_2] < Rank_R[j_1] < Rank[i]$ and $j_1 < j_2 < i$ then definitely $B_<[i] \neq j_2$. Indeed, $LCP(w[1..j_2]^R, w[i..n]) < LCP(w[1..j_1]^R, w[i..n])$, and, moreover, $i - j_2 < i - j_1$. So, if $i - j_2 > g(i)$ then also $i - j_1 > g(i)$, and it follows that $LPrF_g^<[i] \geq LCP(w[1..j_1]^R, w[i..n]) > LCP(w[1..j_2]^R, w[i..n])$, so $B_<[i]$ cannot be $j_2$. This suggests that we could try to construct for each $i$ an ordered list $\mathcal{A}_i$ of all the integers $j \leq n$ such that $Rank_R[j] < i$ and moreover, if $j_1$ and $j_2$ are in $\mathcal{A}_i$ and $j_1 < j_2$ then also $Rank_R[j_1] < Rank_R[j_2]$.

Now we describe how to implement this. Let us now consider $i_1$ and $i_2$ which occur on consecutive positions of the suffix array of $w$, such that $Rank[i_1] < Rank[i_2]$. The list $\mathcal{A}_{i_2}$ can be obtained from $\mathcal{A}_{i_1}$ as follows. We consider one by one, in the order they appear in $Rank_R[\cdot]$, the integers $j$ such that $Rank[i_1] < Rank_R[j] < Rank[i_2]$, and for each of them update a temporary list $\mathcal{A}$, which initially is equal to $\mathcal{A}_{i_1}$. When a certain $j$ is considered, we delete from the right end of the list $\mathcal{A}$ (where $\mathcal{A}$ is ordered increasingly from left to right) all the values $j' > j$; then we insert $j$ in $\mathcal{A}$. When there are no more indices $j$ that we need to consider, we set $\mathcal{A}_{i_2}$ to be equal to $\mathcal{A}$. It is clear that the list $\mathcal{A}_{i_2}$ is computed correctly.

Now, for each $i$ we need to compute the greatest $j \in \mathcal{A}_i$ such that $j < i - g(i)$. As $\mathcal{A}_i$ is ordered increasingly, we could obtain $j$ by performing a predecessor search on $\mathcal{A}_i$ (that is, binary searching the greatest $j$ of the list, which is smaller than $i - g(i)$), immediately after we computed it, and save the answer in $B_<[i]$. However, this would be inefficient. Before proceeding, we note that if we compute the lists $\mathcal{A}_i$ for the integers $i$ in the order they appear in the suffix array of $w$, then it is clear that the time needed to compute all these lists is linear. Indeed, each $j \leq n$ is introduced exactly once in the temporary list, and then deleted exactly once from it. Doing the above mentioned binary searches would add up to a total of $\mathcal{O}(n \log(n))$. We can do better than that.

Now we have reduced the original problem to a data-structures problem. We have to maintain an increasingly ordered (from left to right) list $\mathcal{A}$ of numbers (at most $n$, in the range $\{1, \ldots, 2n\}$, each two different), subject to the following update and query operations. This list can be updated by the following procedure: we are given a number $j$, we delete from the right end of $\mathcal{A}$ all the numbers greater than $j$, then we append $j$ to $\mathcal{A}$. By this update, the list remains increasingly ordered. The following queries can be asked: for a given $\ell$, which is the rightmost number of $\mathcal{A}$, smaller than $\ell$? We want to maintain this list while $n$ updates are executed, and $n$ queries are asked at different moments of time. Ideally, the total time we can spend in processing the list during all the updates should be $\mathcal{O}(n)$, and, after all the updates are processed, we should be able to provide in $\mathcal{O}(n)$ time the correct answer for all the queries (i.e., if a certain query was asked after $k$ update operations were performed on the list, we should return the answer to the query with respect to the state of the list after those $k$ update operations were completed).

Next we describe our solution to this problem.

We use a dynamic tree data-structure to maintain the different stages of $\mathcal{A}$. Initially, the tree contains only one path: the root 0 and the leaf $2n + 1$. When an update is processed, in which a number $j$ is added to the list, we go up the rightmost path of the tree (from leaf to root) until we find a node with a value smaller than $j$. Then $j$ becomes the rightmost child of that node (i.e., $j$ is a leaf). Basically, the rightmost path of a tree after $k$ updates contains the elements of $\mathcal{A}$ after those $k$ updates, preceded by 0. When a query is asked we associate that query with the leaf corresponding to the rightmost leaf of the tree at that moment. In this way, we will be able to identify, after all updates were processed, the contents of the list at the moments of time the queries were, respectively, asked: we just have to traverse the path from the node of the tree associated to that query (this node was a leaf when the query is asked, but after all the updates were processed might have become an internal node) to the root.

The tree can be clearly constructed in linear time: each node is inserted once on the rightmost tree, and it disappears from this rightmost tree (and will not be reinserted there) when a smaller value is inserted in the tree.

In this new setting, the queries can be interpreted as weighted level ancestor queries on the nodes of the constructed tree (where the weight of an edge is the difference between the two nodes bounding it). Considering that the size of the tree is $\mathcal{O}(n)$, all weights are also $\mathcal{O}(n)$, there are $\mathcal{O}(n)$ queries, and these queries are to be answered off-line, it follows (see Preliminaries) that we can return the answers to all these queries in $\mathcal{O}(n)$ time.

This completes the linear solution to our problem. □

To solve Problem 4.3.2(b) we use the following lemma:

**Lemma 4.3.1** *Given a word $w$, let $L[i] = \min\{j \mid j < i, LCP(j, i) \geq LCP(k, i)$ for all $k < i\}$. The array $L[\cdot]$ can be computed in linear time.*

Another lemma shows how the computation of the array $LPF_g[\cdot]$ can be connected to that of the array $L[\cdot]$. For an easier presentation, let $B[i]$ denote the leftmost starting position of the longest factor $x_i$ that occurs both at position $i$ and at a position $j$ such that $j + |x_i| \leq i - g(i)$; if there is no such factor $x_i$, then $B[i] = -1$. In other words, the length of the factor $x_i$ occurring at position $B[i]$ gives us $LPF_g[i]$. In fact, $LPF_g[i] = \min\{LCP(B[i], i), i - g(i) - B[i]\}$.

Now, let $L^1[i] = L[i]$ and $L^k[i] = L[L^{k-1}[i]$, for $k \geq 2$; also, we define $L^+[i] = \{L[i], L[L[i]], L[L[L[i]]], \ldots\}$.

The following lemma shows the important fact that $B[i]$ can be obtained just by looking at the values of $L^+[i]$. More precisely, $B[i]$ equals $L^k[i]$, where $k$ is obtained by looking at the values $L^j[i] \leq i - g(i)$ and taking the one such that the factor starting on it and ending on $i - g(i) - 1$ has a maximal common prefix with $w[i..n]$. Afterwords, Theorem 4.3.4 shows that this check can be done in linear time for all $i$, thus, solving optimally Problem 4.3.2.

**Lemma 4.3.2** *For a word $w$ of length $n$ and all $1 \leq i \leq n$ such that $B[i] \neq -1$, we have that $B[i] \in L^+[i]$.*

**Theorem 4.3.4** *Problem 4.3.2(b) can be solved in linear time.*

*Proof.*

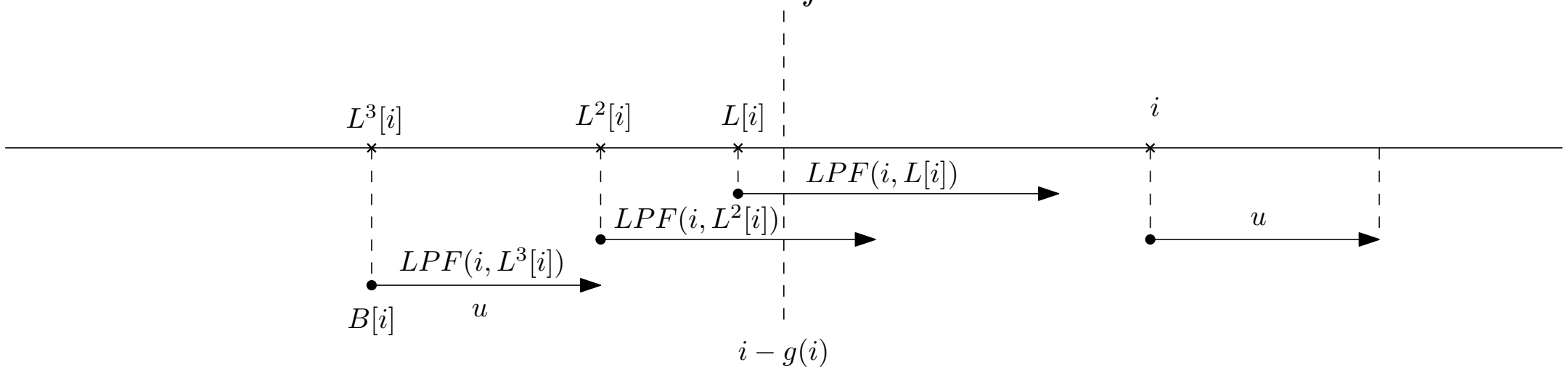

Figure 4.3: Proof of Theorem 4.3.4: We measure the length of the common prefix between each factor starting on $L^k[i]$ and ending on $i - g(i)$ and the suffix $w[i..n]$. The starting position of the factor that produces the longest such common prefix is $B[i]$. Also, this longest common prefix defines the arm $u$ of the gapped repeat $uvu$.

The main idea in this proof is that, to compute $LPF_g[i]$, it is enough to check the elements $j \in L^+[i]$ with $j \leq i - g(i)$, and choose from them the one for which $\min\{LCP(j,i), i - g(i) - j\}$ is maximum; this maximum will be the value we look for. In the following, we show how we can obtain these values efficiently.

First, if $j = L[i]$ for some $i$, we define the value $end[j] = k \leq LCP(L[j], i)$, where $|w[j..k]| \leq \min\{LCP(L[j], i), k - L[j]\}$. Basically, for each position $k' \leq k$, the longest factor $x$ starting at $L[j]$ and ending on $k'$ which also occurs at position $i$ is longer than any factor starting on position $j$ and ending on $k'$ which also occurs at position $i$. Now we note that if $j \in L^+[i]$ is the greatest element of this set such that $end[j] \leq i - g(i) - 1$, then $LPF_g[i] = \min\{LCP(j,i), i - g(i) - j\}$. Clearly, $end[j]$ can be computed in constant time for each $j$.

To be able to retrieve efficiently for some $i$ the greatest element of this set such that $end[j] \leq i - g(i) - 1$ we proceed as follows.

First we define a disjoint-set union-find data structure on the universe $U = [1, n+1)$, where the unions can be only performed between the set containing $i$ and that containing $L[i]$, for all $i$. Initially, each number between 1 and $n$ is a singleton set in this structure. Moreover, our structure fulfills the conditions that the efficient union-find data structure of [31] should fulfill.

Further, we sort in linear time the numbers $i - g(i)$, for all $i$; we also sort in linear time the numbers $end[k]$ for all $k \leq n$. We now traverse the numbers from $n$ to 1, in decreasing order. When we reach position $j$ we check whether $j$ equals $end[k]$ for some $k$; if yes, we unite the set containing $k$ with the set containing $end[k]$ for all $k$ such that $end[k] = j$. Then, if $j = i - g(i)$ for some $i$, we just have to return the minimum of the set containing $i$; this value gives exactly the greatest element $j \in L^+[i]$ such that $end[j] \leq i - g(i) - 1$. So, as described above, we can obtain from it the value of $LPF_g[i]$. The computation of this array follows from the previous remarks.

In order to evaluate the complexity of our approach, note that we do $\mathcal{O}(n)$ union operations and $\mathcal{O}(n)$ find queries on the union-find data structure. By the results in [31], the time needed to construct the union-find data structure and perform these operations on it is also $\mathcal{O}(n)$. From every find query we get in constant time the value of a element $LPF_g[i]$. So the solution of Problem 4.3.2 is linear. □

### Long Armed Repeats and Palindromes

In this section, we solve Problems 4.3.3(a) and 4.3.3(b).

Recall that a long armed palindrome (respectively, repeat) $w[i..j]vw[i'..j']$ is called maximal if the arms cannot be extended to the right or to the left: neither $w[i..j+1]v'w[i'-1..j']$ nor $w[i-1..j]vw[i'..j'+1]$ (respectively, neither $w[i..j+1]v'w[i'..j'+1]$ nor $w[i-1..j]v''w[i'-1..j']$) are long armed palindromes (respectively, repeats). Our solutions rely on the results of [43, 44]. In [43] one proposes an algorithm that, given a word of length $n$, finds the set $S$ of all its factors which are maximal long armed palindromes in $\mathcal{O}(n + |S|)$ time. In [44] the set of maximal long armed repeats with nonempty gap is shown to be of linear size and is computed in $\mathcal{O}(n)$ time for input words over integer alphabets.

In the following, we essentially show that given the set $S$ of all factors of a word which are maximal long armed palindromes (respectively, repeats) we can compute the array $LPal$ (respectively, $LRep$) for that word in $\mathcal{O}(n + |S|)$ time.

**Theorem 4.3.5** *Problem 4.3.3(a) can be solved in $\mathcal{O}(|S| + n)$, where $S$ is the set of all factors of the input word which are maximal long armed palindromes.*

*Proof.*

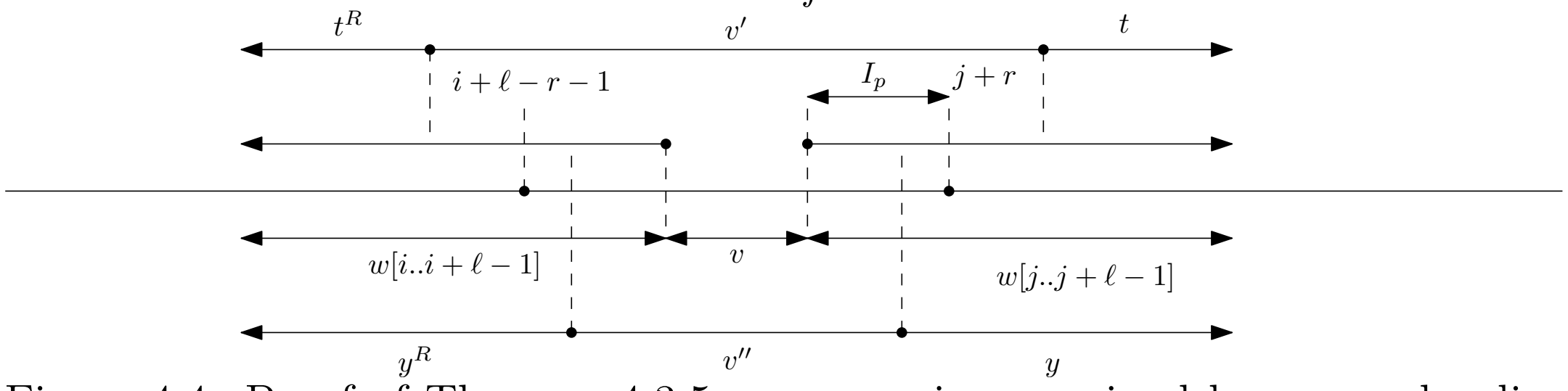

Figure 4.4: Proof of Theorem 4.3.5: $p = uvu$ is a maximal long armed palindrome with $|v| = \delta$ and $|u| = \ell$. We define $r = \frac{\ell-\delta}{3}$. This allows us to define the interval $I_p$, where the right arm of a long armed palindrome obtained from $p$ may start. Further, $y^R v'' y$ is an example of such a long armed palindrome; $t^R v' t$ is a gapped palindrome obtained from $p$, which is not long armed: it's right arm does not start in $I_p$. The interval $I_p$ gets weight $j + \ell - 1$.

We assume that we are given an input word $w$, for which the set $S$ is computed, using the algorithm from [43].

Let us consider a maximal long armed palindrome $w[i..i+\ell-1]vw[j..j+\ell-1]$, with $w[i..i+\ell-1]^R = w[j..j+\ell-1]$. For simplicity, let us denote by $\delta = |v| = j - i - \ell$, the length of the gap; here, $i$ and $j + \ell - 1$ will be called the *outer ends* of this palindrome, while $j$ and $i + \ell - 1$ are the *inner ends.*

It is not hard to see that from a maximal palindrome one can get a family of long armed palindromes whose arms cannot be extended by appending letters simultaneously to their outer ends. We now show how this family of long armed palindromes can be computed. Intuitively, we extend simultaneously the gap in both directions, decreasing in this way the length of the arms of the palindrome, until the gap becomes longer than the arm. The longest possible such extension of the gap can be easily computed. Indeed, let $r = \left\lfloor \frac{\ell-\delta}{3} \right\rfloor$. It is not hard to check that for $r' \leq r$ we have that $w[i..i + \ell - r' - 1]v'w[j + r'..j + \ell - 1]$, with $v' = w[i+\ell-r'-1..i+\ell-1]vw[j..j+r'-1]$, is a long armed palindrome whose left arm cannot be extended by appending letters to their outer ends. For $r' > r$ we have that $w[i..i+\ell-r'-1]v'w[j+r'..j+\ell-1]$, with $v' = w[i+\ell-r'-1..i+\ell-1]vw[j..j+r'-1]$, is still a gapped palindrome, but it is not long armed anymore. So, for a maximal long armed palindrome $p = w[i..i + \ell - 1]vw[j..j + \ell - 1]$, we associate the interval $I_p = [j, j+r]$, and associate to it a weight $g(I_p) = j+\ell-1$. Intuitively, we know that at each position $j' \in I_p$ there exists a factor $u$, ending at position $j + \ell - 1$, such that $u^R v$ is a suffix of $w[1..j' - 1]$ for some $v$.

On the other hand, if $u$ is the longest factor starting at some position $j' \leq n$ such that $u^R v$ is a suffix of $w[1..j-1]$, then the factor $w[i..j] = u^R vu$ is, in fact, a maximal long armed palindrome $x^R yx$(i.e., $u^R$ is a prefix of $x^R$ and $u$ is a suffix of $x$). In other words, $u$ and $u^R$ could be extended simultaneously inside the gap, but not at the outer ends.

Consequently, to compute $LPal[i]$ for some $i \leq n$ we have to find the long armed palindromes $p \in S$ for which the interval $I_p$ contains $i$. Then, we identify which of these intervals has the greatest weight. Say, for instance, that the interval $I_p$m which contains $i$, is the one that weight maximal weight $k$ from all the intervals containing $i$. Then $LPal[j] = k - j + 1$. Indeed, from all the factors $u$ starting at position $j$, such that $u^R v$ is a suffix of $w[1..j'-1]$ for some $v$, there is one that ends at position $k$, while all the other end before $k$ (otherwise, the intervals associated, respectively, to the maximal long armed palindromes containing each of these factors $u^R vu$ would have a greater weight). So, the palindrome ending at position $k$ is the longest of them all.

This allows us to design the following algorithm for the computation of $LPal[j]$. We first use the algorithm of [43] to compute the set $S$ of all maximal long armed palindromes of $w$. For each maximal long armed palindrome $p = w[i..i+\ell-1]w[i+\ell..j-1]w[j..j+\ell-1]$, we associate the interval $I_p = [j, j+r]$, where $r = \left\lfloor \frac{\ell-\delta}{3} \right\rfloor$ and $\delta = j-i-\ell$, and associate to it the weight $g(I_p) = j+\ell-1$. We process these $|S|$ intervals, with weights and bounds in $[1, n]$, in $\mathcal{O}(n + |S|)$ time as in Lemma 2.4.1, to compute for each $j \leq n$ the maximal weight $H[j]$ of an interval containing $j$. Then we set $LPal[j] = H[j] - j + 1$.

The correctness of the above algorithm follows from the remarks at the beginning of this proof. Its complexity is clearly $\mathcal{O}(|S| + n)$. □

The solution of Problem 4.3.3(b) is very similar. Thus, we obtain the following result:

**Theorem 4.3.6** *Problem 4.3.3(b) can be solved in $\mathcal{O}(n)$.*

*Proof.* We first use the algorithm of [44] to compute the set $S$ of all maximal long armed repeats with nonempty gap of $w$. For each maximal long armed repeat $p = w[i..i+\ell-1]w[i+\ell..j-1]w[j..j+\ell-1]$, we associate the interval $I_p = [j, j+r]$, where $r = \left\lfloor \frac{\ell-\delta}{2} \right\rfloor$ and $\delta = j-i-\ell$, and associate to it the weight $g(I_p) = j+\ell-1$. We process these $|S|$ intervals, with weights and bounds in $[1, n]$, in $\mathcal{O}(n + |S|)$ time as in Lemma 2.4.1, to compute for each $j \leq n$ the maximal weight $H[j]$ of an interval containing $j$. Now, we use Lemma 2.4.2 to compute the values $SC[j]$ for each $j \leq n$. We set $LRep[j] = \max\{H[j] - j + 1, SC[j]\}$.

The complexity of this algorithm is $\mathcal{O}(n)$, as $|S| \in \mathcal{O}(n)$ (see [44]).

The correctness of the algorithm follows from the following remark: for a maximal long armed repeat $p = w[i..i + \ell - 1]w[i + \ell..j - 1]w[j..j + \ell - 1]$ let $r = \left\lfloor \frac{\ell-\delta}{2} \right\rfloor$, where $\delta = j - i - \ell$. Then the factors $w[i + r'..i + \ell - 1]w[i + \ell..j + r' - 1]w[j + r'..j + \ell - 1]$ are long armed repeats for all $r' \leq r$, whose right arm cannot be extended anymore to the right. Moreover, the factors $w[i + r'..i + \ell - 1]w[i + \ell..j + r' - 1]w[j + r'..j + \ell - 1]$ are gapped repeats which are not long armed for all $r' > r$. The rest of the arguments showing the soundness of our algorithm are similar to those of Theorem 4.3.5. □

# Chapter 5

# Future Work

## 5.1 Open Problems

There are a lot of open problems that arise from the work in this thesis. They range from formal languages open problems, that are related to duplication languages, to algorithmic and combinatorial problems.

In section 2.4 we prove a few lemmas that provide information about squares in a word using a new solving technique. The following questions seem natural: What other results can we obtain using this technique? Can one for example obtain the Curling number [9] for every prefix of the word in linear time? Can we obtain the longest overlap starting or finishing in each position?

In section 3.1 we discuss about duplication languages obtained through three operations. An interesting direction of research would be considering the languages obtained by the reverse operations of the three. Is $\Omega_k^{-1*}(Reg) \subseteq Reg$, where $\Omega \in \{BPSD, PSD, PSSC\}$?

In section 3.2 many questions remained unanswered. For example, it would be interesting to give a limit for the number of primitive *BPSD* or *PSSC* roots in a word. Other questions are related to the possibility of improving the complexity of the algorithms we obtained: Can we compute in linear time the longest root of a word in relation to the *PSSC* operation (instead of $\mathcal{O}(n\log(n))$? Can we find the common ancestors of two words in relation to the *PSD* operation in $O(n^2)$?

Closely related to the common ancestor problem discussed in the thesis is the common descendant problem:

**Problem 5.1.1** *Given two words $x$ and $y$, find a word $z$ such that $z \in \Theta^*(x)$ and $y$ and $z \in \Theta^*(y)$, where $\Theta \in \{PSD, PSD_k, PSSC\}$.*

This problem seems really difficult to tackle and we have more questions as open problems than answers regarding it.

**Open Problem 5.1.1** *Is it decidable if two words have a common descendant in the case of the prefix-suffix duplication operation?*

**Open Problem 5.1.2** *Is it decidable if two words have a common descendant in the case of the prefix-suffix square completion operation?*

Unlike the two other operations we study in our thesis, the bounded prefix-suffix operation generates regular languages, thus, the following result is an immediate consequence of regular languages properties.

**Lemma 5.1.1** *The common descendant problem is decidable in the case of the bounded prefix-suffix operation and there is an exponential algorithm to find all common descendants.*

*Proof.* We have already showed that $PSD_k^*(x)$ is regular, thus, we can construct an FSA to recognize it, as we can for $PSD_k^*(y)$. We then intersect the two automata to obtain the intersection language. All words in the intersection will be common descendants. □

It would be really interesting to prove the problem is NP-hard, or to find a polynomial algorithm for this problem, thus we give the following open problem:

**Open Problem 5.1.3** *Is the common descendant problem in the case of the prefix-suffix square completion operation NP-hard?*

Another really interesting questions is if we could find all ancestors of a word in relation to the *SD* operation in linear time. We think that some suffix array construction algorithms along with the techniques used in [46, 61] could give a linear solution to this problem. If a linear solution to this problem would appear than we would be able to improve a lot of the algorithms in the thesis.

In section 3.3 we discussed about duplication distances, two open problems remained. One was of algorithmic nature: can we obtain a more efficient algorithm for the prefix-suffix square completion distance problem? The other question was related to the distance between duplication languages. Can the distance between two duplication languages in the case of the *PSD* and *PSSC* languages, be computed? We obtained such a result only for the *BPSD* operation.

In section 4.1 we discussed about infinite word generation using the *PSSC* operation and partly using the *PSD* operation. We gave a conjuncture about generating the infinite Fibonacci word using the *BPSD* operation. A lot of other results in this direction are open to exploring.

In section 4.2 we talked about prefix-suffix square free words. A problem that remained unsolved is computing the number of prefix-suffix square free words of length $n$. Now we give a conjecture about the number of prefix-square free words on length $n$ based on computer simulations.

**Conjecture 5.1.1** *There are $Curling(n, 1)$ prefix-square-free words on length $n$?*

Of course the main question is how many prefix-suffix-square free words of length $n$ exist? In this direction, one might want to find the $l$-th smallest prefix-square or prefix-suffix-square free word.

Another direction of research is computing a lot of the results in the section in relation to the *BPSD* operation. For example, what is the maximum number of bounded prefix-suffix-square free factors of a word?

We also feel that a really interesting problem is the factorization of a word in squares. We have obtained a $\mathcal{O}(n \log n)$ solution for the problem. Can this problem be solved in linear time?

Lastly, in section 4.3 we computed the longest previous factor $(LPrF)$ table, and longest previous reverse factor $(LPrF)$ table in three different settings. Other limitations for the gap can be taken into account, for example, a very natural direction for research is considering the gap to be limited by two functions. That is: searching for factors of the form $uvu$ and $u^r vu$ such that $g(i) < u < G(i)$ and computing the longest previous factor $(LPrF)$ table and the longest previous reverse factor $(LPrF)$ table in this situation.